\newcommand{\<}{\langle}
\renewcommand{\>}{\rangle}
\newcommand{\beq}{\begin{equation}}
\newcommand{\eeq}{\end{equation}}
\newcommand{\beqn}{\begin{eqnarray}}
\newcommand{\eeqn}{\end{eqnarray}}

\newcommand{\pslash}{\not\hspace{-3pt}p}
\documentclass[11pt]{article}
\usepackage{chir_layout}
\begin{document}
\begin{flushright}
Bicocca FT-03-15 \\
ROM2F/2003/15 \\
\end{flushright}
\vspace{1cm}
\centerline{\LARGE{\bf Chirally improving Wilson fermions}}
\smallskip\smallskip
\centerline{\LARGE{\bf I. O($a$) improvement}}
\vskip 1.0cm
\centerline{ {\large{R. Frezzotti}}\,$^{{a)}}$ and
{\large{G.C. Rossi}}\,$^{{b)}}$} \vskip 0.5cm
\centerline{$^{a)}$\small{INFN, Sezione di Milano}}
\centerline{\small{Dipartimento di Fisica, Universit\`a di Milano
``{\it Bicocca}''}}
\centerline{\small{Piazza della Scienza 3 - 20126 Milano (Italy)}}
\smallskip  \centerline{$^{b)}${\small{Dipartimento
di Fisica, Universit\`a di  Roma ``{\it Tor Vergata}''}}}
\centerline{\small{INFN, Sezione di Roma 2}}
\centerline{\small{Via della Ricerca Scientifica - 00133 Roma (Italy)}}
\vskip .5cm
\centerline{\bf ABSTRACT}
\begin{quote}
{\small{We show that it is possible to improve the chiral behaviour
and the approach to the continuum limit of correlation functions
in lattice QCD with Wilson fermions by taking arithmetic averages
of correlators computed in theories regularized with Wilson terms of
opposite sign. Improved hadronic masses and matrix elements can be
obtained by similarly averaging the corresponding physical quantities
separately computed within the two regularizations. To deal with the
problems related to the spectrum of the Wilson--Dirac operator,
which are particularly worrisome when Wilson and mass terms are such as
to give contributions of opposite sign to the real part of the eigenvalues,
we propose to use twisted-mass lattice QCD for the actual computation of
the quantities taking part to the averages. The choice $\pm \pi/2$
for the twisting angle is particularly interesting, as O($a$)
improved estimates of physical quantities can be obtained even without
averaging data from lattice formulations with opposite Wilson terms.
In all cases little or no extra computing power is necessary, compared
to simulations with standard Wilson fermions or twisted-mass lattice QCD.}}
\end{quote}
\newpage

\section{Introduction} \label{sec:INTRO}

The two major limitations of the standard Wilson formulation of lattice
QCD (LQCD)~\cite{W} are related to the non-negligible magnitude of the
systematic errors associated to discretization effects and
the complicated pattern of operator mixing induced by the explicit
breaking of chiral symmetry due to the presence of the Wilson term
in the lattice action~\cite{KS,GHS,BMMRT}. Consequently a lot of effort
has gone in the direction of reducing errors related to the finiteness of
the lattice spacing, $a$, with the full implementation of the Symanzik
program~\cite{SYM} up to O($a$) corrections~\cite{SW,HMPRS,LUS1,LUS2,OTHERS}
and in developing formulations in which chiral symmetry is exact
at the level of the regularized action~\cite{GW,OVER,WALL,HAS}.

Elimination of O($a$) effects can be in principle fully achieved. The standard
approach developed in refs.~\cite{SW,HMPRS,LUS1,LUS2,OTHERS} requires,
besides a suitable improvement of operators, the introduction in the
standard Wilson LQCD action of the so-called Sheikholeslami--Wohlert
(SW) term,
\beq S^{\rm{L}}_{\rm{SW}}=a^5 c_{\rm{SW}}(g^2_0) \sum_x
\,\bar\psi(x) \frac{i}{4}\sigma_{\mu\nu} P_{\mu\nu}(x)\psi(x)
\, , \label{SWSA} \eeq
with an appropriate value of the coefficient function $c_{\rm{SW}}(g^2_0)$
depending on the detailed form of the pure gauge action and
Wilson term~\footnote{In eq.~(\ref{SWSA}) $P_{\mu\nu}$ is any
lattice discretization of the gauge field strength~\cite{BGO}, $g_0$ is the
bare gauge coupling and we define $\sigma_{\mu\nu} = i[\gamma_\mu ,
\gamma_\nu ]/2$, with $\{ \gamma_\mu , \gamma_\nu \} = 2 \delta_{\mu\nu}$
and $\gamma_\mu^\dagger = \gamma_\mu$.}. All the above modifications,
particularly in the unquenched case, make O($a$) improved computations
significantly more CPU-time demanding than standard Wilson simulations.
This is even more so if the evaluation of off-shell or gauge non-invariant
correlators is attempted~\cite{OTHERS}.

Fully chiral invariant lattice actions have been also
constructed~\cite{OVER,WALL,HAS}, where the Ginsparg--Wilson
relation~\cite{GW} is exactly implemented and a first exploration of
their numerical potentiality has been carried
out~\cite{NUMCHIO,NUMCHIDW,NUMCHIFP}. Overlap, domain-wall and
fixed-point fermions have been put at work with rather encouraging
results, though the extra computational effort required with respect
to the standard Wilson case is in most cases very large. Of course this
comparison is not totally fair because the quality of data produced
with fermions obeying the Ginsparg--Wilson relation looks superior to
that usually obtainable with standard Wilson fermions.

If one does not want to make use of chirally invariant fermions, because of
their relatively heavy computational cost, a nice alternative is to use
twisted-mass lattice QCD (tm-LQCD)~\cite{FGSW,FREZ}, which represents a
cheap but effective solution to the problem of exceptional configurations
occurring in quenched (or partially quenched) simulations. Recently it has
been shown, in fact, that tm-LQCD can be judiciously used to alleviate both
the numerical difficulties associated to O($a$) discretization
errors~\cite{FSW,FS} and the mixing problem of Wilson fermions in the
calculation of $B_K$ and the CP-conserving $\Delta S=1$ weak matrix
elements~\cite{FGSW,GHPSV,PSV}.

The present one is the first of two companion papers in which we propose a
simple strategy to get from simulations employing Wilson fermions (or rather,
as we shall see, twisted Wilson fermions) lattice data that are free
of O($a$) discretization effects and have a somewhat smoother and more
chiral behaviour near the continuum limit than data obtained in unimproved
standard Wilson or tm-LQCD simulations.

In this paper we show that, if quarks are arranged in $SU(2)_{\rm{f}}$ flavour
doublets, O($a$) discretization effects are absent from the average of
correlators (Wilson average, $WA$) computed with lattice actions having
Wilson terms of opposite sign and a common value of the subtracted
(unrenormalized) lattice quark mass
\beq
m_q=M_0-M_{\rm{cr}}\, ,\label{QMASS} \eeq
where $M_0$ is the bare mass. For short in the following
we will occasionally call $m_q$ the ``excess" quark mass.

Absence of O($a$) discretization errors in $WA$'s is proved by referring to
the Symanzik expansion of (connected, on-shell) lattice correlators in terms
of continuum Green functions and exploiting the relations derived by matching
the ``${\cal{R}}_5$-parity'' of lattice correlators under the transformation
\begin{equation}
{\cal{R}}_5 : \left \{\begin{array}{ll}
\psi&\rightarrow\psi'=\gamma_5 \psi  \\\\
\bar{\psi}&\rightarrow\bar{\psi}'=-\bar{\psi}\gamma_5
\end{array}\right . \label{PSIBAR} \end{equation}
to the ${\cal{R}}_5$-parities of the related continuum Green functions.
${\cal{R}}_5$ is an element  of the chiral $SU(2)_L\otimes SU(2)_R$ group,
for it can be expressed as the product of the following three transformations
$u_V^{(1)}(\pi)=\exp(i\pi\tau_1/2)$,
$u_V^{(2)}(\pi)=\exp(i\pi\tau_2/2)$,
$u_A^{(3)}(\pi)=\exp(i\pi\gamma_5\tau_3/2)$,
with $\tau_1,\tau_2,\tau_3$ the Pauli matrices. Obviously
$[{\cal{R}}_5]^2=1\!\!1$.

The improvement of $WA$'s is the result of the use of simple dimensional and
symmetry considerations combined with certain algebraic identities holding
between pairs of correlators computed with opposite signs of $M_0$ and $r$.
These identities come from a generalized spurion analysis, which in
turn follows from the fact that the lattice fermionic action is left
invariant if the field transformation~(\ref{PSIBAR}) is accompanied by a
simultaneous change of sign of $M_0$ and the coefficient, $r$, sitting
in front of the Wilson term. {}From these considerations one can prove that
the Symanzik coefficients necessary to match lattice correlators to the
continuum ones have such properties under $r\to-r$ that all O($a$)
corrections to the continuum result cancel in $WA$'s.

An important point to stress is that, in order for the formal counting
of powers of $a$ yielded by the relevant Symanzik expansion to be
meaningful, it is necessary that one is dealing with expectation values
of multiplicatively renormalizable (m.r.) lattice operators.

The obvious problem with the approach we have described is the existence
of real eigenvalues of the massless Wilson--Dirac (WD) operator
with magnitude much smaller than the inverse lattice cutoff. For realistic
values of the lattice spacing, the existence of such spurious eigenvalues
(in particular those that are negative at $m_q=0$)
has been shown to be at the origin of the occurrence of zero modes
of the massive ($m_q >0$) WD operator~\footnote{For a first hint to the
existence of such a problem see ref.~\cite{PPR}.} on the so-called
``exceptional'' configurations~\cite{BETAL,SETAL}. The problem becomes more
severe when the relative sign of the coefficients of Wilson and quark mass
terms is such that they give opposite contributions to the real part of the
spectrum of the WD operator. The remedy to this situation, which we will for
short refer to as the problem of ``spurious fermionic modes'', is to use
tm-LQCD~\cite{FGSW,FREZ}. Fully O($a$) improved lattice data
for energy levels (hence hadronic masses), matrix elements and
renormalization constants can be obtained, without the need of computing
anyone of the usual lattice improvement coefficients, if one takes averages
of correlators evaluated with tm-LQCD actions having twisted
Wilson terms of opposite sign.

The procedure we propose is very flexible as it is based on the
freedom of regularizing different flavours with Wilson terms of
different chiral phases. Its extension to mass non-degenerate
quarks is discussed in ref.~\cite{CAIRNS}. In the second of this
series of paper~\cite{FRTWO} we show that, without loosing O($a$)
improvement, this freedom can be appropriately exploited to have a
real and positive fermionic determinant, while at the same time
canceling all finite and power divergent contributions due to
mixing with operators of ``wrong chirality'' (i.e.\ those due to
the breaking of chiral symmetry induced by the presence of the
Wilson term in the lattice action) in the calculation of the
CP-conserving matrix elements of the effective weak Hamiltonian.
These results confirm and largely extend arguments given in
refs.~\cite{FGSW,PSV}.

The present paper is logically divided in two parts. In the first part
(sections~\ref{sec:SF} and~\ref{sec:IQCD}) we derive our results
momentarily ignoring problems with the spectrum of the WD operator. This key
issue will be taken up in the second part (sections~\ref{sec:TWM1}
to~\ref{sec:ASC}) where, after an introduction to tm-LQCD, we prove that
with appropriate choices of twisting none of the nice results derived for
Wilson fermions is lost. On the contrary, if the special choice $\pm \pi/2$
for the twisting angle is made, simple symmetry considerations imply
that many interesting physical quantities can be extracted from lattice
data that are automatically O($a$) improved with no need of taking any
$WA$~\footnote{We are indebted to M. Della~Morte for his precious help in
numerical tests concerning the problem with spurious fermionic modes of the
WD operator and the question of O($a$) improvement of hadronic masses.}.

The detailed plan of the paper is as follows. In section~\ref{sec:SF}, after
briefly reviewing the standard Wilson formulation of LQCD, we show that the
critical mass, $M_{\rm{cr}}$, changes sign if the sign in front of the Wilson
term is reversed. This is done by providing an argument both in perturbation
theory (PT) and in the presence of spontaneous chiral symmetry breaking
(S$\chi$SB). In section~\ref{sec:IQCD} we outline the basic ideas of the
present approach by showing that averages of lattice correlators
computed in theories with Wilson terms of opposite sign and a
common value of $m_q$ are O($a$) improved. We also prove that similar
averages of hadronic masses, matrix elements and renormalization constants
analogously lead to O($a$) improved quantities. We present
in section~\ref{sec:TWM1} a general discussion of tm-LQCD with special
attention to the non-singlet Ward-Takahashi identities (WTI's) of the
theory. In section~\ref{sec:TWM2} the idea of improvement through $WA$'s
is extended to tm-LQCD and the partial breaking of parity invariance that
occurs in this regularization is carefully analyzed. We show that it is
possible to attribute to the eigenstates of the transfer matrix a parity
label that in the continuum limit turns out to coincide with the standard
parity quantum number. In sect.~\ref{sec:ASC} we discuss the remarkable
simplifications that occur if the choice $\pm\pi/2$ for the twisting angle
is made and we prove that many interesting physical quantities are
automatically O($a$) improved. Conclusions and an outlook of future lines
of investigations can be found in section~\ref{sec:CONCL}. We leave in
a number of appendices certain more technical arguments and a detailed
discussion of the improvement of the non-singlet axial WTI's for standard
Wilson fermions.

\section{Wilson fermions and $\gamma_5$-chirality}
\label{sec:SF}

LQCD for one flavour doublet ($N_{\rm{f}}=2$) of quarks, as formulated by
Wilson~\cite{W}, is expressed in terms of the (Euclidean) partition function
\beq  Z_{\rm{LQCD}} = \int {\cal {D}} \mu (\psi, \bar\psi, U)
\, e^{-[S_{\rm{YM}}^{\rm{L}}(U)+S_{\rm{F}}^{\rm{LW}}(\psi, \bar\psi, U)]}\, ,
\label{GENFUN} \eeq
where ${\cal {D}}  \mu (\psi, \bar\psi, U)$ is the functional integration
measure, $S_{\rm{YM}}^{\rm{L}}(U)$ is any sensible lattice regularization
of the Yang-Mills action and
\beqn
&&S_{\rm{F}}^{\rm{LW}}(\psi, \bar\psi, U)=
a^4 \sum_x \Bigl[-{1\over 2a} \sum_\mu
[\bar\psi (x) U_\mu (x) (r-\gamma_\mu) \psi (x+\mu) + \nonumber
 \\ && +\bar\psi (x+\mu) U_\mu^{\dagger} (x)
(r+\gamma_\mu) \psi(x)]+ \bar\psi (x)(M_0 + {4r\over a})\psi
(x)\Bigr]\equiv\nonumber\\ &&\equiv a^4
\sum_x\,\bar\psi (x)
\Big{[}\frac{1}{2} \sum_\mu \gamma_\mu(\nabla^\star_\mu+ \nabla_\mu)-
a\frac{r}{2}\sum_\mu\nabla^\star_\mu\nabla_\mu+M_0\Big{]}\psi(x)
\label{FERACT}\eeqn
is the standard Wilson fermion action. In the last line of
eq.~(\ref{FERACT}) we have introduced the usual definitions of forward
and backward lattice covariant derivatives
\begin{eqnarray} \label{LATCOVDER}
&& \nabla_\mu \psi(x) \equiv \frac{1}{a}
\left[ U_\mu(x) \psi(x+a\hat\mu) - \psi(x) \right]
\nonumber \\
&& \nabla_\mu^* \psi(x) \equiv \frac{1}{a}
\left[ \psi(x) - U^\dagger_\mu(x-a\hat\mu) \psi(x-a\hat\mu) \right] \, .
\end{eqnarray}
In all our formulae the sum over flavour indices is understood.
In this paper we restrict to the case in which $M_0$ is taken to be
proportional to the unit matrix in flavour space. It is, however,
possible to extend the following arguments to encompass the case of
non-degenerate quark masses. This will be done in refs.~\cite{CAIRNS,FRTWO}.

The piece of the action proportional to $r$ is to the so-called
Wilson term. It is an ``irrelevant'' operator of mass dimension 5,
needed to solve the fermion doubling problem. Its presence
explicitly breaks the $SU(2)_L\otimes SU(2)_R$ chiral symmetry of
the (naive discretization of the) fermionic action. The symmetry
is only recovered, at tree level, when $M_0$ and $a$ go to $0$. In
higher orders of PT powers of the factor $a$ in front of the
Wilson term may be compensated by $1/a^p$ divergencies in loops,
leading to finite or even divergent $r$-dependent contributions.
This means that the formal chiral properties of continuum QCD are
lost: the limit $M_0 \rightarrow 0$ does not correspond to the
chiral limit and operators naively belonging to different chiral
representations mix among themselves. A complicated
renormalization procedure is needed in order to recover
perturbatively (and non-perturbatively) chiral symmetry in the
continuum limit~\cite{BMMRT}.

These are intrinsic problems that do not depend on how fast the
continuum is reached. The same kind of problems appears, in
fact, in the improved version of LQCD~\cite{SW,HMPRS,LUS1,LUS2}, even if,
in the limit $a \to 0$, O($a$) corrections are absent from lattice
estimates of physical quantities.

We are interested in computing expectation values of the type
($x_1\neq x_2\neq\ldots\neq x_n$)
\beqn &&\langle O(x_1,x_2,\ldots ,x_n) \rangle\Big{|}_{(r,M_0)}=
\label{EXPVAL}\\ &&=\frac{1}{Z_{\rm{LQCD}}} \int {\cal {D}} \mu
(\psi, \bar\psi, U)\, e^{- [S_{\rm{YM}}^{\rm{L}}+S_{\rm{F}}^{\rm{LW}}]} \,
O(x_1,x_2,\ldots ,x_n)\nonumber\, , \eeqn
with $O(x_1,x_2,\ldots ,x_n)$ a gauge invariant, multi-local and m.r.\
operator built in terms of the fundamental fields of the theory (quarks and
gluons). We will also assume that $O$ has in the continuum limit a Lorentz
(rather, in Euclidean metric, an $SO(4)$) covariant structure.
The subscript $(r,M_0)$ on the l.h.s.\ of eq.~(\ref{EXPVAL}) means that the
expectation value is taken with the specified values of the Wilson parameter,
$r$, and bare quark mass, $M_0$~\footnote{Throughout the paper to lighten
the notation we will drop, unless necessary for clarity, the dependence upon
the bare coupling constant, $g_0$. Also with a little abuse of notation we
will use the same symbol $O$ for the operator on the Hilbert space and the
function of the elementary fields that appear under the functional
integral.}. Here and in the rest of the paper we will always restrict to
connected correlation functions, though this will not be explicitly
indicated in our subsequent equations.

The key observation of this work is that under the ${\cal{R}}_5$
transformation~(\ref{PSIBAR}) the fermionic action~(\ref{FERACT})
goes into itself, if at the same time we change sign to the Wilson
term (i.e.\ to $r$) and $M_0$. In the spirit of the spurion
analysis, a quick way of studying the situation is to consider $r$
and $M_0$ momentarily as fictitious fields and consider the
combined transformation \beq
{\cal{R}}_5^{\rm{sp}}\equiv{\cal{R}}_5\times [r\rightarrow
-r]\times [M_0\rightarrow -M_0] \label{SPURM0}\eeq as a symmetry
of the lattice theory. In this situation it is easy to prove that
any m.r.\ operator will be either even or odd under
${\cal{R}}_5^{\rm{sp}}$. The argument goes as follows. Given the
bare operator, $O_B$, one can always assume that the latter is
either even or odd under ${\cal{R}}_5$, because
$[{\cal{R}}_5]^2=1\!\!1$. Since ${\cal{R}}_5^{\rm{sp}}$ is a
spurionic symmetry of the action, the renormalization procedure
will necessarily give rise to a m.r.\ operator, $O$, which under
${\cal{R}}_5^{\rm{sp}}$ has exactly the same parity $O_B$ has
under ${\cal{R}}_5$. This parity will be denoted in the following
as $(-1)^{P_{{\cal{R}}_5}[O]}$ and called the ${\cal{R}}_5$-parity
of $O$, while $P_{{\cal{R}}_5}[O]$ will be referred to as the
$\gamma_5$-chirality of $O$.

In general $O$ will be given by a linear combination of bare operators with
finite or power divergent coefficients having a parametric dependence on $r$
and $M_0$. In this linear combination operators that under ${\cal{R}}_5$ have
a parity opposite to that of $O_B$ (if any) will necessarily appear
multiplied by mixing coefficients odd under
$[r\rightarrow -r]\times [M_0\rightarrow -M_0]$.

This argument implies that the expectation value of a m.r.\
operator must satisfy the relation \beq \langle O(x_1,x_2,\ldots
,x_n) \rangle\Big{|}_{(r,M_0)}= (-1)^{P_{{\cal{R}}_5}[O]}\langle
O(x_1,x_2,\ldots ,x_n) \rangle\Big{|}_{(-r,-M_0)} \, ,
\label{KEYRELM0} \eeq where, we recall, the sign inversion of $r$
and $M_0$ has to be performed both in the action and in the mixing
coefficients. This result can be proved by performing the change
of fermionic integration variables induced by~(\ref{PSIBAR}) in
the functional integral defining the l.h.s.\ of
eq.~(\ref{KEYRELM0}).

{}From this discussion one also concludes that the renormalization constant
of $O$, $Z_O$, must be even under $r\rightarrow -r$ in order for the
renormalized operator, $\hat O= Z_O O$, to have the same $\gamma_5$-chirality
as $O$ (and $O_B$). We will see in sect.~\ref{sec:RC} how it is possible
to extract from lattice data non-perturbative estimates of renormalization
constants that are even in $r$ and O($a$) improved.

The interest of eq.~(\ref{KEYRELM0}) lies in the fact that from it one can
prove that discretization effects associated to the presence of the Wilson
term in the action, induce in correlators O($a$) corrections having
well defined properties under $r\rightarrow -r$ ($r$-parity for short).
The latter are such that averages of Green functions computed in
lattice theories with opposite values of $r$ have a faster approach
to the continuum limit and a better chiral behaviour than the two
correlators separately.

Ideas in this direction were already put forward in the literature long
time ago in the papers of ref.~\cite{J,AOK,DH}~\footnote{We thank S. Aoki
and S. Sint for having brought these papers to our attention.}.

To give a precise basis to the line of arguments sketched above,
it is necessary to start by establishing the $r$-parity property
of $M_{\rm{cr}}(r)$. Since Wilson's lattice
theory~(\ref{GENFUN}) is invariant under the spurionic
transformation ${\cal R}_5^{\rm sp}$ (see eq.~(\ref{SPURM0})), the
mass action counterterm $a^4\sum_x \bar\psi(x) M_{\rm{cr}}(r)
\psi(x)$ must be naturally chosen so as to maintain this
spurionic invariance.
From this simple but crucial remark, given the definition of
${\cal R}_5$, it is immediately concluded that the critical mass
must be an odd function of $r$ 
\beq
M_{\rm{cr}}(-r)=-M_{\rm{cr}}(r)\, . \label{MCR} \eeq For the
reader's convenience, we illustrate how this this key property is
born out by the definitions of $M_{\rm{cr}}$ that are commonly
employed either in PT or at the non-perturbative level, where
S$\chi$SB is expected to take place in the large volume regime.

To avoid mixing up different issues we will present the whole discussion
in this and next section by temporarily ignoring the problems with the
spectrum of the WD operator. We will take up this crucial question in the
second part of this paper beginning from sect.~\ref{sec:TWM1}.

\subsection{Critical mass in PT}
\label{sec:CRMASSPT}

We start from the formal definition of the lattice quark propagator,
$\Sigma^{-1}$, in momentum space
\begin{equation} [\Sigma (p;r,M_0)]^{-1}\equiv a^4\sum_x e^{ip\cdot x}
\<\psi(0)\bar\psi(x)\>\Big{|}_{(r,M_0)} \, .
\label{QPROP} \end{equation}
Performing the change of fermionic functional integration variables induced
by the transformation~(\ref{PSIBAR}), one gets the lattice identity
\begin{equation} \Sigma (p;r,M_0)=-\gamma_5\Sigma (p;-r,-M_0)\gamma_5 \, .
\label{QPROPT} \end{equation}
In the continuum inspired decomposition (where we ignore obvious
logarithmic factors irrelevant for this argument)
\begin{equation} \Sigma(p;r,M_0) = i \Sigma_1(p^2;r,M_0) \pslash +
\Sigma_2(p^2;r,M_0) +{\rm{O}}(a)\, , \label{DECOMP} \end{equation}
the critical mass is defined as the value of $M_0$ which solves the
equation
\begin{equation} \Sigma_2 (0;r,M_0)=0\, . \label{SMCR} \end{equation}
Introducing the decomposition~(\ref{DECOMP}) in~(\ref{QPROPT}), we
derive the simple relations
\beqn
&&\Sigma_1(p^2;r,M_0)=\Sigma_1(p^2;-r,- M_0)\\
&&\Sigma_2(p^2;r,M_0)=-\Sigma_2(p^2;-r,-M_0)\,  . \label{PSIG}
\eeqn To any order in perturbation theory $\Sigma_1$ and
$\Sigma_2$ are smooth functions of $M_0$ and $r\neq 0$, which are
defined for arbitrary values of $M_0$ and $|r|\leq 1$. It follows
that, if a certain value of $M_0$ is a solution of~(\ref{SMCR})
for a given $r$, then $-M_0$ solves the same equation with $r$
replaced by $-r$. In other words, barring unexpected patterns of
multiple solutions to eq.~(\ref{SMCR}), we recover
eq.~(\ref{MCR}), in agreement with the argument given in
ref.~\cite{AOK}.

\subsection{Critical mass beyond PT}
\label{sec:CRMASSB}

In the presence of S$\chi$SB $M_{\rm{cr}}$ can be defined as the value of
$M_0$ at which the pion mass vanishes. This definition is implemented
numerically by studying the large $t>0$ behaviour of the two-point correlator
\beq G_{\pi}(t;r,M_0)=a^3\sum_{\rm{\bf x}} \langle P^b({\bf x},t)
P^b(0)\rangle\Big{|}_{(r,M_0)}  \, , \label{P5P5} \eeq
where
\beq P^b=\bar{\psi}\gamma_5\frac{\tau_b}{2}\psi \label{DEFP} \eeq
is the pseudo-scalar quark density with flavour index $b$ ($b=1,2,3$).
Performing in the expectation value~(\ref{P5P5}) the change of functional
integration variables~(\ref{PSIBAR}), one finds
\beq G_{\pi}(t;r,M_0)=G_{\pi}(t;-r,-M_0)\, . \label{TRP5P5} \eeq
As implied by reflection positivity (see sect.~\ref{sec:CRP} below
and Appendix~B), we assume that the mass, $m_\pi$, of the lightest state
contributing to~(\ref{TRP5P5}) can be non-perturbatively, equally well
defined for positive and negative values of $M_0$ and for $|r|\leq 1$. Then
to comply with the behaviour implied by eq.~(\ref{TRP5P5}) in the
$t \rightarrow \infty $ limit, we must  have
\beq m_{\pi}(-r,-M_0)=m_{\pi}(r,M_0)\, . \label{MPIPI} \eeq
{}From the definition of $M_{\rm{cr}}$ the relation~(\ref{MCR})
again follows.

In deriving this non-perturbative result we have implicitly
assumed that, once $g_0^2$ and $r$ are given, there always exists
a {\em unique} value of $M_0$ where the pion mass vanishes.
Careful analyses of cutoff effects in lattice QCD with Wilson
fermions~\cite{AOK,SHSI} actually suggest that, depending on
details of the lattice theory and the value of $g_0^2$, the mass
of the neutral pion can either vanish at two ``critical'' values
of $M_0$ (with the charged pions being massless for $M_0$ in the
interval between them)~\footnote{We restrict attention to the branch 
of the Aoki phase closest to $M_0=0$ (see Fig.~9 of ref.~\cite{AOK}).} or
never vanish and just have a minimum at some value of $M_0$ (in
this situation charged pions remain degenerate with the neutral
state). In the latter case, the value of $M_0$ at which $m_\pi$ is
minimal can be defined as $M_{\rm{cr}}(r)$. This definition
satisfies eq.~(\ref{MCR}). In the former case, instead, the two
``critical values'' of $M_0$ turn out to differ by an amount that
is O($a^2$) in the scaling limit~\footnote{
We thank the referee for drawing our attention to this point.}. 
In this case any of the two
``critical'' values can be chosen as $M_{\rm{cr}}(r)$: e.g.\ the
larger for $r>0$ and the smaller for $r<0$. This choice is
consistent with eq.~(\ref{MPIPI}) and thus with eq.~(\ref{MCR}).

As expected, we find that for the purpose of making the subtracted
Wilson term in the action, $\bar\psi[-a(r/2)\sum_\mu \nabla^*_\mu
\nabla_\mu + M_{\rm{cr}}(r)]\psi$, a truly dimension five operator
the quantity $M_{\rm{cr}}(r)$ can always be taken odd in $r$.

It goes without saying that, if PCAC is used to define
$M_{\rm{cr}}$, an equally good and $r$-odd determination is
obtained (see Appendix~A), only differing by O($a$) terms from
that based on the behaviour of the pion mass, described before.

\subsection{A comment on reflection positivity}
\label{sec:CRP}

An important question in the framework of the computational strategy we are
proposing is whether reflection positivity~\cite{OSCH,OS} remains valid
for the Wilson lattice theory in which the sign of $r$ is reversed.
Reflection positivity is a crucial property for being able to extract
energies and matrix elements from lattice correlators, because it ensures
that a consistent Hamiltonian formalism can be set up at finite lattice
spacing (even though the explicit construction of a transfer matrix may
be technically difficult or not available).

It has been shown in refs~\cite{OS,IY} that the ``link'' reflection
positivity of the lattice action is a sufficient condition to define a
Hilbert space of states with positive metric and prove
the existence of a positive, self-adjoint operator that can be viewed as the
square of the transfer matrix. We recall that link reflection, as opposed
to ``site'' reflection~\cite{MON,MP}, refers to reflection of lattice
points across a time-hyperplane not containing lattice sites.

The standard fermionic Wilson action~(\ref{FERACT}) is known to be link
reflection positive for all (real) values of $g_0^2$, $M_0$ and
$|r|\leq 1$~\cite{IY}. Actually in refs.~\cite{OS,IY} a stronger result
was proved which implies that even the general tm-LQCD action of
eq.~(\ref{TMQCDPSI}) is link reflection positive for $|r|\leq 1$. This fact
is of special relevance for our work, where the problems with the spectrum
of the standard WD operator are avoided by making recourse to tm-LQCD
(see sect.~\ref{sec:TWM1}).

In Appendix~B for completeness we give some useful definitions and summarize
a few results about reflection positivity and lattice transfer matrix.

\section{Removing O($a$) cutoff effects}
\label{sec:IQCD}

In this section we propose a way of extracting O($a$) improved estimates
of physical quantities from lattice correlation functions, which does not
require the knowledge of anyone of the usual improvement coefficients.

Let us consider the expectation value of a gauge invariant, multi-local,
m.r.\ operator which, as we discussed in sect.~\ref{sec:SF}, without loss
of generality can be taken to have a definite $\gamma_5$-chirality.
In fact, while multi-local operators made out only of bosonic local factors
are necessarily eigenstates of ${\cal R}_5^{\rm sp}$, the general
case of multi-local operators containing also baryonic local factors
can be treated by splitting the operator into a ${\cal R}_5$-even
and a ${\cal R}_5$-odd part and applying separately to each of them the
arguments we present in the following.

We start by noting that, since $M_{\rm{cr}}(r)$ is odd in $r$, the excess quark
mass $m_q$, eq.~(\ref{QMASS}), changes sign under $M_0 \rightarrow - M_0$ and
$r \rightarrow -r$. The transformation ${\cal R}_5^{\rm sp}$ under which
the Wilson action is invariant can then be also written as
\beq
{\cal{R}}_5^{\rm{sp}} = {\cal{R}}_5\times [r\rightarrow -r]\times
[m_q\rightarrow -m_q] \, .
\label{SPUR}\eeq
Correspondingly, the relation~(\ref{KEYRELM0}), which expresses the
implications of the spurionic symmetry ${\cal R}_5^{\rm sp}$ on lattice
correlators, takes the form
\beq
\langle O\rangle\Big{|}_{(r,m_q)}= (-1)^{P_{{\cal{R}}_5}[O]}\langle
O\rangle\Big{|}_{(-r,-m_q)} \, , \label{KEYREL} \eeq
where, by slightly changing the notation employed in the previous section,
we have indicated the parameters that specify the fermionic action by
using, besides $r$, the excess quark mass, $m_q=M_0-M_{\rm{cr}}$,
instead of the bare mass $M_0$.

In order to discuss the issue of O($a$) improvement we need to make
reference to the notion of effective theory introduced by Symanzik and
use the related Symanzik expansion of lattice correlators in terms of
correlators of the continuum theory~\cite{SYM}. Schematically up to O($a$)
terms, one can write for the lattice expectation value of a m.r.\ operator
the expansion
\beqn
&&\langle O \rangle\Big{|}_{(r,m_q)} =
[\zeta^{O}_{O}(r)+am_q\xi^{O}_{O}(r)]\langle O \rangle
\Big{|}^{\rm{cont}}_{(m_q)}+
\nonumber\\
&&+ a\,\sum_{\ell}(m_q)^{n_{\ell}}\eta^{O}_{O_{\ell}}(r) \langle
O_{\ell}\rangle\Big{|}^{\rm{cont}}_{(m_q)}+{\rm{O}}(a^2)\, .
\label{SCHSYM}\eeqn The label $|^{\rm{cont}}_{(m_q)}$ in the
correlators appearing on the r.h.s.\ of eq.~(\ref{SCHSYM}) is
meant to recall that they are continuum Green functions
renormalized at the scale $a^{-1}$. They are computed employing
the continuum QCD action, regularized by using, e.g., a second
lattice regularization with lattice spacing much smaller than $a$.
Consistently, the parameter $m_q$ on the r.h.s.\ is to be
interpreted as the continuum value of the quark mass at the
subtraction point $a^{-1}$. Even in the limit of vanishing lattice
spacing, the previous equation has a logarithmically divergent
$a$-dependence. This divergency can be disposed of by multiplying
both sides of eq.~(\ref{SCHSYM}) by the renormalization constant
of the operator $O$. The coefficients $\zeta$, $\xi$ and $\eta$
are finite functions of $r$ and $g_0^2$. All these coefficients
are necessary to match the lattice correlator on the l.h.s.\ to
the continuum Green functions on the r.h.s.\ of~(\ref{SCHSYM}).

The sum over $\ell$ is extended over all operators, $O_{\ell}\neq O$, with
(mass) dimension ($d[O]$ is the dimension of $O$)
\beq
d[O_{\ell}]=d[O]-n_{\ell}+1\, ,\label{DIMO}\eeq
having the same unbroken quantum numbers of $O$. The obvious positivity
of $d[O]$ and $d[O_{\ell}]$ puts an upper bound, $n_{\rm{max}}$, to the
possible values of $n_{\ell} \geq 0$. Two types of correlators will contribute
to the sum over $\ell$: correlators in which products of local operators are
inserted and correlators where $O$ is inserted with space-time integrated
local densities. The former are related to the improvement of the single local
factors making up $O$, while the latter owe their origin from the need of
improving the action~\cite{LUS1}.

A careful discussion of the structure of the Symanzik expansion will be
given in Appendix~A, where we present a detailed analysis of the improvement
of flavour non-singlet axial WTI's.

In the light of eq.~(\ref{KEYREL}) we want to compare the
expansion~(\ref{SCHSYM}) with the similar one obtained after changing
sign to $r$ and $m_q$. To this end we observe that
\begin{enumerate}
\vspace{-.2cm}
\item in the r.h.s.\ of eq.~(\ref{SCHSYM}) $m_q$ appears not only explicitly
in some of the factors that multiply the continuum correlators, but, as
we have explicitly indicated in the formulae, also through the dependence
of the continuum action on $m_q$;
\vspace{-.2cm}
\item the transformation~(\ref{PSIBAR}) is part of the chiral group and,
acting on the continuum action, is equivalent to changing the sign of $m_q$;
since $[{\cal{R}}_5]^2=1\!\!1$, the operators $O_\ell$ can always be taken
to have well defined $\gamma_5$-chiralities, $P_{{\cal{R}}_5}[O_{\ell}]$;
consequently their continuum expectation values will have a definite parity
under $m_q\rightarrow -m_q$, namely
\beq
\langle O_{\ell}\rangle\Big{|}^{\rm{cont}}_{(m_q)}=
(-1)^{P_{{\cal{R}}_5}[O_{\ell}]}
\langle O_{\ell}\rangle\Big{|}^{\rm{cont}}_{(-m_q)}\, .
\label{PARPRO}\eeq
\end{enumerate}
Using~(\ref{PARPRO}), one gets from~(\ref{SCHSYM})
\beqn
&&\langle O \rangle\Big{|}_{(-r,-m_q)} = (-1)^{P_{{\cal{R}}_5}[O]}
[\zeta^{O}_{O}(-r)-am_q\xi^{O}_{O}(-r)]\langle O \rangle
\Big{|}^{\rm{cont}}_{(m_q)}+
\nonumber\\
&&+a\,\sum_{\ell}(-1)^{P_{{\cal{R}}_5}[O_{\ell}]+n_{\ell}}(m_q)^{n_{\ell}}
\eta^{O}_{O_{\ell}}(-r)
\langle O_{\ell}\rangle\Big{|}^{\rm{cont}}_{(m_q)}+{\rm{O}}(a^2)\, .
\label{SCHSYMR}\eeqn
Inserting the expansions~(\ref{SCHSYM}) and~(\ref{SCHSYMR}) in
eq.~(\ref{KEYREL}) and equating O($a^0$) and O($a$) terms, one obtains
the relations
\beqn
&&\hspace{-.8cm}\zeta^{O}_{O}(r)-\zeta^{O}_{O}(-r)=0\, ,\label{ZETA}\\
\nonumber\\
&&\hspace{-.8cm}am_q[\xi^{O}_{O}(r)+\xi^{O}_{O}(-r)]\langle O \rangle
\Big{|}^{\rm{cont}}_{(m_q)}+\label{XIETA}\\
&&\hspace{-.8cm}+a\,\sum_{\ell}(m_q)^{n_{\ell}}\Big{[}
\eta^{O}_{O_\ell}(r)-(-1)^{P_{{\cal{R}}_5}[O]+
P_{{\cal{R}}_5}[O_{\ell}]+n_{\ell}} \eta^{O}_{O_\ell}(-r)\Big{]}
\langle O_{\ell}\rangle\Big{|}^{\rm{cont}}_{(m_q)}=0\nonumber\, .
\eeqn We now want to show that eqs.~(\ref{ZETA}) and~(\ref{XIETA})
imply the formula (Wilson average - $WA$) \beq \langle O
\rangle\Big{|}_{(m_q)}^{WA}\equiv \frac{1}{2}\Big{[}\langle O
\rangle\Big{|}_{(r,m_q)}+ \langle O
\rangle\Big{|}_{(-r,m_q)}\Big{]}= \zeta^{O}_{O}(r) \langle O
\rangle\Big{|}^{\rm{cont}}_{(m_q)}+{\rm{O}}(a^2)\, . \label{WILAV}
\eeq We recall again than in the above equation the prescription
of changing the sign of $r$ must be extended to the whole
$r$-dependence present in the expectation value, i.e.\ to the
$r$-dependence of the action and of the mixing coefficients making
up $O$ in terms of bare operators.

To prove eq.~(\ref{WILAV}) we first observe that from
eq.~(\ref{SCHSYM}), by changing only sign to $r$, one gets
\beqn
&&\langle O \rangle\Big{|}_{(r,m_q)}+
\langle O \rangle\Big{|}_{(-r,m_q)}=
\Big{[}\zeta^{O}_{O}(r)+\zeta^{O}_{O}(-r)\Big{]}\langle O\rangle
\Big{|}^{\rm{cont}}_{(m_q)}+\nonumber\\
&&+a\,m_q\Big{[}\xi^{O}_{O}(r)+\xi^{O}_{O}(-r)\Big{]}\langle O\rangle
\Big{|}^{\rm{cont}}_{(m_q)}+\nonumber\\
&&+ a\,\sum_{\ell}(m_q)^{n_{\ell}}
\Big{[}\eta^{O}_{O_{\ell}}(r)+
\eta^{O}_{O_{\ell}}(-r)\Big{]}\langle O_{\ell}
\rangle\Big{|}^{\rm{cont}}_{(m_q)}+{\rm{O}}(a^2)\, .
\label{INTERM}
\eeqn
In Appendix~C we prove that simple symmetry considerations imply
the validity of the relation
\beq
P_{{\cal{R}}_5}[O]+P_{{\cal{R}}_5}[O_{\ell}]+n_{\ell} =1\,\,{\mbox{mod}}\,(2)
\, .\label{PPAR}\eeq
Consequently all O($a$) terms on the r.h.s.\ of eq.~(\ref{INTERM}) vanish
owing to the relation~(\ref{XIETA}), while the O($a^0$) terms add up to
the desired continuum result thanks to eq.~(\ref{ZETA}), thus proving
our statement.

It is worth noting that a compact way of summarizing the results obtained
above is to say that the Symanzik coefficients $\zeta$ are even functions
of $r$, while all the others ($\xi$ and $\eta$) are odd. These properties
immediately follow from the vanishing of the sum of terms in the last two
lines of eq.~(\ref{INTERM}) (owing to eqs.~(\ref{XIETA}) and~(\ref{PPAR})),
thanks to the fact that in each term $r$ and space-time dependences are
completely factorized.

Making use of eq.~(\ref{KEYREL}) in the second term of the
formula~(\ref{WILAV}), we can rewrite the latter in the equivalent
form (mass average - $MA$)~\footnote{We thank M. Testa for a very useful
discussion on this point.}
\beqn
&&\langle O \rangle\Big{|}_{(m_q)}^{MA}\equiv
\frac{1}{2}\Big{[}\langle O \rangle\Big{|}_{(r,m_q)}+
(-1)^{P_{{\cal{R}}_5}[O]}\langle O \rangle
\Big{|}_{(r,-m_q)}\Big{]}=\nonumber\\&&=\zeta^{O}_{O}(r)
\langle O \rangle\Big{|}^{\rm{cont}}_{(m_q)}+{\rm{O}}(a^2)\, ,
\label{MASSAV}
\eeqn
which says that an O($a$) improved lattice correlator can be obtained
by taking the sum with an appropriate relative sign of the two correlators
computed within the same lattice regularization (same value of $r$) but
with opposite sign of the excess quark mass, $m_q$.

Since, as we have seen, eqs.~(\ref{WILAV}) and~(\ref{MASSAV}) are
identical, in the following we will often refer to either one of them
as Wilson average.

\subsection{Comments}
\label{sec:TC}

A few observations on the general strategy we have outlined above are in
order here. A first remark concerns the key property characterizing the
operators to which our method is applicable. Two others have to do with the
question of whether O($a$) discretization effects in $M_{\rm{cr}}$ may
jeopardize the results we have obtained and the delicate problem of taking
the $m_q\rightarrow 0$ limit in eqs.~(\ref{WILAV}) and~(\ref{MASSAV}).

\subsubsection{Multiplicative renormalizability and improvement}
\label{sec:RMI}

At the beginning of sect.~\ref{sec:IQCD} we restricted our
consideration to m.r.\ (gauge invariant) operators, $O$. Indeed
the Symanzik expansion~(\ref{SCHSYM}), in the way we have written
and exploited it, is only valid if $O$ is (apart from a possible
logarithmically divergent renormalization factor) a finite
operator. If this is not the case, negative powers of the lattice
spacing would appear in the r.h.s.\ of the expansion, together
with dreadful terms arising from the interference of O($a$) (and
higher) discretization effects with power divergencies. In this
situation the naive counting of powers of $a$ is not anymore valid
and the Symanzik expansion is not immediately useful for our
purposes~\footnote{We will take back this question
in~\cite{FRTWO}, when dealing with power divergent operators, like
the effective weak Hamiltonian, and show how one can go around
this problem.}.

\subsubsection{O($a$) discretization effects in $M_{\rm{cr}}$}
\label{sec:OADEIMCR}

While in PT, to any order in $g^2_0$, $M_{\rm{cr}}$ can be defined
to an arbitrary accuracy in the lattice spacing, this is not so
non-perturbatively. Beyond PT, $M_{\rm{cr}}$ can be determined, in
the presence of S$\chi$SB, by looking at the behaviour of the
pion masses (as discussed in sect.~\ref{sec:CRMASSB}) or, 
more in general, as the
value where the axial currents are conserved (PCAC). In any case the critical
mass can only be known up to discretization errors whose magnitude depends
on the condition chosen to determine it and the details of its computational
implementation. Then an obvious question to ask is whether these
ambiguities may invalidate the argument about improvement presented above.

We show in Appendix~D that, although $M_{\rm{cr}}$ can only be known
with O($a$) discretization errors (unless the theory is fully improved \`a la
Symanzik, something which obviously we do not imagine to be doing in the
present context), the formulae for Wilson (eq.~(\ref{WILAV})) - or mass
(eq.~(\ref{MASSAV})) - averages lead to O($a$) improvement as claimed above.

\subsubsection{The limit $m_q\rightarrow 0$}
\label{sec:LMQO}

In the presence of S$\chi$SB the chiral symmetry is not realized
{\it \`a la} Wigner and, as the mass goes to zero, the chiral phase of the
vacuum is driven by the phase of the quark mass term. The same must be true
on the lattice, thus ideally the continuum limit should be taken before
letting $m_q\rightarrow 0$. As a result, taking the chiral limit is a
numerically delicate matter (see also sect.~\ref{sec:CPV})~\footnote{We wish
to thank R. Sommer for drawing our attention to this important point.}.

In practice in order to ensure that on the lattice the chiral phase of the
vacuum is determined by the quark mass term proportional to $m_q$, and not by
the Wilson term, it is necessary to work with lattice parameters satisfying
\beq \label{COND1}
m_q \Lambda_{\rm QCD}^{-1} \, \gg \, a \Lambda_{\rm QCD} \, ,
\eeq
in the unimproved case, or possibly under the weaker condition
\beq \label{COND2}
m_q \Lambda_{\rm QCD}^{-1} \, \gg \, a^2 \Lambda_{\rm QCD}^2 \, ,
\eeq
if one is dealing with O($a$) improved correlators. When conditions
like~(\ref{COND1}) or~(\ref{COND2}) are not fulfilled, renormalized lattice
quantities may strongly differ from their continuum limit counterpart, thus
resulting of little use in extracting information about the underlying
theory.

To analyze the subtleties of taking the limit $m_q\rightarrow 0$ in our
approach it is convenient to first discuss the non-physical case in which
there is no S$\chi$SB.

$\bullet$ In the absence of S$\chi$SB there would be no problem in
setting $m_q=0$ and from~(\ref{KEYREL}) one would get \beq \langle
O\rangle\Big{|}_{(r,m_q=0)}=  (-1)^{P_{{\cal{R}}_5}[O]}\langle
O\rangle\Big{|}_{(-r,m_q=0)} \, . \label{KEYREL1} \eeq This
relation is telling us that expectation values of operators with
$P_{{\cal{R}}_5}[O]=0$~mod\,(2) are automatically O($a$) improved.
Indeed all the operators $O_\ell$ that are left over after taking
the limit $m_q\rightarrow 0$ (i.e.\ those with $n_\ell=0$) in the
Symanzik expansion of $O$ are necessarily ${\cal{R}}_5$-odd (see
eq.~(\ref{PDPDL}) with $d[O_\ell]=d[O]+1$), hence have a vanishing
continuum expectation value. So there are actually no non-zero
O($a$) contributions in the Symanzik expansion of the lattice
expectation value of $O$ if $P_{{\cal{R}}_5}[O]=0$~mod\,(2). On
the other hand for operators having
$P_{{\cal{R}}_5}[O]=1$~mod\,(2) the $WA$ yields a quantity which
is identically zero, in agreement with the expected continuum
result.

$\bullet$ In the presence of S$\chi$SB the chiral phase of the vacuum matters
and, since the theory is non-analytic at $m_q=0$, eq.~(\ref{KEYREL1}) does
not make sense anymore, rather the limit $m_q\rightarrow 0$ should be
approached under the condition~(\ref{COND1}), or~(\ref{COND2}) if correlators
are O($a$) improved. In this situation a direct analysis of the
$m_q$-dependence of $\langle O \rangle |_{(r,m_q)} $, based on the parities
of the continuum expectation values under $m_q \rightarrow -m_q$
(see eq.~(\ref{PARPRO})), confirms the validity of eq.~(\ref{MASSAV}).

The remarks of this section suggest that in practice, for each
value of $g_0^2$, the cutoff effects on quantities sensitive to
S$\chi$SB may become quite large when $m_q$ goes below some
threshold value. This threshold value should be determined by
numerical experiments and will depend in general on the observable
under consideration. As a consequence, performing a chiral fit of
lattice data obtained at fixed $g_0^2$ becomes difficult and
possibly inappropriate, if data at too small values of $m_q$ are
included.  The safe region
is determined by the order of magnitude inequality~(\ref{COND1})
or rather, for improved quantities,~(\ref{COND2}). It is
reassuring to note that for realistic lattice spacings, say $a
\leq 0.1$~fm, the bound~(\ref{COND2}) is satisfied down to pretty
low values of $m_q$, of the order of $10$~Mev.

\subsection{Matrix elements and hadronic masses}
\label{sec:HSME}

An immediate consequence of the absence of O($a$) discretization
effects in Wilson averaged correlators is the observation that
improved hadronic masses and matrix elements can be similarly
obtained by taking averages of the corresponding quantities,
extracted from the individual terms in the l.h.s.\ of equations
like~(\ref{WILAV}) (or equivalently~(\ref{MASSAV})), i.e.\ by
averaging the values of the masses and matrix elements computed in
separate simulations performed with standard Wilson actions having
opposite values of $r$ and fixed $m_q$ (or fixed $r$ and opposite
values of $m_q$).

This remark is very important as it yields a simple and practical
prescription to extract improved physical quantities from Monte
Carlo simulations.
It also solves an apparent paradox of the approach we propose~\footnote{We
would like to thank R. Sommer for a useful discussion on this question.}.
The problem has to do with the observation that different regularizations,
as the ones provided by Wilson lattice actions with opposite Wilson terms,
will lead at finite $a$ to slightly different values for the masses of
intermediate states. Now suppose that we consider the spatially integrated
two-point Green function of some local, m.r.\ (bosonic) operator, $\Phi_h$,
with definite $\gamma_5$-chirality and the quantum numbers appropriate
to create the state $h$ ($|h\>$) from the vacuum ($|\Omega\>$). We have
proved that to get O($a$) improved quantities, one has to average correlators
computed with opposite values of $r$. In particular we expect to be able to
extract improved masses from the average
\beq
G_{hh}(t)\Big{|}^{WA}_{(m_q)}=\frac{1}{2}\Big{[}G_{hh}(t;r,m_q)
+G_{hh}(t;-r,m_q)\Big{]}\, ,\label{GHWA}\eeq
with
\beq
G_{hh}(t;r,m_q)=a^3\sum_{\rm{\bf x}} \langle \Phi_h({\bf x},t)
\Phi_h^{\dagger}(0)\rangle\Big{|}_{(r,m_q)} \, .   \label{HH}
\eeq
Here the operator $\Phi_h^{\dagger}$ is defined in terms of the lattice
reflection operator, $\Theta$, appropriate to Euclidean metric (see
Appendix~B for details), so as to guarantee the validity of the relation
$\<\alpha|\Phi_h^{\dagger}|\beta\>=\<\beta|\Phi_h|\alpha\>^\star$ for
generic states $|\alpha\>$ and $|\beta\>$.

If we take at finite $a$ the limit $t\rightarrow\infty$ of
$G_{hh}(t)|^{WA}_{(m_q)}$, we seem to run into a paradox. On the
one hand, in fact, as $t\rightarrow\infty$ one of the two
correlators, namely that in which the lowest lying state has the
lighter mass, will dominate over the second. On the other hand,
one single correlator is not in general O($a$) improved by itself.
Moreover at finite lattice spacing it is not obvious that one can
extract hadronic masses from the large time behaviour of the
Wilson averaged correlator, $G_{hh}(t)|^{WA}_{(m_q)}$, as the
latter does not admit a spectral decomposition in the usual sense.

The solution of this problem lies in the remark that for the
purpose of getting O($a$) improvement of physical quantities via
$WA$'s the limits $t\rightarrow\infty$ and $a\rightarrow 0$ do not
really commute. In fact, if we let $|t| \to \infty$ at fixed $a$,
the corrections of order $a |t|$ are no longer small. As a result
the expansion in $a$ breaks down and the continuum limit cannot be
taken. This simply means that the continuum limit $a \to 0$ must
be taken first. This observation, however, is not immediately
useful for the purpose of actually extracting improved masses and
matrix elements from lattice data.

To see how one should proceed in practice let us 
fix the attention on the spatial Fourier Transform (FT) \beq
G_{hh}({\bf k},t;\pm r,m_q)=a^3\sum_{\rm{\bf x}}e^{i{\bf {k\cdot
x}}} \langle \Phi_h({\bf x},t)
\Phi_h^{\dagger}(0)\rangle\Big{|}_{(\pm r,m_q)}  \, . \label{HHFT}
\eeq {}From the general principles of quantum mechanics and the
existence of a lattice Hamiltonian for any $|r| \leq 1$ (see
sect.~\ref{sec:CRP}), it follows~\cite{OSCH,OS} that the following
spectral representation ($t>0$) must hold \beq G_{hh}({\bf
k},t;\pm r,m_q)=\sum_{n} \frac{e^{-E_{h,n}({\bf k};\pm r,m_q)
t}}{2E_{h,n}({\bf k};\pm r,m_q)} |R_{h,n}({\bf k};\pm r,m_q)|^2 \,
,\label{QM_PM}\eeq where the $E_{h,n}({\bf k};\pm r,m_q)$'s denote
the energies of the (covariantly normalized) states $|h,n,{\bf
k}\>|_{(\pm r,m_q)}$ in the two regularizations we are comparing
and \beq R_{h,n}({\bf k};\pm r,m_q)= \<\Omega|\Phi_h|h,n,{\bf
k}\>|_{(\pm r,m_q)}\, .\label{RNPM}\eeq For combinations of
operators and states such that $R_{h,n}({\bf k};\pm r,m_q)$ is
non-vanishing in the continuum limit, we expect for sufficiently
small $a$ \beqn &&\hspace{-1.8cm}E_{h,n}({\bf k};\pm r,m_q)=E^{\rm
cont}_{h,n}({\bf k},m_q)+ \delta^\pm_{h,n}\, , \quad
\delta^\pm_{h,n} = {\rm O}(a) \, ,
\label{DELDEF}\\
&&\hspace{-1.8cm}|R_{h,n}({\bf k};\pm r,m_q)|^2 =
|\zeta^{\Phi_h}_{\Phi_h}(r)|^2
|R^{\rm cont}_{h,n}({\bf k},m_q)|^2 (1 + \epsilon^\pm_{h,n}) \, ,
\quad \epsilon^\pm_{h,n} = {\rm O}(a) \, .
\label{EPSDEF}\eeqn
{}From the absence of O($a$) terms in the average
\beq
G_{hh}({\bf k},t)\Big{|}^{WA}_{(m_q)}=
\frac{1}{2}\Big{[}G_{hh}({\bf k},t;r,m_q)
+G_{hh}({\bf k},t;-r,m_q)\Big{]}\label{GHWAK}\eeq
and the arbitrariness of $t$, one gets
\beqn
&&\delta^+_{h,n} + \delta^-_{h,n} = {\rm O}(a^2)  \label{DELAV}\\
&&\epsilon^+_{h,n} + \epsilon^-_{h,n} = {\rm O}(a^2) \label{EPSAV}\eeqn
and consequently
\beq
\hspace{-1.5cm}\frac{1}{2} \Big{[}E_{h,n}({\bf k};r,m_q)+[r\rightarrow -r]
\Big{]}=E^{\rm cont}_{h,n}({\bf k},m_q)+{\rm O}(a^2) \, ,\label{MASSIMP}\\
\eeq \beq \frac{1}{2} \Big{[}|R_{h,n}({\bf
k};r,m_q)|^2+[r\rightarrow -r]
\Big{]}=|\zeta^{\Phi_h}_{\Phi_h}(r)|^2 |R^{\rm cont}_{h,n}({\bf
k},m_q)|^2 +{\rm O}(a^2)\, .\label{MATIMP}\eeq The practical
recipe suggested by these equations is to first extract energies
(hence masses, $E_{h,n}({\bf 0};r,m_q)=M_{h,n}(r,m_q)$) and matrix
elements separately from the two terms in the r.h.s.\ of
eq.~(\ref{GHWAK}) and then take the $WA$ of the corresponding
quantities to get rid of O($a$) cutoff effects.

We show in Appendix~E that the argument leading to eq.~(\ref{MASSIMP})
directly applies to the energy of baryonic states, with no need of
splitting the interpolating operators into ${\cal R}_5$-even and -odd parts.

A somewhat more elaborate argument can be made to show that it is possible
to take the ``square root'' of eq.~(\ref{EPSDEF}) before computing the
average, thus getting the simpler relations
\beq
\frac{1}{2}\Big{[}R_{h,n}({\bf k};r,m_q)+[r\rightarrow -r]\Big{]}=
\zeta^{\Phi_h}_{\Phi_h}(r)R^{\rm cont}_{h,n}({\bf k},m_q)+
{\rm O}(a^2)\, , \label{RPRM}\eeq
which say that $R_{h,n}({\bf k};r,m_q)$ and $R_{h,n}({\bf k};-r,m_q)$
only differ by O($a$) terms that are odd in $r$~\footnote{If there is only
one operator that can create from the vacuum the state
$|h,n,{\bf k}\>|_{(r,m_q)}$, a suitable choice for its phase is sufficient
to obtain eq.~(\ref{RPRM}). When there are several such operators,
consideration of all the related two-point correlators is necessary.}.

\subsubsection{Generalization to three-point correlators}
\label{sec:AG}

The above line of arguments can be generalized to encompass the case of
three-point correlators. Let us consider the double FT
\beqn
&&G_{hBh'}({\bf k},t,{\bf k'},t';\pm r,m_q)=\nonumber\\
&&=a^6\sum_{\rm{\bf x},{\rm{\bf x}}'}
e^{i{\bf {k\cdot x}}} e^{-i{\bf {k}}'{\bf{\cdot x}}'}
\langle \Phi_h({\bf x},t) B(0)\Phi_{h'}^\dagger({\bf x}',t')\rangle
\Big{|}_{(\pm r,m_q)} \, , \label{HBHFT} \eeqn
where, like $\Phi_h$, also $B$ is a local, m.r.\ operator.
With an obvious extension of the previous notations
we can write for the spectral decomposition of $G_{hBh'}$ ($t>0$, $t'<0$)
\beqn
&&\hspace{-.5cm}G_{hBh'}({\bf k},t,{\bf k'},t';\pm r,m_q)=\sum_{n,n'} \,
\frac{e^{-E_{h,n}({\bf k};\pm r,m_q)t}}{2E_{h,n}({\bf k};\pm r,m_q)}\,
\frac{e^{E_{h',n'}({\bf k}';\pm r,m_q)t'}}{2E_{h',n'}({\bf k}';\pm r,m_q)}
\cdot\nonumber\\&&\hspace{-.5cm}\cdot R_{h,n}({\bf k};\pm r,m_q)
\langle h,n,{\bf k}|B|h',n',{\bf k}'\rangle\Big{|}_{(\pm r,m_q)}
R_{h',n'}^\star({\bf k}';\pm r,m_q)\, . \label{SPHBH} \eeqn
O($a$) improvement of the correlator~(\ref{HBHFT}) via $WA$, together
with eqs.~(\ref{MASSIMP}) and~(\ref{MATIMP}) (or rather~(\ref{RPRM})),
imply the set of relations
\beqn
&&\hspace{-1.5cm}\frac{1}{2}\Big{[}
\langle h,n,{\bf k} |B|h',n',{\bf k}'\rangle\Big{|}_{(r,m_q)} +
[r\rightarrow -r]\Big{]}
=\nonumber\\&&\hspace{-1.5cm}= \zeta^{B}_{B}(r)
\langle h,n,{\bf k} |B|h',n',{\bf k}'\rangle\Big{|}^{\rm{cont}}_{(m_q)}
+{\rm{O}}(a^2)\, . \label{FINHBH} \eeqn
Eqs.~(\ref{FINHBH}) prove that $WA$'s of matrix elements of local
operators with definite $\gamma_5$-chirality between pairs of states
having arbitrary three-momenta are O($a$) improved.

\subsection{Renormalization constants}
\label{sec:RC}

In agreement with the general argument about the $r$-dependence of
renormalization costants given in sect.~\ref{sec:SF} (below
eq.~(\ref{KEYREL})), we explicitly show that, in a mass independent
renormalization scheme, the lattice renormalization constant, $Z_B$,
of a m.r.\ local operator, $B$, can (and should) be defined as an even
function of $r$. Furthermore we prove that $Z_B$ can be determined
non-perturbatively with only O($a^2$) discretization errors.

To define the renormalization constant of the m.r.\ operator, $B$,
let us fix our attention on the correlator
\beq
G_B(x;r,m_q)=\<\Phi_B(x)B(0)\>\Big{|}_{(r,m_q)}\, ,
\label{CORBP}\eeq
with $\Phi_B$ a m.r.\ operator located at $x\neq 0$, with the same unbroken
quantum numbers and ${\cal{R}}_5$-parity as $B$. Schematically, mimicking
the standard continuum construction, one can get a non-perturbative estimate
of the lattice renormalization constant of $B$ using the formula
\beqn
&&Z_B(a\mu,r)\frac{\<\Phi_B(x)B(0)\>\Big{|}_{(r,0)}}
{[\<\Phi_B(x)\Phi_B(0)\>\Big{|}_{(r,0)}]^{1/2}}\Big{|}_{|x|=\mu^{-1}}=
\nonumber\\&&=\frac{\<\Phi_B(x)B(0)\>\Big{|}_{(r,0)}}
{[\<\Phi_B(x)\Phi_B(0)\>\Big{|}_{(r,0)}]^{1/2}}
\Big{|}_{|x|=\mu^{-1}}^{\rm{tree}}\, ,\label{ZBDEF}\eeqn
where $\mu$ plays the role of subtraction point. At the cost of introducing
some infrared regulator, all correlators in~(\ref{ZBDEF}) are evaluated at
$m_q=0$. This is not in conflict with what we said in sect.~(\ref{sec:LMQO}),
because the phenomenon of S$\chi$SB is not expected to affect in any way the
value of renormalization constants.

Eq.~(\ref{ZBDEF}) automatically implies that $Z_B(a\mu,r)$ is even in $r$,
because each correlator in~(\ref{ZBDEF}) is ${\cal{R}}_5$-even. Any other
definition of renormalization constant will have to be even in $r$, otherwise
the renormalized lattice operator, $\hat B=Z_B B$, could not have the same
${\cal{R}}_5$ (hence chiral) transformation properties as the corresponding
continuum operator.

In actual simulations one has to relax the condition $m_q=0$, unless the
Schr\"odinger functional setup is employed. But one may define at
non-vanishing mass a renormalization constant, $Z_B(a\mu,am_q,r)$, still
employing a formula like~(\ref{ZBDEF}). To avoid introducing a new symbol,
we gave to this determination of the renormalization constant the same name
we used before, in view of the fact that $Z_B(a\mu,0,r)=Z_B(a\mu,r)$. The
important point is that $Z_B(a\mu,am_q,r)$ is no longer even in $r$ and near
to the continuum limit differs from $Z_B(a\mu,r)$ by O($am_q$) terms.

An O($a$) improved $r$-even estimate, $Z_B^{\rm{I}}$, can be however
obtained either from the equation
\beqn
&&Z_B^{\rm{I}}(a\mu,am_q,r)\frac{\<\Phi_B(x)B(0)\>\Big{|}_{(m_q)}^{WA}}
{[\<\Phi_B(x)\Phi_B(0)\>\Big{|}_{(m_q)}^{WA}]^{1/2}}\Big{|}_{|x|=\mu^{-1}}=
\nonumber\\&&=\frac{\<\Phi_B(x)B(0)\>\Big{|}_{(m_q)}^{WA}}{[\<\Phi_B(x)
\Phi_B(0)\>\Big{|}_{(m_q)}^{WA}]^{1/2}}\Big{|}_{|x|=\mu^{-1}}^{\rm{tree}}\, ,
\label{ZBDEFI}\eeqn
or, as we did in the case of masses and matrix elements, by taking the $WA$ of
the unimproved estimates computed with fermionic actions having Wilson
terms of opposite signs. In formulae
\beq
Z_B^{\rm{I}}(a\mu,am_q,r)=\frac{1}{2}\Big{[}Z_B(a\mu,am_q,r)+
Z_B(a\mu,am_q,-r)\Big{]}+{\rm{O}}(a^2)\, .
\label{ZBLI}\eeq

The arguments developed in this section apply, in particular,
to the renormalization constants of the pseudo-scalar and scalar
quark densities, $Z_P$ and $Z_{S^0}$ (we recall that
$z_m Z_P^{-1}=Z_{S^0}^{-1}$~\cite{MMPT,LUS1} is the quark mass
renormalization constant, see eq.~(\ref{CONLATM}) of Appendix~A), as
well as the renormalization constants of the vector and axial vector
currents, $Z_V$ and $Z_A$.

Also the renormalization constant of the gauge coupling, $Z_g$, is even in
$r$. A simple way of seeing it is to observe that the whole $r$-dependence
of $Z_g$ is through the fermionic determinant, which at $m_q=0$ is even
in $r$.

Concretely $Z_g$ could be computed by making reference to the
quark-antiquark potential, $V_{q\bar{q}}$, using the equations
\beqn
&&Z_g(a/d,r) g_0^2=g_{q\bar{q}}^2(a/d,r)\, ,\\
&&\frac{C_F}{4\pi}g_{q\bar{q}}^2(a/d,r)=
\frac{ \partial V_{q\bar{q}}(d/a,r)}{\partial d} \, ,
\label{QBQF}\eeqn
where $C_F = 4/3$ for the gauge group $SU(3)_{\rm{c}}$ and $V_{q\bar{q}}$
can be extracted from the expectation value of two Polyakov loops located
at spatial distance $d$ ($d^{-1}$ plays the r\^ole of subtraction
point, $\mu$). In a mass independent renormalization scheme,
we can go in the chiral limit and observe that the Polyakov loop
correlation function is even in $r$. The same holds for
$\partial V_{q\bar{q}}(d/a,r)/\partial d$ and consequently for
$Z_g(a/d,r)$. At $m_q\neq 0$ an O($a$) improved $r$-even
determination of $Z_g$ can be obtained through an average of the
type~(\ref{ZBLI}).

We close this section by remarking that, in the context of the
standard O($a$) improvement \`a la Symanzik, the improvement coefficients
$c_{\rm{SW}}$~\cite{SW}, $b_m$ and $b_g$~\cite{LUS1,LUS2} are necessarily odd
under $r\rightarrow -r$. Similarly odd must be the improvement coefficients of
quark bilinears~\cite{LUS2,OTHERS,BGLS}. These results follow from the
observation that the structure of O($a$) terms in the unimproved
Wilson theory is dictated by the spurionic invariance of the lattice Wilson
action under ${\cal{R}}_5^{\rm{sp}}$. In particular, since these
terms must be invariant under ${\cal{R}}_5^{\rm{sp}}$, they
can only be canceled by similarly ${\cal{R}}_5^{\rm{sp}}$ invariant
improvement counterterms added to either the action or the operators.
To conclude that $c_{SW}$, $b_m$, $b_g$ are $r$-odd, it is thus sufficient
to observe that they multiply O($a$) improvement terms in the action which
are odd under ${\cal{R}}_5 \times (m_q \rightarrow -m_q)$. Similarly quark
bilinear improvement coefficients are all $r$-odd, because the O($a$) terms
they go along with have transformation properties under
${\cal{R}}_5 \times (m_q \rightarrow -m_q)$ opposite to that of the quark
bilinear in which they appear. This argument can be extended to a generic
operator with the result that all improvement coefficients are $r$-odd.

Actually one can go one step further and make use of this kind of
spurion analysis to deduce the $r$-parity properties of the
coefficients of ``wrong chirality'' mixings. This topic will be
central in the discussion of ref.~\cite{FRTWO}.

\section{Twisted mass LQCD}
\label{sec:TWM1}

In this and the following sections we want to extend the idea of Wilson
averaging to tm-LQCD. For completeness and to set our notations, we
start by giving a short introduction to tm-LQCD which extends the
presentation of ref.~\cite{FREZ}.

\subsection{The action}
\label{sec:ACT}

For the purpose of this paper it is convenient to begin with a slightly more
general expression of the lattice regularized action than the ones discussed
in refs.~\cite{FGSW,FREZ}. We consider a fermionic action of the form
\beqn
&&\hspace{-1.cm}S_{\rm{F}}^{\rm{L-tm}}[\psi,\bar\psi,U]=
a^4 \sum_x\,\bar\psi (x)
\Big{[}\frac{1}{2}\sum_\mu\gamma_\mu(\nabla^\star_\mu+ \nabla_\mu)+
\nonumber\\&&\hspace{-1.cm}
-a\frac{r}{2}\sum_\mu\nabla^\star_\mu\nabla_\mu
\exp(-i\omega_r\gamma_5\tau_3) +
M'_0+M''_0(-i\gamma_5\tau_3)\Big{]}\psi(x)\, ,\label{TMQCDPSI}
\eeqn
describing an $SU(2)_{\rm{f}}$ flavour doublet of mass degenerate
quarks~\footnote{The case of non-degenerate masses requires some special
considerations and will be discussed in refs.~\cite{CAIRNS,FRTWO}.}. Besides
$r$, three more real parameters, $M'_0, M''_0$ and $\omega_r$, appear in
eq.~(\ref{TMQCDPSI}), though, as we shall see, only two are actually relevant.

As for the question of reflection positivity of tm-LQCD, it was shown in
ref.~\cite{IY} that the most general form of a (link) reflection positive
(bilinear) fermionic action is such as to include the
expression~(\ref{TMQCDPSI}) for all values of $g^2_0$, $M_0'$,
$M_0''$ and $|r|\leq 1$ (see Appendix~B for more details).

The standard Wilson action (in the notations of eq.~(\ref{FERACT})) is
recovered setting $\omega_r=0, M''_0=0$, while tm-LQCD as formulated
in~\cite{FGSW} is obtained for $\omega_r=0, M'_0=m_0$ and $M''_0=-\mu_q$.
The spurionic symmetry of $S_{\rm{F}}^{\rm{L-tm}}$ analog
of~(\ref{SPURM0}) takes the form
\beq
{\cal{R}}_5^{\rm{sp}}\equiv{\cal{R}}_5\times [r\rightarrow -r]\times
[M'_0\rightarrow -M'_0]\times [M''_0\rightarrow -M''_0]\, .
\label{SPURM0P}\eeq

It is immediately seen that under the (non-anomalous) chiral change
of variables
\begin{equation}
\left \{\begin{array}{ll}
&\hspace{-.3cm}\psi(x)\rightarrow \psi'(x)=T(\theta)^{-1} \psi(x) \\\\
&\hspace{-.3cm}\bar{\psi}(x)\rightarrow\bar\psi'(x)=
\bar{\psi}(x)T(\theta)^{-1}
\end{array}\right . \label{PSICHI} \end{equation}
with
\beq
T(\theta)=\exp(i\theta\gamma_5\tau_3/2)\, ,
\label{SOM}\eeq
the fermionic action~(\ref{TMQCDPSI}) transforms as follows
\beqn
&&\hspace{-1.cm}S_{\rm{F}}^{\rm{L-tm}}[\psi,\bar\psi,U] \; \longrightarrow \;
a^4 \sum_x\,\bar\psi (x)
\Big{[}\frac{1}{2}\sum_\mu\gamma_\mu(\nabla^\star_\mu+ \nabla_\mu)
+\nonumber\\
&&\hspace{-1.cm}-a\frac{r}{2}\sum_\mu\nabla^\star_\mu\nabla_\mu
\exp[-i(\omega_r+\theta)\gamma_5\tau_3]
+ M'_0\cos\theta-M''_0\sin\theta+\nonumber\\&&\hspace{-1.cm}+
(M''_0\cos\theta+M'_0\sin\theta)(-i\gamma_5\tau_3)\Big{]}\psi(x)\, .
\label{TMQCDP}\eeqn
With an eye to the standard expression of the formal continuum Dirac action,
one can without loss of generality use this freedom to somewhat simplify the
form of the regularized lattice action~(\ref{TMQCDPSI}). It goes without
saying that in computing expectation values of the kind~(\ref{EXPVAL}),
the fields in the action as well as those in $O$ must be simultaneously
rotated, should one be willing to make use of~(\ref{PSICHI}).

At $M'_0=M''_0=0$ the action~(\ref{TMQCDPSI}) is invariant under the
discrete transformation ($x_P=(-{\bf{x}},t)$)
\begin{equation}
{\cal{P}}_{\omega_r}:\left \{\begin{array}{ll}
&\hspace{-.3cm} U_0(x)\rightarrow U_0(x_P)\qquad U_k(x)\rightarrow
U_k^{\dagger}(x_P-a\hat{k})\quad k=1,2,3\\\\
&\hspace{-.3cm}\psi(x)\rightarrow T(\omega_r)\gamma_0 T(\omega_r)^{-1}
\psi(x_P)  \\\\
&\hspace{-.3cm}\bar{\psi}(x)\rightarrow\bar{\psi}(x_P)
T(\omega_r)^{-1}\gamma_0 T(\omega_r)
\end{array}\right . \label{PTILDE} \end{equation}
and the continuous ones ($b=1,2,3$, no sum over $b$)
\begin{equation}
{\cal{I}}^{b}_{\omega_r}(\theta):\left \{\begin{array}{ll}
&\hspace{-.3cm}\psi(x)\rightarrow T(\omega_r)\exp(i\theta_b\tau_b/2)
T(\omega_r)^{-1}\psi(x)  \\\\
&\hspace{-.3cm}\bar{\psi}(x)\rightarrow\bar{\psi}(x)T(\omega_r)^{-1}
\exp(-i\theta_b\tau_b/2)T(\omega_r)
\end{array}\right . \label{ITILDE} \end{equation}
In connection with these formulae we note that ${\cal{P}}_{\omega_r}$
is the ``rotated'' version of the standard parity operation
(${\cal{P}}_{\omega_r=0}={\cal{P}}$, see eq.~(\ref{PAROP}) below) and that
the factors $T$ and $T^{-1}$ in~(\ref{ITILDE}) cancel if $b=3$.

The (1-point split) currents associated with the flavour
transformations (\ref{ITILDE}) are
\beqn
&&{J}_\mu^b(x)=\frac{1}{2}\Big{[}\bar\psi(x)\frac{\tau_b}{2}
[\gamma_\mu T(\omega_r)^{-2}-r]U_\mu(x)
\psi(x+a\hat{\mu})+\nonumber\\
&&+\bar\psi(x+a\hat{\mu})\frac{\tau_b}{2}[\gamma_\mu T(\omega_r)^{-2}+r]
U^\dagger_\mu(x)\psi(x)\Big{]}\, ,\qquad b=1,2 \label{OPSC2}\eeqn
\beqn
&&{J}_\mu^3(x)=\frac{1}{2}\Big{[}\bar\psi(x)\frac{\tau_3}{2}
[\gamma_\mu-rT(\omega_r)^{-2}]U_\mu(x)
\psi(x+a\hat{\mu})+\nonumber\\
&&+\bar\psi(x+a\hat{\mu})\frac{\tau_3}{2}[\gamma_\mu+rT(\omega_r)^{-2}]
U^\dagger_\mu(x)\psi(x)\Big{]}\, .\label{OPSC1}\eeqn
Of course ${J}_\mu^3$ is a vector current, which is conserved even when
the condition $M'_0=M''_0=0$ is not fulfilled. The other two currents
are mixtures of vector and axial currents with flavour indices 1 and/or 2,
whose relative magnitude depends on the value of the angle $\omega_r$.
We will discuss in sect.~\ref{sec:NSWTI} the structure of the non-singlet
WTI's of tm-LQCD.

The action~(\ref{TMQCDPSI}) is also invariant under charge conjugation
\begin{equation}
{\cal{C}}:\left \{\begin{array}{ll}
&\hspace{-.3cm} U_\mu(x)\rightarrow U_\mu^*(x)  \\\\
&\hspace{-.3cm}\psi(x)\rightarrow i \gamma_0 \gamma_2 \bar\psi^T\!(x) \\\\
&\hspace{-.3cm}\bar{\psi}(x)\rightarrow - \psi^T\!(x) i \gamma_0 \gamma_2
\end{array}\right . \label{CHARGE} \end{equation}
and the two discrete transformations
\begin{equation}
{\cal{P}}_{F}^{1,2}:\left \{\begin{array}{ll}
&\hspace{-.3cm} U_0(x)\rightarrow U_0(x_P)\qquad U_k(x)\rightarrow
U_k^{\dagger}(x_P-a\hat{k})\quad k=1,2,3\\\\
&\hspace{-.3cm}\psi(x)\rightarrow i \gamma_0 \tau_{1,2}\psi(x_P)  \\\\
&\hspace{-.3cm}\bar{\psi}(x)\rightarrow -i \bar{\psi}(x_P)\tau_{1,2}\gamma_0
\end{array}\right . \label{POT} \end{equation}
Charge conjugation and ${\cal{P}}_{F}^2$ invariances rule out action
counterterms of the form
$a^4\sum_x\bar\psi(x)\sum_\mu\gamma_\mu
(\nabla_\mu^* + \nabla_\mu)\,\gamma_5\,\tau\,\psi(x) $
with $\tau$ any matrix in flavour space. In analogy with the
definitions~(\ref{POT}), we could also introduce the transformation
${\cal{P}}_{F}^3$. The latter, however, as soon as $\omega_r \neq 0$
or $M''_0 \neq 0$, is not a symmetry of the action~(\ref{TMQCDPSI}).

The existence, at vanishing $M'_0$ and $M''_0$, of the
symmetries~(\ref{PTILDE}) and~(\ref{ITILDE}) implies
in particular that radiative corrections can only generate
(possibly divergent) mass terms invariant under ${\cal{P}}_{\omega_r}$ and
${\cal{I}}^{1,2}_{\omega_r}$. Consequently the critical mass will contribute
the lattice action a single term having the form
$\bar\psi M_{\rm{cr}}\exp(-i\omega_r\gamma_5\tau_3)\psi$ with $M_{\rm{cr}}$
the critical mass of the standard Wilson theory. To prove this statement
it is convenient to first rewrite the action~(\ref{TMQCDPSI}) in the
form
\beqn
&&\hspace{-1.cm}S_{\rm{F}}^{\rm{L-tm}}[\psi,\bar\psi,U]=
a^4 \sum_x\,\bar\psi (x)
\Big{[}\frac{1}{2}\sum_\mu\gamma_\mu(\nabla^\star_\mu+ \nabla_\mu)+
\label{TMQCDPSICH}\\&&\hspace{-1.cm}
+\Big{(}-a\frac{r}{2}\sum_\mu\nabla^\star_\mu\nabla_\mu
+M_{\rm{cr}}(r)\Big{)}\exp(-i\omega_r\gamma_5\tau_3)+
m_q \exp(i\omega_m\gamma_5\tau_3)\Big{]}\psi(x)\, ,\nonumber\eeqn
where both Wilson and mass term are twisted. To get eq.~(\ref{TMQCDPSICH})
from~(\ref{TMQCDPSI}) we have introduced besides the excess quark mass
\beq
M'_0-M_{\rm{cr}}\cos\omega_r=m_q\cos\omega_m\, ,
\label{EXQM}\eeq
also the excess twisted quark mass
\beq
M''_0-M_{\rm{cr}}\sin\omega_r=-m_q\sin\omega_m
\label{EXTQM}\eeq
with $m_q\geq 0$. Physics is unaffected if we subject~(\ref{TMQCDPSICH})
to the transformation~(\ref{PSICHI}) with $\theta=-\omega_r$ (change of
variables in the functional integral). In the new variables the fermionic
action reads
\beqn
&&\hspace{-1.cm}S_{\rm{F}}^{\rm{tm-LQCD}}[\chi,\bar\chi,U]=
a^4 \sum_x\,\bar\chi (x)
\Big{[}\frac{1}{2}\sum_\mu\gamma_\mu(\nabla^\star_\mu+ \nabla_\mu)+
\label{TMQCDCHI}\\&&\hspace{-1.cm}
+\Big{(}-a\frac{r}{2}\sum_\mu\nabla^\star_\mu\nabla_\mu
+M_{\rm{cr}}(r)\Big{)}+m_q\exp[i(\omega_m+\omega_r)\gamma_5\tau_3]\Big{]}
\chi(x)\, .\nonumber\eeqn
We gave it the name $S_{\rm{F}}^{\rm{tm-LQCD}}$ because with the
identifications
\beqn
&&m_q\cos(\omega_m+\omega_r)\rightarrow m_q\\
&&m_q\sin(\omega_m+\omega_r)\rightarrow \mu_q
\label{IDENTM}\eeqn
eq.~(\ref{TMQCDCHI}) is precisely the tm-LQCD action of ref.~\cite{FGSW}.
In the form~(\ref{TMQCDCHI}) it is clear that $M_{\rm{cr}}$ cannot
depend on $\omega_r$ (nor on $\omega_m$), because the whole $\omega_r$
(and $\omega_m$) dependence is brought by the excess quark mass term. The
argument also shows that the value of the critical mass is the same as
for the standard Wilson theory.

Going back for a while to the form~(\ref{TMQCDPSICH}) of the tm-LQCD action,
it is worth noting that 1) the transformations ${\cal{P}}_{\omega_r}$ and
${\cal{I}}^{b}_{\omega_r}$, $b=1,2$, are exact symmetries if $m_q=0$ for any
$\omega_r$ and, if $\omega_r=-\omega_m$, even when $m_q\neq 0$, provided
$m_q$ is a multiple of the identity in flavour space; 2) since the
currents given by the eqs.~(\ref{OPSC2}) and~(\ref{OPSC1}) are conserved
in the massless limit, they do not suffer of any renormalization. We will
make use of this observation in sect.~\ref{sec:ASA}.

Starting from the action~(\ref{TMQCDPSICH}), we might equally well set
the phase of $m_q\bar\psi\psi$ to zero by bringing it fully to the Wilson
term with a rotation~(\ref{PSICHI}) of an angle $\omega_m$. Precisely because
in this basis the quark mass term has the canonical form, we will call it the
``physical'' basis. In this basis the action takes the form
\beqn
&&\hspace{-1.cm}S_{\rm{F}}^{\rm{L,ph}}[\psi_{\rm{ph}},\bar\psi_{\rm{ph}},U]=
a^4 \sum_x\,\bar\psi_{\rm{ph}} (x)
\Big{[}\frac{1}{2}\sum_\mu\gamma_\mu(\nabla^\star_\mu+ \nabla_\mu)+
\label{PHYSCHI}\\&&\hspace{-1.cm}
+\Big{(}-a\frac{r}{2}\sum_\mu\nabla^\star_\mu\nabla_\mu+M_{\rm{cr}}(r)\Big{)}
\exp(-i\omega\gamma_5\tau_3)+m_q\Big{]}\psi_{\rm{ph}}(x)
\nonumber\eeqn
and, to avoid any possible confusion, the quark fields have been denoted by
$\psi_{\rm{ph}},\bar\psi_{\rm{ph}}$. Eq.~(\ref{PHYSCHI}) shows that physics
can all be extracted from lattice quantities that only depend on three real
parameters, $m_q$, $r$ and the combination
\beq
\omega=\omega_m+\omega_r\,\,{\mbox{mod}}\,(2\pi)\, .
\label{OMEGA}\eeq
In particular the fermion determinant, ${\cal{D}}_F$, has the expression
\beq
{\cal{D}}_F={\rm{det}}\Big{[}(D_{\rm{W}}^{\rm{cr}}+m_q\cos\omega)^\dagger
(D_{\rm{W}}^{\rm{cr}}+m_q\cos\omega)+m_q^2\sin^2\omega\Big{]}\, ,
\label{FERDET}\eeq
\beq
D_{\rm{W}}^{\rm{cr}}=\sum_\mu \Big{[}\frac{1}{2}\gamma_\mu(\nabla^\star_\mu+
\nabla_\mu)-a\frac{r}{2}\nabla^\star_\mu\nabla_\mu\Big{]}+M_{\rm{cr}}(r)\, .
\label{CHIROP}\eeq
Eq.~(\ref{FERDET}) shows that the fermion determinant is real and
positive~\cite{AOGO} and no zero modes can occur if $m_q, \omega\neq 0$.

Ending this section, we wish to remark that the tm-LQCD
action~(\ref{PHYSCHI}) with $|\omega|=\pi/2$ is closely related to the
action first proposed by Osterwalder and Seiler (OS)~\cite{OS}~\footnote{The
use of the OS formulation to solve the problem of exceptional configurations
was first suggested in ref.~\cite{SETAL}.}. There is, however,
a key difference between the OS proposal and tm-LQCD. While
in the action of ref.~\cite{OS} quarks of different flavours are all
twisted with the same angle, in tm-LQCD quarks of different flavours appear
combined in pairs and the twisting involves a traceless flavour matrix,
which for degenerate quarks is conveniently chosen to be $\tau_3$ (besides,
tm-LQCD is defined for any twisting angle in the 0-2$\pi$ range).
This specific pairing of quarks is crucial to guarantee
the validity of two related properties, which are not satisfied
by the OS action~\cite{K,SEST}: positivity of the fermionic determinant
(see eq.~(\ref{FERDET})) and independence from the twisting angle
of all the continuum renormalized correlation functions.

The second property can be proved in different ways. One way is to rely on
universality arguments~\cite{FGSW} and notice that for GW fermions a twist
in the mass term of the non-singlet type, like the one we are considering
here, can always be let to disappear from the lattice action by a
non-anomalous chiral transformation. This is the same as saying that in the
continuum limit physics will not be dependent on the value of the twisting
angle. An equivalent, perhaps more direct, way is to notice that in
tm-LQCD, through a chiral change of fermionic variables in the path integral,
one can always bring the whole twist on the ``subtracted'' Wilson term (see
eq.~(\ref{PHYSCHI})). The latter on the basis of symmetry and power
counting arguments can only mix with dimension 5 operators (in particular
it cannot mix with $i{\rm{Tr}}[F_{\mu\nu}\tilde{F}_{\mu\nu}]$, because of the
different transformation properties under ${\cal{P}}_F^{1,2}$
(eq.~(\ref{POT})), thus it appears as a truly irrelevant operator. On the
contrary, if one considers a fermion regularization with a flavour singlet
twisted Wilson term, like the one proposed in~\cite{OS}, a term proportional
to $i\int d^4x\,{\rm{Tr}}[F_{\mu\nu}\tilde{F}_{\mu\nu}](x)$
multiplied by the twisting angle will occur in the effective gauge
action. In this situation the corresponding continuum theory will
explicitely depend on the twisting angle, with the latter playing
the role of the $\theta$-angle of the vacuum~\cite{SEST}.

\subsection{Non-singlet Ward-Takahashi identities}
\label{sec:NSWTI}

In this section we want to write down explicitly the expression of the
renormalized operators (currents and quark densities) entering the flavour
non-singlet WTI's of the theory described by the general fermionic
action~(\ref{TMQCDPSICH}), where both the Wilson and the mass term are
twisted. The results that we present below can be derived in various
equivalent ways. One possibility is to follow the general procedure of
ref.~\cite{BMMRT} and apply it to the present situation. A shortcut is to
start from what has been already proved in ref.~\cite{FGSW} and derive the
required formulae by performing the chiral change of fermionic variables that
brings from the standard tm-LQCD action to the action~(\ref{TMQCDPSICH}).

Our formulae are written in terms of (bare) local currents and quark density
operators. To make easy contact with the existing literature the relevant
renormalization coefficients are expressed in terms of the ($r$-even)
renormalization constants, $Z_V$, $Z_A$, $Z_P$ and $Z_{S^0}$, of the local
operators of the standard Wilson theory (see Appendix~A for notations and
definitions).

The renormalized vector and axial currents can be taken to be
\beqn
&&\hat{V}_\mu^1=Z_{VV}(\omega_r,\omega)\,
\bar\psi\gamma_\mu\frac{\tau_1}{2}\psi+Z_{VA}(\omega_r,\omega)\,
\bar\psi\gamma_\mu\gamma_5\frac{\tau_2}{2}\psi\label{GENCUR1}\\
&&\hat{V}_\mu^2=Z_{VV}(\omega_r,\omega)\,
\bar\psi\gamma_\mu\frac{\tau_2}{2}\psi-Z_{VA}(\omega_r,\omega)
\,\bar\psi\gamma_\mu\gamma_5\frac{\tau_1}{2}\psi \label{GENCUR2}\\
&&\hat{V}_\mu^3=Z_V\,\bar\psi \gamma_\mu \frac{\tau_3}{2}\psi
\label{GENCUR3}\\
&&\hat{A}_\mu^1=Z_{AA}(\omega_r,\omega)\,
\bar\psi\gamma_\mu\gamma_5\frac{\tau_1}{2}\psi+
Z_{AV}(\omega_r,\omega)\,
\bar\psi\gamma_\mu\frac{\tau_2}{2}\psi\label{GENCUR4} \\
&&\hat{A}_\mu^2=Z_{AA}(\omega_r,\omega)\,
\bar\psi\gamma_\mu\gamma_5\frac{\tau_2}{2}\psi-
Z_{AV}(\omega_r,\omega)\,
\bar\psi\gamma_\mu\frac{\tau_1}{2}\psi\label{GENCUR5}\\
&&\hat{A}_\mu^3=Z_A\,\bar\psi \gamma_\mu \gamma_5\frac{\tau_3}{2}\psi\, ,
\label{GENCUR6}\eeqn
where $Z_{VV}$, $Z_{VA}$, $Z_{AV}$, $Z_{AA}$ are finite functions of $g^2_0$,
$r$, $\omega_r$ and $\omega$, given by the formulae
\beqn
&&\hspace{-1.5cm}Z_{VV}(\omega_r,\omega)=
\frac{1}{Z_M(\omega)}\Big{[}Z_V z_m\cos\omega_r\cos\omega+
Z_A\sin\omega_r\sin\omega\Big{]}
\label{GENAV1}\\
&&\hspace{-1.5cm}Z_{VA}(\omega_r,\omega)=
\frac{1}{Z_M(\omega)}\Big{[}-Z_Vz_m\sin\omega_r\cos\omega
+Z_A\cos\omega_r\sin\omega\Big{]}
\label{GENAV2}\\
&&\hspace{-1.5cm}Z_{AA}(\omega_r,\omega)=
\frac{1}{Z_M(\omega)}\Big{[}Z_Az_m\cos\omega_r\cos\omega
+Z_V\sin\omega_r\sin\omega\Big{]}
\label{GENAV3}\\
&&\hspace{-1.5cm}Z_{AV}(\omega_r,\omega)=
\frac{1}{Z_M(\omega)}\Big{[}-Z_Az_m\sin\omega_r\cos\omega+
Z_V\cos\omega_r\sin\omega\Big{]}\, .
\label{GENAV4}\eeqn
In eqs.~(\ref{GENAV1}) to~(\ref{GENAV4}) we used the definitions~(\ref{OMEGA})
and~(\ref{ZMZPZS}) and we set
\beq
Z_M(\omega)=\Big{[}z^2_m\cos^2\omega+\sin^2\omega\Big{]}^{1/2}\, .
\label{RENH}\eeq
With reference to the currents~(\ref{GENCUR1}) to~(\ref{GENCUR6}), the WTI's
with the insertion of the renormalized (multi-local) operator $\hat{O}(y)$
($ y\neq x$) take the expected form ($b=1,2,3$)
\beqn
&&\hspace{-1.5cm}
\langle{\partial}^\star_\mu\hat{V}^b_\mu(x)
\hat{O}(y)\rangle\Big{|}^{(\omega_r,\omega_m)}_{(r,m_q)}={\rm{O}}(a)
\label{WTITWV}\\
&&\hspace{-1.5cm}
\langle{\partial}^\star_\mu\hat{A}^b_\mu(x)\hat{O}(y)\rangle
\Big{|}^{(\omega_r,\omega_m)}_{(r,m_q)}=2\hat{m}_q\langle\hat{P}^b(x)
\hat{O}(y)\rangle
\Big{|}^{(\omega_r,\omega_m)}_{(r,m_q)}+{\rm{O}}(a)\label{WTITWA}
\eeqn
with
\beq
\hat{m}_q = Z_M(\omega) Z_P^{-1}m_q\, ,
\label{RENMA}\eeq
provided we define in terms of bare quantities the renormalized
pseudo-scalar operators, $\hat{P}^b$, to be
\beqn
&&\hspace{-1.cm}\hat{P}^b=Z_P\, \bar\psi\frac{\tau_b}{2}\gamma_5\psi
\qquad b=1,2\label{PTM1}\\
&&\hspace{-1.cm}\hat{P}^3=Z_P\Big{[}Z_{PP}(\omega_r,\omega)
\bar\psi\frac{\tau_3}{2}\gamma_5\psi+
Z_{PS^0}(\omega_r,\omega) \frac{i}{2}\bar\psi\psi+
Z_{P1\!\!1}(\omega)1\!\!1\frac{i}{a^3}\Big{]}
\label{PTM2}\eeqn
with
\beqn
&&\hspace{-.5cm}Z_{PP}(\omega_r,\omega)=\frac{1}{Z_M(\omega)}
\Big{[}z_m\cos\omega_r\cos\omega+
z_m^{-1}\sin\omega_r\sin\omega\Big{]} \label{ZPP}\\
&&\hspace{-.5cm}Z_{P{S^0}}(\omega_r,\omega)= \frac{1}{Z_M(\omega)}
\Big{[}-z_m\sin\omega_r\cos\omega
+z_m^{-1}\cos\omega_r\sin\omega\Big{]}\label{ZPS}\\
&&\hspace{-.5cm}Z_{P1\!\!1}(\omega)=\frac{\sin\omega}{Z_M(\omega)}
\rho_P(am_q,\omega)\label{Z1}\, .\eeqn
Equations from~(\ref{WTITWV}) to~(\ref{Z1}) need a number of qualifications.
First of all, the set of labels $(\omega_r,\omega_m,r,m_q)$ in
eqs.~(\ref{WTITWV}) and~(\ref{WTITWA}) is there to remind that, if the general
lattice action~(\ref{TMQCDPSICH}) is used, the formulae~(\ref{GENCUR1})
to~(\ref{GENCUR6}) for the currents and~(\ref{PTM1}), (\ref{PTM2}) for the
pseudo-scalar quark densities have to be consequently employed. Secondly, the
unusual mixing of $\bar\psi\frac{\tau_3}{2}\gamma_5\psi$ with
$\bar\psi\psi$ and the identity operator is due to the breaking of parity
and isospin induced in the action~(\ref{TMQCDPSICH}) by the twisting
of the Wilson and mass term. Finally $\rho_P(am_q,\omega)$ is a coefficient
function even in $\omega$ (and odd in $r$) which admits a simple polynomial
expansion in $am_q$. Details about the structure of the power divergent
subtraction in eq.~(\ref{PTM2}) can be found in ref.~\cite{DMFH}.

In conclusion in this section we have recollected for completeness the
general formulae (equations from~(\ref{GENCUR1}) to~(\ref{Z1})), that can be
used to smoothly interpolate between the situation in which the whole twist
is carried by the Wilson term (physical quark basis with
$\omega_r=\omega$, $\omega_m=0$, eq.~(\ref{PHYSCHI})) and the tm-LQCD
formulation of ref.~\cite{FGSW}, where $\omega_r=0$, $\omega_m=\omega$
(eq.~(\ref{TMQCDCHI})).

\section{Improved physics from tm-LQCD}
\label{sec:TWM2}

To make the discussion of improvement transparent it is convenient
to use the form~(\ref{PHYSCHI}) of the tm-LQCD action.

With the help of the formalism developed in sect.~\ref{sec:IQCD} it is
straightforward to show that $WA$'s of pairs of tm-LQCD correlators computed
with opposite values of $r$ and a common value of $m_q$ are free of O($a$)
corrections at fixed value of $\omega$. Exactly as in the case of the
standard Wilson formulation, an identical result is obtained if
correlators computed with a fixed value of $r$ but mass terms of
opposite sign are combined (mass average, $MA$) with a relative sign equal
to the ${\cal{R}}_5$-parity of the operator $O$ whose expectation value is
being computed. Moreover, an analysis of the properties of the (unimproved)
lattice correlators and derived quantities under $\omega \to -\omega$ leads
to a deeper understanding of the practical consequences of the parity and
isospin breaking that occurs at $\omega \neq 0$. {}From this analysis it
turns out that the tm-LQCD action~(\ref{PHYSCHI}) with $\omega =\pm\pi/2$
is particularly convenient for physical applications.

\subsection{Improving correlators: the general argument}
\label{sec:ICGA}

The proof of improvement of $WA$'s (or $MA$'s) of correlators in tm-LQCD
exactly parallels the argument presented in sect.~\ref{sec:IQCD}, observing
that tm-LQCD action~(\ref{PHYSCHI}) is invariant under the spurionic
transformation~(\ref{SPUR}). We only remark for future use that changing
the sign of the Wilson term is equivalent to shifting the twisting angle
by $\pi$. In other words the combined transformation
$[r\rightarrow -r]\times[\omega\rightarrow\omega+\pi]$ leaves the
action~(\ref{PHYSCHI}) invariant.

The proof proceeds by showing that all the steps we went through in the
untwisted, Wilson case remain valid also here. One starts with the Symanzik
expansion
\beqn
&&\langle O \rangle\Big{|}^{(\omega)}_{(r,m_q)} =
[\zeta^{O}_{O}(\omega,r)+am_q\xi^{O}_{O}(\omega,r)]\langle O \rangle
\Big{|}^{\rm{cont}}_{(m_q)}+
\nonumber\\
&&+ a\,\sum_{\ell}(m_q)^{n_{\ell}}\eta^{O}_{O_{\ell}}(\omega,r)
\langle O_{\ell}\rangle\Big{|}^{\rm{cont}}_{(m_q)}+{\rm{O}}(a^2)\, ,
\label{SCHSYMT}\eeqn
where, as in sect.~\ref{sec:IQCD}, $O$ is a multi-local, m.r.\ operator.
Lattice expectation values in eq.~(\ref{SCHSYMT}) are characterized by
$r$, $m_q$ and $\omega$. The parameter $\omega$ is kept fixed in all the
arguments of this section and the dependence on $g_0^2$ is always understood.
As in the standard Wilson case we are concerned
with the $r$-dependence of the Symanzik coefficients $\zeta$, $\xi$ and
$\eta$. It should be noted that for generic values of $\omega$ more
operators, $O_\ell$, contribute to the expansion~(\ref{SCHSYMT}), as
compared to the standard Wilson case. The new operators arise due to the
breaking of parity and isospin induced by the chiral twist of the Wilson
term in eq.~(\ref{PHYSCHI}). This fact, however, does not harm the argument
that follows. In fact, as we prove in Appendix~C, the relation
\beq
P_{{\cal{R}}_5}[O]+P_{{\cal{R}}_5}[O_{\ell}]+n_{\ell} =1\,\,{\mbox{mod}}\,(2)
\label{PPART}\eeq
also holds in the tm-LQCD case, because the symmetry considerations upon
which it is based remain valid for the whole set of operators $O_\ell$
entering the expansion~(\ref{SCHSYMT}). Thus all the formulae we proved
in sect.~\ref{sec:IQCD} maintain their validity also here at any fixed value
of $\omega$. In particular all the results concerning the $WA$ improvement of
correlators (eq.~(\ref{WILAV})) and derived quantities (eqs.~(\ref{MASSIMP}),
(\ref{RPRM}) and~(\ref{FINHBH})), as well as the property of renormalization
constants of being even in $r$ and the formula for the O($a$) improvement
of their determination (sect.~\ref{sec:RC}) remain true in the framework
of tm-LQCD.

As in the case of standard Wilson fermions, O($a$) improvement via $WA$
is equivalent to the observation that the Symanzik coefficients
$\zeta$, $\xi$ and $\eta$ appearing in the r.h.s.\ of eq.~(\ref{SCHSYMT})
have definite $r$-parity properties
\beqn
&&\zeta^{O}_{O}(\omega,r)=\zeta^{O}_{O}(\omega,-r)=
\zeta^{O}_{O}(\omega+\pi,r)\nonumber\\
&&\xi^{O}_{O}(\omega,r)=-\xi^{O}_{O}(\omega,-r)=-\xi^{O}_{O}(\omega+\pi,r)
\label{PAROMR}\\
&&\eta^{O}_{O_{\ell}}(\omega,r)=-\eta^{O}_{O_{\ell}}(\omega,-r)=
-\eta^{O}_{O_{\ell}}(\omega+\pi,r)\, .\nonumber\eeqn
In eqs.~(\ref{PAROMR}) the second equality of each line follows from
the invariance of the action~(\ref{PHYSCHI}) under
$[r\rightarrow -r]\times[\omega\rightarrow\omega+\pi]$.

\subsection{The $\omega$-dependence of lattice correlators}
\label{sec:LATCORPROP}

While working with $\omega\neq 0$ in the action~(\ref{PHYSCHI}) solves
all the problems related to spurious modes of the WD operator, it implies,
as we repeatedly said, a breaking of the parity (and isospin) symmetry.
It should be noted, however, that if the parity operation is accompanied
by a change of sign of $\omega$, the action remains invariant.

It is, in fact, immediate to verify that under the combined transformation
${\cal P}\times (\omega\to-\omega)$, where (remember the definition
$x_P=(-{\bf{x}},t)$)
\begin{equation}
{\cal{P}}:\left \{\begin{array}{ll}
&\hspace{-.3cm} U_0(x)\rightarrow U_0(x_P)\, ,\qquad U_k(x)\rightarrow
U_k^{\dagger}(x_P-a\hat{k})\, ,\qquad k=1,2,3\\\\
&\hspace{-.3cm}\psi_{\rm ph}(x)\rightarrow \gamma_0 \psi_{\rm ph}(x_P)\\\\
&\hspace{-.3cm}\bar{\psi}_{\rm ph}(x)
\rightarrow\bar{\psi}_{\rm ph}(x_P)\gamma_0
\end{array}\right . \label{PAROP}  \end{equation}
is the physical parity of the theory, the action~(\ref{PHYSCHI}) goes
into itself. In the spirit of spurion analysis this invariance and the further
property $[{\cal P}\times (\omega\to-\omega)]^2=1\!\!1$ imply that
a definite parity label, $p$, can be attributed to any multi-local
operator on the basis of the relation (see also Appendix~F)
\beqn
&&\langle O^{(p)}(x_{1},x_{2},\ldots ,x_{n})
\rangle\Big{|}^{(\omega)}_{(r,m_q)}= \nonumber\\&&=\,(-1)^p \,
\langle O^{(p)}(x_{1P},x_{2P},\ldots ,x_{nP})
\rangle\Big{|}^{(-\omega)}_{(r,m_q)}\, .\label{OPARGEN}
\eeqn
We remark that the renormalization of tm-LQCD is such as to respect
eq.~(\ref{OPARGEN}). In particular, this means that, whenever two bare
local operators with opposite transformation properties under~(\ref{PAROP})
mix, the relative mixing coefficient must be odd in $\omega$. Examples of
this phenomenon are found in eqs.~(\ref{GENCUR1}) - (\ref{GENCUR2}),
(\ref{GENCUR4}) - (\ref{GENCUR5}) and~(\ref{PTM2}). Clearly the
situation for ${\cal P}$ here is identical to the one we had before for
${\cal R}_5$. From now on, for short, we will call $(-1)^p$ the
formal parity of the operator $O^{(p)}$. Notice that what we are calling
formal parity coincides with the parity of the operator when parity
invariance is restored ($\omega =0$ and continuum limit).

Inserting the appropriate form of Symanzik expansion~(\ref{SCHSYMT})
in the two sides of the relation~(\ref{OPARGEN}), one can prove the following
parity properties of the Symanzik coefficients under $\omega\to-\omega$
\beqn
&&\zeta^{O^{(p)}}_{O^{(p)}}(\omega,r)=\zeta^{O^{(p)}}_{O^{(p)}}(-\omega,r)
\nonumber\\
&&\xi^{O^{(p)}}_{O^{(p)}}(\omega,r)=\xi^{O^{(p)}}_{O^{(p)}}(-\omega,r)
\label{OMPAR}\\&&\eta^{O^{(p)}}_{O_\ell}(\omega,r)=
(-1)^{p+p_\ell}\eta^{O^{(p)}}_{O_\ell}(-\omega,r)\, .\nonumber\eeqn
In the last of the eqs.~(\ref{OMPAR}) $(-1)^{p_\ell}$ is the physical
parity of the operator $O_\ell$ in the formal continuum theory, defined
by the formula (analogous to~(\ref{OPARGEN}))
\beq
\langle O_\ell(x_{1},x_{2},\ldots ,x_{n})
\rangle\Big{|}^{\rm{cont}}_{(m_q)}=(-1)^{p_\ell} \,
\langle O_\ell(x_{1P},x_{2P},\ldots ,x_{nP})
\rangle\Big{|}_{(m_q)}^{\rm{cont}}\, .\label{OPARGENCONT}
\eeq
{}From eqs.~(\ref{OMPAR}) one can derive a number of interesting consequences
concerning the nature and the magnitude of parity violating terms in lattice
correlators.

First of all, we observe that at $\omega=0$ (no twisting), where the
physical parity~(\ref{PAROP}) is an exact symmetry of the lattice theory,
the last of the relations~(\ref{OMPAR}) implies
\beq
\eta^{O^{(p)}}_{O_\ell}(0,r)=(-1)^{p+p_\ell}\eta^{O^{(p)}}_{O_\ell}(0,r)
\, .\label{ETAPAR}\eeq
As expected, we find that only operators, $O_\ell$, with the
same parity as $O^{(p)}$ ($p_\ell=p$~mod\,(2)) can contribute to the
Symanzik expansion of the latter, because $p+p_\ell=1\,\,{\mbox{mod}}\,(2)$
implies $\eta^{O^{(p)}}_{O_\ell}(0,r)=0$.

If $\omega\neq 0$, parity is broken and also operators, $O_\ell$, with
$p_\ell=p+1$~mod\,(2) may contribute. These terms are odd in $\omega$ and,
consistently with universality (they vanish at $\omega=0$), appear
in the Symanzik expansion multiplied by an explicit factor of $a$.

The results of the whole analysis can be summarized by saying that
\begin{enumerate}
\vspace{-.2cm}
\item all physically relevant quantities are contained in the $\omega$-even
parts of lattice correlators (see the first of eqs.~(\ref{OMPAR})),
\vspace{-.2cm}
\item $\omega$-odd parts of lattice correlators are O($a$),
\vspace{-.2cm}
\item contributions from continuum operators with $p_\ell=p+1$~mod\,(2)
enter lattice correlators only at O($a$) or higher.
\end{enumerate}

\subsection{The $\omega$-parity of energies and matrix elements}
\label{sec:ODMME}

Exploiting the invariance of the tm-LQCD action~(\ref{PHYSCHI}) under the
${\cal{P}}\times (\omega\rightarrow -\omega)$, it is possible to
study the transformation properties under $\omega\rightarrow -\omega$
of energy eigenvalues and matrix elements of m.r.\ local operators
between pairs of eigenstates of the transfer matrix. From this analysis it
follows that hadron masses and matrix elements between zero momentum
eigenstates have definite parity under $\omega\to -\omega$.

Furthermore we show that one can introduce a parity label for the
eigenstates of the transfer matrix that turns
out to coincide with the usual parity quantum number when parity is restored
as a symmetry of the theory, i.e.\ either at $\omega=0$ or in the continuum
limit. This notion is interesting in itself as it allows to control the
magnitude of parity violating contributions in lattice correlators and
proves to be useful in discussing O($a$) improvement in the special case
$\omega=\pm\pi/2$.

\subsubsection{Matrix elements, energy eigenvalues and masses}
\label{sec:OPTR}

It is not difficult to analyse the transformation properties of energy
eigenvalues and matrix elements under $\omega\rightarrow -\omega$ and prove
that masses are $\omega$-even quantities.

Let us start by considering the two-point correlation function
\beq
G_{hh}({\bf x},t;\omega,r,m_q)=\langle \Phi_h({\bf x},t)\,
\Phi_h^{\dagger}({\bf 0},0)\rangle\Big{|}^{(\omega)}_{(r,m_q)}
\, ,\label{HHKO}\eeq
where $\Phi_h$ is a m.r.\ local operator with definite formal parity,
$(-1)^{p_{\Phi_h}}$. The spatial FT of~(\ref{HHKO})
\beq
G_{hh}({\bf k},t;\omega,r,m_q)=a^3\sum_{\rm{\bf x}}e^{i{\bf {k\cdot x}}}
\langle \Phi_h({\bf x},t)\,
\Phi_h^{\dagger}({\bf 0},0)\rangle\Big{|}^{(\omega)}_{(r,m_q)}\label{HHFTK}
\eeq
admits the spectral representation ($t>0$)
\beq
G_{hh}({\bf k},t;\omega,r,m_q)=\sum_{n}
\frac{e^{-E_{h,n}({\bf k};\omega,r,m_q) t}}{2E_{h,n}({\bf k};\omega,r,m_q)}
|R_{h,n}({\bf k};\omega,r,m_q)|^2 \, ,\label{QMKPP}\eeq
where
\beq
R_{h,n}({\bf k};\omega,r,m_q)=
\<\Omega|\Phi_h({\bf 0},0)|h,n,{\bf k}\>\Big{|}^{(\omega)}_{(r,m_q)}
\label{RNPMK}\eeq
and $|h,n,{\bf k}\>|_{(r,m_q)}^{(\omega)}$ is the eigenstate of
the transfer matrix, $\widehat{T}(\omega,r,m_q)$, that belongs to
the eigenvalue $E_{h,n}({\bf k};\omega,r,m_q)$, according to the
relation~\footnote{Actually a transfer matrix has been constructed only
for $|r|=1$. For $|r|<1$ one should make reference to the transfer matrix
over two lattice spacings (see Appendix~B). For simplicity we shall ignore
this subtlety in the following.}
\beq
\widehat{T}(\omega,r,m_q)|h,n,{\bf k}\>\Big{|}^{(\omega)}_{(r,m_q)} =
e^{-a E_{h,n}({\bf k};\omega,r,m_q)}|h,n,{\bf k}\>\Big{|}^{(\omega)}_{(r,m_q)}
\, .\label{EVTM}\eeq

The key formula is the relation
\beq
G_{hh}({\bf k},t;\omega,r,m_q)=
G_{hh}(-{\bf k},t;-\omega,r,m_q)\, ,\label{PKO}
\eeq
which follows by performing in the functional integral defining the
expectation value in~(\ref{HHFTK}) the change of integration variables
induced by the parity transformation~(\ref{PAROP}).  Eq.~(\ref{PKO})
implies similar relations for energy eigenvalues and the modulus of
matrix elements, namely
\beqn
&&E_{h,n}({\bf k};\omega,r,m_q) = E_{h,n}(-{\bf k};-\omega,r,m_q) \, .
\label{EPOMKAP}\\
&&|R_{h,n}({\bf k};\omega,r,m_q)|=|R_{h,n}(-{\bf k};-\omega,r,m_q)|\, ,
\label{MTOMKAP}\eeqn
At ${\bf{k}}={\bf{0}}$ eq.~(\ref{EPOMKAP}) becomes the announced result
about the invariance of masses under a sign change of $\omega$, i.e.
\beq
M_{h,n}(\omega,r,m_q) = M_{h,n}(-\omega,r,m_q) \, .
\label{EPOMEGA}
\eeq
In Appendix~E we sketch how this result can be extended to the baryonic
sector of the theory.

\subsubsection{The parity of the transfer matrix eigenstates}
\label{sec:PTME}

As we already remarked, the symmetry under parity of the (continuum
and) standard Wilson theory is replaced in tm-LQCD by the invariance
of the action under the transformation
${\cal{P}}^{\rm{sp}}\equiv{\cal{P}}\times (\omega\rightarrow -\omega)$.

Since at $\omega\neq 0$ parity is broken, a natural question to ask is how
the parity operator, $\widehat{\cal P}$, acts on the eigenstates of the
transfer matrix. The answer is surprisingly simple and it is contained in
the formula
\beq
\widehat{\cal{P}}\,|h,n,{\bf k}\>\Big{|}^{(\omega)}_{(r,m_q)}=
\eta_{h,n}|h,n,-{\bf k}\>\Big{|}^{(-\omega)}_{(r,m_q)}\, ,
\quad\quad\eta_{h,n}^2=1 \, .\label{PARSTATT}\eeq
This equation, which is proved in a costructive way in Appendix~F, allows us
to uniquely associate a sign, $\eta_{h,n}=\pm 1$, to each eigenstate of the
transfer matrix. With a little abuse of language we will simply call it
the ``parity'' of the state, in view of the fact that at $\omega=0$
the above formula precisely yields the usual definition of parity.
Eq.~(\ref{PARSTATT}) implies for the matrix elements of a
m.r.\ operator $B$ with definite formal parity $(-1)^{p_B}$ the
important relation
\beq
\<h,n,{\bf k}|B|h',n',{\bf k}'\>\Big{|}^{(\omega)}_{(r,m_q)}=\eta_{hn,h'n'}^B
\<h,n,-{\bf k}|B|h',n',-{\bf k}'\>\Big{|}^{(-\omega)}_{(r,m_q)}
\, ,\label{PARMAT}\eeq
where
\beq
\eta_{hn,h'n'}^B=\eta_{h,n}(-1)^{p_B}\eta_{h',n'}\, .
\label{PARETA}\eeq
Eq.~(\ref{PARMAT}) confirms that the label we have introduced coincides
in value with the usual parity quantum number of the continuum states.
In fact, at zero spatial momenta, ${\bf k}={\bf k}'={\bf 0}$, the matrix
element $\<h,n,{\bf 0}|B|h',n',{\bf 0}\>|_{(r,m_q)}^{(\omega)}$ is odd
in $\omega$ if the product $\eta_{h,n}\eta_{h',n'}$ does not match the
formal parity of $B$, thus it vanishes at $\omega=0$ and by universality
it must also do so as $a \to 0$ for generic $\omega$. This means that for
sufficiently small $a$ it is O($a$) or higher. If instead
$\eta_{hn,h'n'}^B=+1$, eq.~(\ref{PARMAT}) at vanishing three-momenta simply
tells us that the matrix elements of $B$ are even in $\omega$, thus not
necessarily vanishing in the continuum limit.

\subsubsection{Deciding the parity of transfer matrix eigenstates}
\label{sec:DWPC}

We want to schematically describe how one can in practice decide what is
the parity label, as defined by eq.~(\ref{PARSTATT}), of a given transfer
matrix eigenstate.

Let $\Phi_1$ and $\Phi_2$ be two local, m.r.\ operators with
the same unbroken lattice quantum numbers and definite opposite
formal parities, $\eta_{1}=(-1)^{p_1}$ and $\eta_{2}=(-1)^{p_2}$
($\eta_1=-\eta_2$) under ${\cal{P}}$.
As a concrete example one may think of the currents in eqs.~(\ref{GENCUR2})
and~(\ref{GENCUR4}). Consider the four correlators ($t>0$)
\beq
G_{ij}(t;\omega,r,m_q)=a^3\sum_{\rm{\bf x}} \langle \Phi_i({\bf x},t)
\Phi_j^{\dagger}({\bf 0},0)\rangle\Big{|}^{(\omega)}_{(r,m_q)}
\quad i,j=1,2 \, . \label{GIJ} \eeq
Performing the change of fermionic integration variables induced by
the transformation~(\ref{PAROP}), it is easily proved that the
correlators with $i=j$ are even in $\omega$, hence of O(1) in $a$,
while the correlators with $i\neq j$ are odd and by universality of
O($a$). The spectral decomposition of the correlators $G_{ij}$ reads
\beq
G_{ij}(t;\omega,r,m_q)=
\sum_{n}\frac{e^{-M_{h,n}(\omega,r,m_q)t}}{2M_{h,n}(\omega,r,m_q)}
\<\Omega|\Phi_i|h,n\>\<h,n|\Phi_j^{\dagger}|\Omega\>
\Big{|}^{(\omega)}_{(r,m_q)}\, ,\label{SPDC}\eeq
where the states $|h,n\>|^{(\omega)}_{(r,m_q)}$~\footnote{We drop the
momentum label ${\bf k}={\bf 0}$ as it is of no relevance for the present
argument.} are characterized by the set of conserved quantum numbers $h$ and
the integer $n$ which labels them in order of increasing mass.

Consider now the state lying at level $n$ and compare the quantities
\beq
\Big{|}\<h,n|\Phi_1|\Omega\>|^{(\omega)}_{(r,m_q)}\Big{|}^2
\quad {\mbox{and}}\quad
\Big{|}\<h,n|\Phi_2|\Omega\>|^{(\omega)}_{(r,m_q)}\Big{|}^2\, ,
\label{CMEL}\eeq
that appear in the spectral expansions~(\ref{SPDC}) with $i=j$.
Since the cross product $\<\Omega|\Phi_1|h,n\>
\<h,n|\Phi_2^{\dagger}|\Omega\>|^{(\omega)}_{(r,m_q)}$ is necessarily
of O($a$), as this quantity (or its c.c.) is the $n^{th}$ coefficient in the
expansions~(\ref{SPDC}) with $i\neq j$, we find that for each eigenstate
one must have either
\beq
\Big{|}\<h,n|\Phi_1|\Omega\>|^{(\omega)}_{(r,m_q)}\Big{|}^2\gg
\Big{|}\<h,n|\Phi_2|\Omega\>|^{(\omega)}_{(r,m_q)}\Big{|}^2\, ,
\label{MLARG}\eeq
or
\beq
\Big{|}\<h,n|\Phi_2|\Omega\>|^{(\omega)}_{(r,m_q)}\Big{|}^2 \gg
\Big{|}\<h,n|\Phi_1|\Omega\>|^{(\omega)}_{(r,m_q)}\Big{|}^2\, .
\label{MSMAL}\eeq
The reason is simply that in order for the four correlators~(\ref{GIJ})
to have the right ``order of magnitude'' in $a$, one has to have either
\beq
\<h,n|\Phi_1|\Omega\>
\Big{|}^{(\omega)}_{(r,m_q)}={\mbox{O}}(1) \quad {\mbox{and}}\quad
\<h,n|\Phi_2|\Omega\>\Big{|}^{(\omega)}_{(r,m_q)}={\mbox{O}}(a)\, ,
\label{HYP1}\eeq
or
\beq
\<h,n|\Phi_1|\Omega\>\Big{|}^{(\omega)}_{(r,m_q)}=
{\mbox{O}}(a) \quad {\mbox{and}}\quad
\<h,n|\Phi_2|\Omega\>\Big{|}^{(\omega)}_{(r,m_q)}={\mbox{O}}(1)\, .
\label{HYP2}\eeq
In case the inequality~(\ref{MLARG}) holds, with ${\eta_{h,n}}$
the parity of the state $|h,n\>$, we will set
\beq
\eta_{h,n}=(-1)^{p_{1}}\, .
\label{DEFPAR1}\eeq
If viceversa the inequality~(\ref{MSMAL}) is satisfied, we will set
\beq
\eta_{h,n}=(-1)^{p_{2}}\, .
\label{DEFPAR2}\eeq
Notice that the conditions~(\ref{MLARG}) and~(\ref{MSMAL}) are not
difficult to analyze numerically in the scaling region, as one is
comparing O(1) with O($a^2$) quantities. For the same reason,
sufficiently close to the continuum limit the value assigned to $\eta_{h,n}$
will be independent from the choice of the operators $\Phi_1$ and $\Phi_2$
and the details of their lattice discretization. This must be so, if
$\eta_{h,n}$ has to be interpreted as the parity label for the state
$|h,n\>^{(\omega)}_{(r,m_q)}$.

We end with an observation about isospin. Although tm-LQCD breaks isospin,
``partial'' isospin labels that become conserved quantum numbers
in the continuum limit can be attached to transfer matrix eigenstates.
This possibility results from the fact that the transformations
${\cal{P}}_{F}^{1,2}$ can be written in the form
\beq
{\cal{P}}_{F}^{1,2}={\cal{P}}\times {\cal{T}}^{1,2}\, ,
\label{PI}\eeq
where
\begin{equation}
{\cal{T}}^{1,2} : \left \{\begin{array}{ll}
&\hspace{-.3cm}\psi_{\rm{ph}}\rightarrow\,
e^{i\pi\tau_{1,2}/2}\,\psi_{\rm{ph}}  \\\\
&\hspace{-.3cm}\bar{\psi}_{\rm{ph}}\rightarrow \, \bar{\psi}_{\rm{ph}}\,
e^{-i\pi\tau_{1,2}/2} \end{array}\right . \label{ISO12} \end{equation}
are discrete isospin transformations~\footnote{While the transformations
${\cal{P}}_{F}^{1,2}$ have the form~(\ref{POT}) in all the quark bases
we have considered, ${\cal{P}}$ and ${\cal{T}}^{1,2}$ have only been
reported in the text in the physical basis.}. We remark that the combined
transformation ${\cal{T}}^{1,2}\times (\omega\rightarrow -\omega)$ sends the
action~(\ref{PHYSCHI}) into itself, just like the product
${\cal{P}}\times (\omega\rightarrow -\omega)$ does.

Eq.~(\ref{PI}) proves our statement. In fact, as soon as a parity
has been attributed to a state, definite values for the isospin labels
associated to the discrete transformations ${\cal{T}}^{1,2}$ are in turn
uniquely determined, as the transformations ${\cal{P}}_{F}^{1,2}$ are
exact symmetries of the action~(\ref{PHYSCHI}).

\subsection{Some observations about the practical use of tm-LQCD}
\label{sec:CAVEATS}

In this section we want to address two important issues concerning the
practical feasibility of numerical simulations within a lattice
regularization which, like tm-LQCD, besides chiral symmetry, also breaks
parity. The first has to do with the problem of taking the limit
$m_q\rightarrow 0$ and the second with the question of the scaling behaviour
and the numerical importance of contributions with ``wrong parity'' to
lattice correlators.

\subsubsection{The limit $m_q\rightarrow 0$}
\label{sec:CPV}

The subtleties of the limit $m_q\rightarrow 0$ we have emphasized
in the case of Wilson fermions show up also in the case of
tm-LQCD. Their discussion does not reveal any new feature here and
we refer the reader to sect.~\ref{sec:LMQO}. In particular (in the
presence of S$\chi$SB) to ensure that the chiral phase of the
vacuum is determined by the quark mass term proportional to $m_q$,
and not by the twisted Wilson term, it is necessary to work with
lattice parameters satisfying the bound~(\ref{COND1}) in the
unimproved case, or possibly under the weaker
condition~(\ref{COND2}) if one is dealing with O($a$) improved
correlators. When these order of magnitude conditions are violated,
cutoff effects on lattice observables may become numerically
large. As discussed in sect.~\ref{sec:LMQO}, it remains the fact that
the continuum limit should in principle be taken before the chiral
limit.

\subsubsection{``Wrong parity'' contributions}
\label{sec:WPC}

The discussion of sect.~\ref{sec:ODMME} is relevant for the analysis of
actual simulation data, where energy eigenvalues and operator matrix
elements are evaluated (at finite lattice spacing $a$, usually for
$m_q\neq 0$) by parameterizing lattice expectation values in terms of a
spectral representation. As a consequence of the breaking of parity
at $\omega \neq 0$,
contributions from states with both values of the parity label
we have introduced above will appear in the spectral representation.

To illustrate the situation in a simple instance and show how one can
possibly control these effects, let us consider the typical case of a
two-point correlator with operators inserted at zero three-momentum
\beq
G_{\Phi\Xi}(t;\omega,r,m_q)=a^3\sum_{\rm{\bf x}} \langle \Phi({\bf x},t)
\Xi^{\dagger}({\bf 0},0)\rangle\Big{|}^{(\omega)}_{(r,m_q)}
\label{GPX} \eeq
with $\Phi$ and $\Xi$ m.r.\ local operators having equal formal parity.
{}From all we said in sect.~\ref{sec:ODMME} it should be clear that
$G_{\Phi\Xi}$ is even in $\omega$. As for the matrix elements of the local
operators between the vacuum and the eigenstates of the transfer matrix,
there are two possibilities: 1) if the intermediate state in consideration
has the same parity as the operators $\Phi$ and $\Xi$, both matrix elements
are $\omega$-even and O(1) in the continuum limit; 2) if the
intermediate state has the opposite parity, the two matrix elements are
$\omega$-odd and vanish in the continuum limit at least as $a$. Thus
``wrong parity'' contributions, necessarily appearing in pairs, lead
to O($a^2$) (or higher) effects.

Arguments of this kind can be constructed for more complicated
$\omega$-even correlators, as for instance the three-point correlators
with two of the operators inserted at zero three-momentum. As intermediate
state projectors occur in pairs in the spectral representation, contributions
from states of ``wrong parity'' (compared to the naively expected one),
necessarily contain an even number of $\omega$-odd matrix elements. Thus
again states with ``wrong parity'' lead to contributions of O($a^2$) or
higher. When operators are inserted at non-vanishing three-momentum, the
situation is governed by eq.~(\ref{PARMAT}). No new special problems arise
due to the breaking of parity. Though a number of specific results could be
derived on the subject, we will not pursue any further their discussion
in this paper.

Notwithstanding all the control we have on ``wrong parity'' contributions,
one should not forget that, if intermediate states with naively
unexpected parity come into play, they cannot be ignored altogether,
because some of them may have masses smaller, or not much larger, than the
state(s) of interest for the problem under study. Thus, for some range of
values of the relevant time separations, the contributions from states with
``wrong parity'' can be numerically non-negligible. In this
situation, if one does not want to end up with intolerably large lattice
discretization errors on the final result~\footnote{Non-large O($a^2$) errors
on physical quantities extracted from $\omega$-even correlators are however
perfectly tolerable.}, a disentangling procedure of the type one is used to
in the case of mixings among low-lying and excited states has to be carried
out. This task requires the evaluation of several correlators differing
among each other by the formal parity of the interpolating operators.

\section{A special case: $\omega=\pm\pi/2$}
\label{sec:ASC}

The choice $\omega=\pm\pi/2$ (full twist) in the tm-LQCD
action~(\ref{PHYSCHI}) is especially worth mentioning, because all
quantities that are even under $\omega\rightarrow -\omega$ are O($a$)
improved with no need of any averaging. This result is a consequence of
the fact that for the particular value $\omega=\pi/2$ (or $\omega=-\pi/2$)
of the twisting angle a sign inversion of the latter is equivalent
(mod\,($2\pi)$) to a shift by $\pi$. This operation is in
turn the same as inverting the sign of $r$. The quantities that are
even in $\omega$ are hence also even in $r$. In this situation the two
terms entering the Wilson average are just identical, therefore performing
the average is unnecessary to get up O($a$) improved lattice quantities.

The list of physical quantities that are automatically O($a$) improved
includes among others the following important physical quantities (see
sect.~\ref{sec:ODMME})
\begin{enumerate}
\vspace{-0.2cm}
\item hadronic masses
\vspace{-0.2cm}
\item matrix elements
$\<h,n,{\bf 0}|B|h',n',{\bf 0}\>|^{(\pm{\pi}/{2})}_{(r,m_q)}$,
of operators whose formal parity is equal to the product of the
parities of the external states.
\end{enumerate}
Actually one can prove more than that. One can prove that taking
appropriate linear combinations of quantities computed with opposite values
of the external momenta leads to O($a$) improved lattice data. This
observation is very important as it shows that O($a$) improved energies and
matrix elements can be obtained from simulations carried out within a single
lattice regularization, i.e.\ without the need to consider averaging over
simulations performed with opposite values of $r$.

\subsection{O($a$) improvement of energy eigenvalues and matrix elements}
\label{IEEEME}

The O($a$) improvement of averages of energy eigenvalues with opposite
momenta at, say, $\omega=\pi/2$ follows from the chain of equalities
\beq
E_{h,n}({\bf{k}};\frac{\pi}{2},r,m_q)=
E_{h,n}({\bf{k}};-\frac{\pi}{2},-r,m_q)=
E_{h,n}(-{\bf{k}};\frac{\pi}{2},-r,m_q)\, ,
\label{CEQ}\eeq
where the first equality is a consequence of the fact that
$(\omega\to\omega+\pi)\times (r\to -r)=1\!\!1$ and the second of
eq.~(\ref{EPOMKAP}). The same result could be obtained by using the
spurionic symmetry (only valid at $\omega=\pm\pi/2$)
${\cal{P}}\times (r\to -r)$.

At this point it is enough to observe that the $WA$ formula for energy
eigenvalues (analogous to the eq.~(\ref{MASSIMP}) valid in the
Wilson case)
\beq
\frac{1}{2} \Big{[}E_{h,n}({\bf k};\frac{\pi}{2},r,m_q)+
[r\rightarrow -r]\Big{]}=E^{\rm cont}_{h,n}({\bf k},m_q)+{\rm O}(a^2)
\, ,\label{ENIMP}\eeq
becomes
\beq
\frac{1}{2} \Big{[}E_{h,n}({\bf k};\frac{\pi}{2},r,m_q)+
[{\bf{k}}\rightarrow -{\bf{k}}]\Big{]}=E^{\rm cont}_{h,n}({\bf k},m_q)+
{\rm O}(a^2)\, .\label{ENIMPK}\eeq

Concerning the matrix elements of local operators between pairs
of transfer matrix eigenstates at non-zero spatial momenta, we recall
that we only have the relation~(\ref{PARMAT}) from which automatic O($a$)
improvement does not follow. Although improvement can always be achieved
via $WA$, at the particular value of the twisting angle we are
considering in this section there is the possibility of getting O($a$)
improved lattice matrix elements without the need of carrying out two
separate computations. Consider, in fact, the linear combination
\beqn
&&\hspace{-.8cm}\<h,n,{\bf k}|B|h',n',{\bf k}'\>
\Big{|}^{(\omega)\,{\rm{sym}}}_{(r,m_q)}=\label{SYMAV}
\\&&\hspace{-.8cm}=\frac{1}{2}\Big{[}\<h,n,{\bf k}|B|h',n',{\bf k}'\>
\Big{|}^{(\omega)}_{(r,m_q)}+
\eta_{hn,h'n'}^B\<h,n,-{\bf k}|B|h',n',-{\bf k}'\>
\Big{|}^{(\omega)}_{(r,m_q)}\Big{]}\, .\nonumber\eeqn
Thanks to eq.~(\ref{PARMAT}), this particular quantity
is symmetric in $\omega$, thus O($a$) improved at $\omega=\pm\pi/2$. {}From
the extension of eq.~(\ref{FINHBH}) to the tm-LQCD case with, say,
$\omega=\pi/2$ one can make contact with the continuum, obtaining
\beqn
&&\hspace{-.8cm}\<h,n,{\bf k}|B|h',n',{\bf k}'\>
\Big{|}_{(r,m_q)}^{(\frac{\pi}{2})\,\rm{sym}}=\label{SYMAVC}
\\&&\hspace{-.8cm}=\frac{1}{2}\Big{[}\<h,n,{\bf k}|B|h',n',{\bf k}'\>
\Big{|}_{(r,m_q)}^{(\frac{\pi}{2})}+\<h,n,{\bf k}|B|h',n',{\bf k}'\>
\Big{|}_{(-r,m_q)}^{(\frac{\pi}{2})}\Big{]}=\nonumber\\
&&\hspace{-.8cm}=  \zeta^{B}_{B}(\frac{\pi}{2},r)
\langle h,n,{\bf k} |B|h',n',{\bf k}'\rangle\Big{|}^{\rm{cont}}_{(m_q)}
+{\rm{O}}(a^2)\, ,\nonumber\eeqn
where in the first equality the spurionic symmetry
${\cal{P}}\times (r\to -r)$ has been used again.

Symanzik expansions too simplify noticeably at $\omega=\pm\pi/2$. Combining
eqs.~(\ref{PAROMR}) with~(\ref{OMPAR}), one obtains for an operator
of formal parity $(-1)^p$ (eq.~(\ref{OPARGEN})) the set of relations
\beqn
&&\hspace{-.5cm}\zeta^{O^{(p)}}_{O^{(p)}}(\pi/2,r)
\stackrel{\rm{eq}.(\ref{OMPAR})}=\zeta^{O^{(p)}}_{O^{(p)}}(-\pi/2,r)
\stackrel{{\rm{eq}}.(\ref{PAROMR})}=\zeta^{O^{(p)}}_{O^{(p)}}(\pi/2,r)
\nonumber\\
&&\hspace{-.5cm}\xi^{O^{(p)}}_{O^{(p)}}(\pi/2,r)
\stackrel{{\rm{eq}}.(\ref{OMPAR})}=\xi^{O^{(p)}}_{O^{(p)}}(-\pi/2,r)
\stackrel{{\rm{eq}}.(\ref{PAROMR})}=
-\xi^{O^{(p)}}_{O^{(p)}}(\pi/2,r)\label{OMPI}\\
&&\hspace{-.5cm}\eta^{O^{(p)}}_{O_\ell}(\pi/2,r)
\stackrel{{\rm{eq}}.(\ref{OMPAR})}=(-1)^{p+p_\ell}
\eta^{O^{(p)}}_{O_\ell}(-\pi/2,r)\stackrel{{\rm{eq}}.(\ref{PAROMR})}=
-(-1)^{p+p_\ell}\eta^{O^{(p)}}_{O_\ell}(\pi/2,r)\, .
\nonumber\eeqn {}From the above equations one concludes that the
coefficients $\xi^{O^{(p)}}_{O^{(p)}}(\pm\pi/2)$ are vanishing and
that the only operators, $O_\ell$, which can contribute to the
Symanzik expansion of $\langle O^{(p)}\rangle|_{(\pm\pi/2,m_q)}$
have $p_\ell=p+1$~mod\,(2), i.e.\ are those which are multiplied
by Symanzik coefficients odd under a sign inversion of the
twisting angle. This analysis confirms that the whole mass
spectrum and all the matrix elements of local operators that are
even under sign inversion of the twisting angle are automatically
O($a$) improved.

The same results could be obtained by noticing that the
action~(\ref{PHYSCHI}) with $\omega=\pm\pi/2$ is invariant under the
combined transformation~\footnote{This transformation is the product of
${\cal{R}}^{\rm{sp}}_5$ (which is valid for any $\omega$) times
${\cal{P}}\times (r\to -r)$ (which only holds at $\omega=\pm\pi/2$).}
\begin{equation}
{\cal R}_5\,\times\,{\cal{P}}\,\times\,(m_q\rightarrow -m_q)\, .\label{PR5}
\end{equation}
Analogously to what we did in sects.~\ref{sec:IQCD} and~\ref{sec:ICGA},
to get eqs.~(\ref{OMPI}) it is enough to equate the Symanzik expansions
of the two sides of the equation
\beqn
&&\langle O^{(p)}(x_1,x_2,\ldots ,x_n)\rangle
\Big{|}_{(r,m_q)}^{(\frac{\pi}{2})}=
\nonumber\\&&=(-1)^{{P_{{\cal{R}}_5}[O]}+p}\langle
O^{(p)}(x_{P1},x_{P2},\ldots ,x_{Pn})\rangle
\Big{|}_{(r,-m_q)}^{(\frac{\pi}{2})} \, ,\label{RPM}\eeqn
which is obtained performing in the functional integral the change of
integration variables induced by the transformation
${\cal R}_5\,\times\,{\cal{P}}$. At this point one has to make use of the
parity and chirality properties of the continuum theory as well as the
relation~(\ref{PPAR}).

We wish to end this section by explicitly noting that, remarkably, the
critical WD operator appearing in the fermion action ~(\ref{PHYSCHI}) at
$\omega=\pm\pi/2$
\beq
D^{(\pm\pi/2)}_{\;\rm{cr}}=\frac{1}{2}\sum_\mu\gamma_\mu(\nabla^\star_\mu+
\nabla_\mu) \pm i\gamma_5\tau_3\Big{(}a\frac{r}{2}
\sum_\mu\nabla^\star_\mu\nabla_\mu -
M_{\rm{cr}}(r)\Big{)}\label{WDCR}\eeq
is anti-Hermitian, so its spectrum is purely imaginary. Consequently the
full WD operator cannot have any vanishing eigenvalue as soon as $m_q\neq 0$
(see eq.~(\ref{FERDET}) at $\omega=\pm\pi/2$).

\subsection{An application: the computation of $F_\pi$}
\label{sec:ASA}

As an example of the peculiarity of the choice $\omega=\pm\pi/2$, we wish
to discuss the computation of the pion decay constant, $F_\pi$. It was
already pointed out in refs.~\cite{FS,FGSW} that within the fully twisted
formulation of LQCD $F_\pi$ can be obtained from a conserved lattice current
(i.e.\ one with trivial renormalization properties) and without O($a$) cutoff
effects, provided the lattice action is improved \`a la Symanzik. Here we
restate this result in our notation and show that, in fact, in tm-LQCD
with $\omega=\pm\pi/2$, there is no need to improve the action.

In the formal continuum theory $F_\pi$ is defined by the
equation~\footnote{Strictly speaking eq.~(\ref{FPI}) is not written in a
completely correct way. To comply with eq.~(\ref{PARPRO}) (and always taking
$F_\pi > 0$) a factor equal to the sign of $m_q$ should appear in the r.h.s.
As it is, eq.~(\ref{FPI}) is valid for $m_q >0$, given our conventions
concerning the relative sign of the kinetic and mass terms in the fermionic
action.}
\beq
\langle\Omega|\hat{A}^1_0|\pi\rangle\Big{|}_{(m_q)}^{\rm cont}
= m_\pi\sqrt{2}F_\pi\, ,
\label{FPI}\eeq
with $\hat{A}_0^1$ the temporal component of the renormalized axial current,
and can be extracted by evaluating at large times, e.g., the ratio ($t>0$)
\beq
\frac{e^{m_\pi t/2}}{\sqrt{m_\pi}}
\frac{a^3\sum_{\rm{\bf x}}\langle\hat{A}_0^1({\bf x},t)
P^1(0)\rangle\Big{|}_{(\frac{\pi}{2},m_q)}}
{\sqrt{a^3\sum_{\rm{\bf x}}\langle P^1({\bf x},t)
P^1(0)\rangle\Big{|}_{(\frac{\pi}{2},m_q)}}}
\stackrel{t\rightarrow\infty}\rightarrow F_\pi\, ,
\label{API}\eeq
where $P^1$ is the pseudo-scalar quark density operator. In~(\ref{API}) we
did not multiply $P^1$ by its renormalization constant, because all $Z_P$
factors would anyway drop out in the ratio. The isospin index has been set
equal to 1, which corresponds to a charged pion. The reason for this
choice is that in this way other contributing intermediate states
are substantially heavier than the charged pion, including states with
``wrong parity'' (see the concluding remark in sect.~\ref{sec:WPC}).

Among the possible choices for the lattice (discretization of the) axial
current to be used in eq.~(\ref{API}), the most naive one is to take the
current $\hat{A}^1_\mu$ of eq.~(\ref{GENCUR4}), which at
$\omega_r=\omega=\pi/2$ simply becomes
\beq
\hat{A}^1_\mu(x)=Z_V\bar{\psi}_{\rm{ph}}(x)\frac{\tau_1}{2}
\gamma_\mu\gamma_5\psi_{\rm{ph}}(x)\, .
\label{AUMU}\eeq
The operator $\hat{A}^1_\mu$ has a definite (positive) ${\cal R}_5$-parity
(in the sense of the definitions~(\ref{KEYRELM0}) or~(\ref{KEYREL})).
At the cost of a preliminary determination of $Z_V$, an O($a$) improved
estimate of $F_\pi$ is then obtained with no need of averaging. We are, in
fact, dealing with zero three-momentum correlators, which are even
under a sign inversion of the twisting angle (sect.~\ref{sec:PTME})).

A remarkable fact about the choice $\pm\pi/2$ is that it is possible to get
an O($a$) improved estimate of $F_\pi$ which requires neither Wilson
averaging nor the computation of any renormalization factor. Let us,
in fact, consider the 1-point split current~\cite{FS,FGSW}
\beqn
&&\hspace{-1.cm}\hat{A}_\mu^1(x)\Big{|}^{1-{\rm{pt}}}=
\frac{1}{2}\Big{[}\bar\psi_{\rm{ph}}(x)\frac{\tau_1}{2}
\gamma_\mu\gamma_5 U_\mu(x)\psi_{\rm{ph}}(x+a\hat{\mu})+\nonumber\\
&&\hspace{-1.cm}+\bar\psi_{\rm{ph}}(x+a\hat{\mu})\frac{\tau_1}{2}
\gamma_\mu\gamma_5 U^\dagger_\mu(x)\psi_{\rm{ph}}(x)\Big{]}+
\nonumber\\
&&\hspace{-1.cm}-\frac{r}{2}\Big{[}\bar\psi_{\rm{ph}}(x)\frac{\tau_2}{2}
U_\mu(x)\psi_{\rm{ph}}(x+a\hat{\mu})-
\bar\psi_{\rm{ph}}(x+a\hat{\mu})\frac{\tau_2}{2}
U^\dagger_\mu(x)\psi_{\rm{ph}}(x)\Big{]}\, .\label{OPSCEX}\eeqn
The current~(\ref{OPSCEX}) has similarly to the current~(\ref{AUMU})
a definite (positive) ${\cal R}_5$-parity. It is obtained from the
eq.~(\ref{OPSC2}) with $b=2$ by setting $\omega_r=\pi/2$. With this choice
the transformation~(\ref{ITILDE}) is an axial transformation with isospin
index equal to 1. Since, as we remarked in sect.~\ref{sec:ACT}, this current
is exactly conserved at $m_q=0$, its renormalization constant is equal to 1.

The net result is that with the choice $\omega=\pm\pi/2$ of the twisting
angle in the action~(\ref{PHYSCHI}) and the use of the (conserved) 1-point
split current~(\ref{OPSCEX}) one can get an automatically O($a$) improved
determination of $F_\pi$ which furthermore does not require the computation
of any renormalization constant. This is certainly a remarkable result,
especially if one observes that no problems can occur with spurious fermionic
modes, while the required computational effort is minimal.

\section{Conclusions and outlook}
\label{sec:CONCL}

In this paper we have shown that in LQCD with Wilson fermions it is possible
to improve the approach to the continuum limit and the chiral behaviour of
correlation functions of gauge invariant m.r.\ operators by taking
arithmetic averages of pairs of correlators computed in theories
regularized with Wilson terms of opposite sign and identical values of $m_q$
(Wilson average), or equivalently the appropriate linear combination of the
correlators computed in a given regularization but with opposite values of
$m_q$ (mass average). Improved hadronic masses and matrix elements can be
similarly obtained by taking the Wilson average of the corresponding
quantities computed in the two different regularizations, or the appropriate
linear combination of quantities computed with opposite values of $m_q$.

To avoid the problems related to the nature of the spectrum of the WD operator
we advocate the use of tm-LQCD for the actual computation of the correlators
taking part to the averages. We have shown that all the nice cancellations
of O($a$) terms that we find in the standard Wilson case extend to tm-LQCD
with mass degenerate quark doublets. Moreover, it turns out that, if the
tm-LQCD formulation with $|\omega| = \pi/2$ is employed, all hadronic
energies and operator matrix elements can be evaluated with no O($a$) lattice
artifacts and without making recourse to $WA$'s, but taking (appropriate)
linear combinations of quantities evaluated with opposite external momenta.

A striking, though still partial, confirmation of the validity of
the approach outlined above and of the remarkable properties of
tm-LQCD at $|\omega|=\pi/2$ comes from the recent work of
ref.~\cite{KSUW}, where a (quenched) study of the scaling
behaviour of the pseudoscalar meson decay constant and the vector
meson mass was carried out down to small values of $\beta$.
Lattice results for these quantities at fixed pseudoscalar meson
mass ($\sim 700$~MeV) show surprisingly small cutoff effects from
$\beta=6.2$ to $\beta=5.85$. Furthermore a nice analytic check
of automatic O($a$) improvement of the pseudoscalar meson mass and
decay constant in tm-LQCD at $|\omega|=\pi/2$ has been obtained in
lattice chiral perturbation theory by the authors of
ref.~\cite{GMetal}. 

Our presentation of tm-LQCD, which follows a perspective complementary
to that of ref.~\cite{FGSW}, provides a simple understanding of the
many simplifications that have been previously observed to occur at
$|\omega|=\pi/2$~\cite{FGSW,FREZ,FSW,FS}, in the renormalization and O($a$)
improvement of the matrix elements of certain local operators, such as the
isotriplet axial current or the isosinglet scalar density.

All the results we have obtained are based on the observation that
$M_{\rm cr}(r)$ is odd under $r \rightarrow -r$, the
property~(\ref{OPTRA}) enjoyed by lattice and continuum
correlators (see Appendix~C), as well as the behaviour of O($a$)
terms under the spurionic symmetry ${\cal{R}}_5^{\rm{sp}}$
(eq.~(\ref{SPUR})). Considerations of this kind can be generalized
without any essential modification to the case of mass
non-degenerate quark doublets~\cite{CAIRNS}. The point will be
further discussed in a forthcoming paper~\cite{FRTWO}, where we
deal with the strategy for computing the matrix elements of the
(CP-conserving) effective weak Hamiltonian.

Before briefly illustrating the potential of the method we have presented
in this paper for further physical applications, we wish to comment on two
important, related aspects of our approach, namely its computational cost
compared to other schemes and the problem of how to most effectively perform
the statistical analysis of available simulation data.

\subsection{Computational cost}
\label{sec:CC}

The computational cost of the approach we propose is in the worst
case (i.e.\ if averaging is required) twice the typical one for
simulations with unimproved standard Wilson fermions or
tm-LQCD~\cite{FREZ,DMFH,BITAR,DMFH_LATT02,MMN}.

The gain offered by our strategy is, however, quite significant in
practical terms, because the systematic removal of O($a$) lattice artifacts,
allowing a better control of cutoff effects, gives the possibility
of working with coarser lattices than in the absence of improvement.
This may lead to a dramatic reduction of the computational cost especially
in the unquenched case, where CPU times have been estimated to increase
like $a^{-k}$ with $k\simeq 7$~\cite{PROC_BERL}. In the quenched case,
assuming ``ideal'' critical slowing down, one finds a similar behaviour
with $k=5$. It should be also mentioned that the efficiency of algorithms
for simulations with dynamical fermions of the multi-boson
type~\cite{LUSCHMB} is expected to significantly benefit from the
absence of the SW term~(\ref{SWSA}) in the lattice action.

\subsection{Statistical analysis of simulation data}
\label{sec:SASD}

A little thinking reveals that the most useful way to make use of
simulation data depends on whether the fermionic determinant~(\ref{FERDET})
(and hence the partition function) is invariant under $r\rightarrow -r$ (or
equivalently under the shift $\omega\rightarrow \omega+\pi$) or not.

If the determinant is not invariant under $r\rightarrow -r$, the computation
of Wilson averages requires to carry out two independent simulations leading
to two different Markov chains of gauge configurations. The statistical
analysis of the corresponding two sets of estimators, $\{X_{j,r}\}$
and $\{X_{j,-r}\}$ ($j=1, \ldots, N_{\rm meas}$), of the physical
quantity $X$ must be carried out separately. Only after this step has been
completed, the Wilson average of the two mean values and the associated
statistical error can be computed.

When the determinant is invariant, the whole functional measure is invariant
too and only one Markov chain of configurations needs to be generated. In
this situation the optimal procedure is to perform the Wilson average at
the level of the individual (possibly ``jackknived'') statistical estimators,
$X_{j,r}$. This is done by first computing the averages
\beq
X_{j}^{WA} = \frac{1}{2} \left(X_{j,r} + X_{j,-r}\right) \, ,
\quad\quad\quad\quad j=1, \ldots, N_{\rm meas} \, , \label{XPQ}\eeq
and then going on with the standard statistical analysis of the $N_{\rm meas}$
Wilson averaged quantities $X_{j}^{WA}$.

As we already noticed, in all cases the extra computational effort
necessary when averaging is required is not larger than a factor
$\sim 2$ as compared to the case of simulations with unimproved
fermions. Most interestingly, this factor reduces essentially to 1 in the
case of simulations with $|\omega|=\pi/2$, because Wilson averaging is
unnecessary.

{}From this discussion it emerges that the best strategy to deal with
the unquenched theory is to ``decouple'' sea and valence quarks, by having
all the sea quarks assembled in flavour doublets and regularized with
twisting angles equal to $\pm\pi/2$, while for valence quarks each flavour is
introduced with the chiral phase of the Wilson term chosen so as to minimize
the complexity of the mixing pattern of (valence) quark composite operators.
Unlike what is usually done, here we are proposing to regularize sea and
valence quarks with Wilson terms of not necessarily identical
structure~\cite{FRTWO}.

\subsection{Outlook}
\label{sec:OUT}

The last considerations bring about the question of the
computation of the matrix elements of the various (valence) quark
operators that are of interest for physical applications, like,
for instance, the chiral order parameter and the matrix elements
of the effective weak Hamiltonian, ${\cal{H}}_{\rm{eff}}$. In
particular for the CP-conserving sector of ${\cal{H}}_{\rm{eff}}$
we will show in~\cite{FRTWO} that, with a clever choice of the
twisting angle entering the Wilson term of the different valence
quarks, it is possible at the same time to have O($a$) improvement
and to get rid of all the ``wrong chirality'' mixings. The virtue
of tm-LQCD in simplifying the mixing pattern of CP-conserving four
fermion operators in both the $\Delta S=2$ and the $\Delta S=1$
sector was already observed in refs.~\cite{FGSW} and~\cite{PSV},
respectively. Besides achieving O($a$) improvement, we will
prove~\cite{CAIRNS,FRTWO} that mass non-degenerate sea quarks can
be regularized so as to also have the resulting fermionic
determinant real and positive.

We would like to close this list of applications by mentioning another
important case, where the strategy advocated in this work can be of help.
This is the computation of the decay constants, semi-leptonic form factors
and masses of heavy-light mesons, which has been recently shown to be
feasible with relativistic O($a$) improved $c$~\cite{SRJ} and, even,
$b$~\cite{PETR} quark, by making recourse in the last case to a clever use
of finite size scaling techniques to bridge the gap between the largely
different scales present in the problem.

\vspace{0.5cm}
{\bf  Acknowledgements} - We would like to thank R.~Sommer for many
useful suggestions and M.~Testa for an illuminating discussion. We
thank M.~Della~Morte for his help in preliminary numerical tests of ideas
developed in this paper and several discussions. Discussions with L.~Giusti,
A.~Pelissetto and A.~Vladikas are also acknowledged. This work was supported
in part by the European Community's Human Potential Programme under contract
HPRN-CT-2000-00145, Hadrons/Lattice QCD.

\appendix
\section*{Appendix A - Improving axial Ward--Takahashi identities}
\renewcommand{\thesection}{A}
\label{sec:APPA}

In this Appendix for illustrative purposes we discuss in full detail
the improvement of the flavour non-singlet axial WTI's in the framework of
the standard Wilson formulation.

{}From the action~(\ref{FERACT}) one derives the lattice identity
\beqn &&\langle {\partial}^\star_\mu
A^b_\mu(x)  P^b(0)\rangle\Big{|}_{(r,M_0)} = \label{IDENT}\\
&&=2M_0\langle  P^b(x) P^b(0)\rangle\Big{|}_{(r,M_0)} + \langle X^b(x)
P^b(0)\rangle\Big{|}_{(r,M_0)}  \quad x\neq 0\, , \nonumber \eeqn
where ${\partial}^\star_\mu$ is the discretized backward
derivative~\footnote{For notational simplicity we write formulae with
derivatives inside expectation values. It is, however, always understood
that derivatives have to be thought as acting outside. As in the main
text, we indicate nowhere the dependence upon the bare coupling constant,
$g^2_0$, unless necessary to remove  possible ambiguities.}. The lattice
expressions of ${\partial}^\star_\mu A^b_\mu$ and  $X^b$, whose explicit
form we do not need here, can be found in ref.~\cite{BMMRT}. $A^b_\mu$ and
$P^b$ are the axial current and the  pseudo-scalar quark density
(eq.~(\ref{DEFP})), respectively. To be concrete the inserted operator
in~(\ref{IDENT}) was taken to be the pseudo-scalar quark density. It would
not be difficult to generalize the arguments that follow by considering
the insertion of a more general m.r.\ operator with definite
$\gamma_5$-chirality.

In ref.~\cite{BMMRT} it is shown how one can recover
from~(\ref{IDENT}) the continuum WTI. Indeed, (\ref{IDENT}) can be
renormalized to yield \beqn &&{Z_A(r)}\langle {\partial}^\star_\mu
A^b_\mu(x) P^b(0)\rangle\Big{|}_{(r,M_0)} =\label{WTI}\\ &&=
{2[M_0-\bar{M}(r,M_0)]}\langle P^b(x)
P^b(0)\rangle\Big{|}_{(r,M_0)}  +{\rm{O}}(a) \quad x\neq 0\, .
\nonumber \eeqn The two sides of eq.~(\ref{WTI}) possess a well
defined continuum limit, as soon as they are multiplied by a
suitable (logarithmically divergent) renormalization constant
(i.e.\ the renormalization constant of the pseudo-scalar quark
density), if the quantities $Z_A$ and $\bar{M}$ are set to their
appropriate values. Comparing the expression~(\ref{WTI}) with the
form of the continuum WTI \beq \langle\partial_\mu
\hat{A}^b_\mu(x)\hat{P}^b(0)\rangle =2\hat{m}_q\langle
\hat{P}^b(x) \hat{P}^b(0)\rangle\, , \label{CWTI}\eeq one
concludes that $Z_A$ is the renormalization constant of the
lattice axial current and the critical mass is the value of $M_0$
which solves the equation $M_0=\bar{M}(r,M_0)$. We note
incidentally that from the transformation properties of the
various terms in eqs.~(\ref{IDENT}) and~(\ref{WTI}) under
${\cal{R}}_5$, one can infer the relations
\beqn &&Z_A(r)=Z_A(-r)\, ,\label{ZA}\\
&&\bar{M}(r,M_0)=-\bar{M}(-r,-M_0)\, .\label{MBAR} \eeqn
Eqs.~(\ref{ZA}) and (\ref{MBAR}) confirm the result that the renormalization
constant of a multiplicative renormalizable operator is generally even under
$r\rightarrow -r$ (see sect.~\ref{sec:RC}), while the critical mass
is odd (eq.~(\ref{MCR})).

The relations between the renormalized continuum quantities indicated with the
symbol $\,{\bf \hat{} }\,$ in eq.~(\ref{CWTI}) and the corresponding bare
lattice ones are
\beqn
&&\hat{A}^b_\mu=Z_A A^b_\mu\label{CONLATA}\\
&&\hat{P}^b=Z_P P^b\label{CONLATP}\\
&&\hat{m}_q=z_m Z_P^{-1} (M_0-M_{\rm{cr}})= Z_{S^0}^{-1} m_q\, ,
\label{CONLATM}\eeqn
where $Z_{S^0}$ is the multiplicative renormalization constant
of the (subtracted) isosinglet scalar quark density and
\beq
z_m(r)=1-\frac{\partial\bar{M}(r,M_0)}{\partial M_0}
\Big{|}_{M_0=M_{\rm{cr}}(r)}\, .
\label{ZM}\eeq
The last equality in eq.~(\ref{CONLATM}) follows from the identity
\beq
z_m=\frac{Z_P}{Z_{S^0}}\, ,
\label{ZMZPZS}\eeq
proved in ref.~\cite{MMPT}.

As we argued in the text, in order to discuss the issue of improvement
one has to make reference to the notion of effective theory introduced
by Symanzik and use the related Symanzik expansion of lattice correlators
in terms of correlators of the continuum theory. Recalling the analysis
carried out in ref.~\cite{LUS1}, one gets up to O($a$) corrections the
explicit expansions ($x\neq 0$)
\beqn &&\langle P^b(x)P^b(0)\rangle\Big{|}_{(r,m_q)} =
[\zeta^{PP}_{PP}(r)+am_q\xi^{PP}_{PP}(r)]\langle
P^b(x) P^b(0)\rangle\Big{|}^{\rm{cont}}_{(m_q)} +\nonumber\\&& + a\,
{\eta}^{PP}_{\rm{SW}}(r)\langle P^b(x)P^b(0) \int d^4y\,{\cal{L}}_{\rm{SW}}(y)
\rangle\Big{|}^{\rm{cont}}_{(m_q)}+\nonumber\\ && + a
m_q\,{\eta}^{PP}_{FF}(r)\langle P^b(x)P^b(0)\int d^4y\,
 FF(y)\rangle\Big{|}^{\rm{cont}}_{(m_q)}+\nonumber\\ && + a m_q^2
\,{\eta}^{PP}_{\bar{\psi}\psi}(r)\langle P^b(x)P^b(0)\, \int
d^4y\,\bar{\psi}\psi(y)\rangle\Big{|}^{\rm{cont}}_{(m_q)}+ {\rm{O}}(a^2)\, ,
\label{PPEXPMS}  \eeqn
\beqn &&\langle\tilde{\partial}_\mu
A^b_\mu(x)P^b(0)\rangle\Big{|}_{(r,m_q)}= [\zeta^{AP}_{AP}(r)+am_q
\xi^{AP}_{AP}(r)]\langle\partial_\mu
 A^b_\mu(x)P^b(0)\rangle\Big{|}^{\rm{cont}}_{(m_q)}+\nonumber\\
&&+a\,\eta_{PP}^{AP}(r)\langle\partial_\mu\partial_\mu
P^b(x) P^b(0)\rangle\Big{|}^{\rm{cont}}_{(m_q)}+\nonumber\\
&& + a\,\eta^{AP}_{\rm{SW}}(r) \langle\partial_\mu A^b_\mu(x)P^b(0) \int
d^4y\,{\cal L}_{\rm{SW}}(y) \rangle\Big{|}^{\rm{cont}}_{(m_q)}+\nonumber\\
&& + a m_q\,\eta^{AP}_{FF}(r) \langle\partial_\mu A^b_\mu(x)P^b(0)\int
d^4y\, FF(y) \rangle\Big{|}^{\rm{cont}}_{(m_q)}+\nonumber\\ && + a m_q^2
\,\eta^{AP}_{\bar{\psi}\psi}(r) \langle\partial_\mu
A^b_\mu(x)P^b(0)\int d^4y\,\bar{\psi}\psi(y)
\rangle\Big{|}^{\rm{cont}}_{(m_q)}+{\rm{O}}(a^2)\, , \label{APEXPMS} \eeqn
where $FF$ is a shorthand for ${\rm{Tr}}[F_{\mu\nu}F_{\mu\nu}]$ and
\beq
{\cal L}_{\rm{SW}}(x)=
\bar\psi\frac{i}{4}\sigma_{\mu\nu}{F}_{\mu\nu}\psi(x)\, .
\label{LSW}\eeq
To avoid trivial O($a$) effects we have replaced the backward lattice
derivative of eq.~(\ref{IDENT}) with the symmetric one,
$\tilde\partial_\mu\equiv(\partial_\mu+\partial^\star_\mu)/2$.
As in the text, by slightly changing the notation employed in
eqs.~(\ref{IDENT}) and~(\ref{WTI}), we have indicated the parameters
that specify the fermionic action using, besides $r$, the excess
quark mass, $m_q=M_0-M_{\rm{cr}}$, instead of $M_0$. In order to get the
Symanzik expansions in the form given above, use has been made of
the equations of motion and the continuum WTI~(\ref{CWTI}) in the r.h.s.\
of eqs.~(\ref{PPEXPMS}) and~(\ref{APEXPMS}).

As explained in the text (sect.~\ref{sec:IQCD}), the coefficients
$\zeta$, $\xi$ and $\eta$ are all finite functions of $r$ and
$g_0^2$ and are necessary to match the lattice correlators on the
l.h.s.\ to the continuum Green functions on the r.h.s.\ of
eqs.~(\ref{PPEXPMS}) and~(\ref{APEXPMS}). There is an obvious
correspondence between the appearance of the terms with the
coefficients $\xi$ and $\eta$ in front in the above Symanzik
expansions and the need to introduce the coefficients
$c_{\rm{SW}}$, $b_m$, $b_g$ and $c_A$, $b_A$, $b_P$ to improve the
action and the operators $A^b_\mu$, $P^b$.

Even in the limit of vanishing lattice spacing, eqs.~(\ref{PPEXPMS})
and~(\ref{APEXPMS}) have a logarithmically divergent $a$-dependence which
can be easily disposed of by multiplying the two sides of the equations
by the appropriate renormalization constants.

We remark that in the expansions~(\ref{PPEXPMS})
and~(\ref{APEXPMS}), besides correlators of genuine multi-local
operators (``fully localized operators'' in the following), one
has also to include terms where the multi-local operator under
consideration (be it $P^b(x)P^b(0)$ or $\tilde{\partial}_\mu
A^b_\mu(x)P^b(0)$) is inserted together with space-time integrated
local densities. The former are related to the improvement of the
various local factors making up the operator in the l.h.s., while
the latter owe their origin from the need to improve the
action~\cite{LUS1}. Integrated terms require a little discussion,
as the integration variable, $y$, will inevitably hit the points
where the bi-local (in the present instance) operator in
consideration is localized (i.e.\ $x$ and 0). To analyze the
situation one may imagine to split the space-time integral in
three different regions: one around $x$, a second one around 0
plus the rest. Terms coming from integration regions concentrated
around the points $x$ and 0 can be shown, by an analysis based on
the O.P.E., to lead to O($a$) contributions of the same form as
those of the fully localized multi-local operators already present
in the expansion. They only have the effect of redefining the
corresponding Symanzik coefficients. The integration over the
remaining portion of the space-time is finite and gives rise to
genuine O($a$) contributions.

Starting from the expansions~(\ref{PPEXPMS}) and~(\ref{APEXPMS}),
there are two equivalent ways to construct O($a$)
improved lattice quantities, which we now want to make fully explicit:
the Wilson average and the mass average method.

\subsection*{The Wilson average method} \label{sec:WILAV}

Under the change of variables~(\ref{PSIBAR}), we obtain the lattice
formulae (valid for any value of $a$)
\beqn &&\langle P^b(x) P^b(0)\rangle\Big{|}_{(r,M_0)}=
\langle P^b(x) P^b(0)\rangle\Big{|}_{(-r,-
M_0)} \label{IDENTPP}\\ &&\langle \tilde{\partial}_\mu A^b_\mu
(x) P^b(0)\rangle\Big{|}_{(r,M_0)}=  -\langle
\tilde{\partial}_\mu A^b_\mu(x) P^b(0)\rangle\Big{|}_{(-r,-M_0)} \, .
\label{IDENTAP} \eeqn
Using~(\ref{MCR}), these relations can be rewritten, with the same
little abuse of notation as before,  in the more expressive form
\beqn &&\langle P^b(x)
P^b(0)\rangle\Big{|}_{(r,m_q)}=
\langle P^b(x) P^b(0)\rangle\Big{|}_{(-r,-m_q)} \label{IDENTPPM}\\
&&\langle \tilde{\partial}_\mu
A^b_\mu (x) P^b(0)\rangle\Big{|}_{(r,m_q)}=
-\langle\tilde{\partial}_\mu A^b_\mu(x) P^b(0)\rangle\Big{|}_{(-r,-
m_q)}\, . \label{IDENTAPM} \eeqn

To write down the equations that are obtained from the
expansions~(\ref{PPEXPMS}) and~(\ref{APEXPMS}) when both $r$ and $m_q$
change sign we refer to the analysis presented after eq.~(\ref{DIMO})
in sect.~\ref{sec:IQCD}.

For completeness we list below the obvious parities of the
continuum correlators appearing in the r.h.s.\ of eqs.~(\ref{PPEXPMS})
and~(\ref{APEXPMS}) under $m_q\rightarrow -m_q$, which are
implied by the transformation~(\ref{PSIBAR}):
\beqn \langle P^b(x) P^b(0)\rangle\Big{|}^{\rm{cont}}_{(m_q)} &&\quad
\rm{even} \label{PP}\\ \langle P^b(x)P^b(0)\int d^4y\,{\cal{L}}_{\rm{SW}}(y)
\rangle\Big{|}^{\rm{cont}}_{(m_q)} &&\quad \rm{odd}
\label{PPSW}\\ \langle P^b(x)P^b(0)\int d^4y\,
 FF(y)\rangle\Big{|}^{\rm{cont}}_{(m_q)}&&\quad \rm{even}
\label{PPFF}\\ \langle P^b(x)P^b(0)\int
d^4y\, \bar{\psi}\psi(y)\rangle\Big{|}^{\rm{cont}}_{(m_q)}&&\quad \rm{odd}
 \label{PPPSI} \eeqn
\beqn \langle\partial_\mu A^b_\mu(x)
P^b(0)\rangle \Big{|}^{\rm{cont}}_{(m_q)}&&\quad \rm{odd}
\label{AP}\\ \langle\partial_\mu\partial_\mu
P^b(x) P^b(0)\rangle\Big{|}^{\rm{cont}}_{(m_q)}
&&\quad\rm{even}\label{MQPAR}\\
\langle\partial_\mu A^b_\mu(x) P^b(0)
\int d^4y\,{\cal {L}}_{\rm{SW}}(y) \rangle\Big{|}^{\rm{cont}}_{(m_q)}&&\quad
\rm{even} \label{APSW}\\ \langle\partial_\mu A^b_\mu(x)P^b(0)\int
d^4y\, FF(y) \rangle\Big{|}^{\rm{cont}}_{(m_q)}&&\quad \rm{odd}
\label{APFF}\\ \langle\partial_\mu A^b_\mu(x)P^b(0)\int d^4y\,
\bar{\psi}\psi(y) \rangle\Big{|}^{\rm{cont}}_{(m_q)}&&\quad \rm{even}
\label{APPSI} \eeqn
Using equations from~(\ref{PP}) to~(\ref{PPPSI}), the
identity~(\ref{IDENTPPM}) leads to the relations
\beqn &&\hspace{-1cm}\zeta^{PP}_{PP}(r)=\zeta^{PP}_{PP}(-r)\, ,\label{ZPPPP}\\
\nonumber\\
&&\hspace{-1cm}am_q\,[\xi^{PP}_{PP}(r)+\xi^{PP}_{PP}(-r)]
\langle P^b(x)P^b(0)\rangle\Big{|}^{\rm{cont}}_{(m_q)}+\nonumber\\
&&\hspace{-1cm}+a\,[{\eta}^{PP}_{\rm{SW}}(r)+
{\eta}^{PP}_{\rm{SW}}(-r)]\langle P^b(x)P^b(0)
\int d^4y\,{\cal{L}}_{\rm{SW}}(y)
\rangle\Big{|}^{\rm{cont}}_{(m_q)}+\nonumber\\
&&\hspace{-1cm}+ a m_q\,[{\eta}^{PP}_{FF}(r)+{\eta}^{PP}_{FF}(-r)]
\langle P^b(x)P^b(0)\int
d^4y\, FF(y)\rangle\Big{|}^{\rm{cont}}_{(m_q)}+\nonumber\\&&\hspace{-1cm}
+am_q^2\,[{\eta}^{PP}_{\bar{\psi}\psi}(r)+{\eta}^{PP}_{\bar{\psi}\psi}(-r)]
\langle P^b(x)P^b(0)\, \int
d^4y\,\bar{\psi}\psi(y)\rangle\Big{|}^{\rm{cont}}_{(m_q)}=0
\label{PPEXPMSID}\eeqn
from terms of O($a^0$) and O($a$), respectively. Similarly, using equations
from (\ref{AP}) to~(\ref{APPSI}), the identity~(\ref{IDENTAPM}) leads
to the relations
\beqn &&\hspace{-1.5cm}\zeta^{AP}_{AP}(r)=\zeta^{AP}_{AP}(-r)\, ,\label{ZAPAP}
\\\nonumber\\
&&\hspace{-1.5cm}am_q\,[\xi^{AP}_{AP}(r)+\xi^{AP}_{AP}(-r)]
\langle\partial_\mu
A^b_\mu(x)P^b(0)\rangle\Big{|}^{\rm{cont}}_{(m_q)}+\nonumber \\
&&\hspace{-1.5cm}+a\,
[\eta_{PP}^{AP}(r)+\eta_{PP}^{AP}(-r)] \langle\partial_\mu\partial_\mu
P^b(x) P^b(0)\rangle\Big{|}^{\rm{cont}}_{(m_q)}+\nonumber\\
&&\hspace{-1.5cm} + a \,[\eta^{AP}_{\rm{SW}}(r)+\eta^{AP}_{\rm{SW}}(-r)]
\langle\partial_\mu A^b_\mu(x)P^b(0) \int d^4y\,{\cal {L}}_{\rm{SW}}(y)
\rangle\Big{|}^{\rm{cont}}_{(m_q)}+\nonumber\\ &&\hspace{-1.5cm} + a
m_q\,[\eta^{AP}_{FF}(r)+\eta^{AP}_{FF}(-r)] \langle\partial_\mu
A^b_\mu(x)P^b(0)\int d^4y\,
FF(y) \rangle\Big{|}^{\rm{cont}}_{(m_q)}+\nonumber\\&&\hspace{-1.5cm}+am_q^2
\,[\eta^{AP}_{\bar{\psi}\psi}(r)+\eta^{AP}_{\bar{\psi}\psi}(-r)]
\langle\partial_\mu A^b_\mu(x)P^b(0)\int
d^4y\,\bar{\psi}\psi(y) \rangle\Big{|}^{\rm{cont}}_{(m_q)}=0\, .
\label{APEXPMSID}\eeqn
{}From the above four equations one immediately comes to the conclusion
that the average of correlators computed with lattice actions having
Wilson terms of opposite signs and the same value of $m_q=M_0-M_{\rm{cr}}$
are free of O($a$) discretization effects. In formulae one gets, in
fact
\beqn &&\frac{1}{2}\Big{[}\langle
P^b(x)P^b(0)\rangle\Big{|}_{(r,m_q)} + \langle
P^b(x)P^b(0)\rangle\Big{|}_{(-r,m_q)}\Big{]}=\nonumber\\ &&= \;
\zeta^{PP}_{PP}(r)\langle P^b(x)
P^b(0)\rangle\Big{|}^{\rm{cont}}_{(m_q)} +{\rm{O}}(a^2)\, , \label{WAVPP}
\\\nonumber\\
&&\frac{1}{2}\Big{[}\langle\tilde{\partial}_\mu
A^b_\mu(x)P^b(0)\rangle\Big{|}_{(r,m_q)}
+\langle\tilde{\partial}_\mu A^b_\mu(x)P^b(0)
\rangle\Big{|}_{(-r,m_q)}\Big{]}= \nonumber\\&&= \;
\zeta^{AP}_{AP}(r)\langle\partial_\mu
A^b_\mu(x)P^b(0)\rangle\Big{|}^{\rm{cont}}_{(m_q)}+{\rm{O}}(a^2)\, .
\label{WAVAP}\eeqn
Though unnecessary to obtain the improved formulae~(\ref{WAVPP})
and~(\ref{WAVAP}), we observe that the vanishing of the sum of terms in
eqs.~(\ref{PPEXPMSID}) and~(\ref{APEXPMSID}) for arbitrary values of $x$
implies that all the Symanzik coefficients $\eta$ and $\xi$ are odd
functions of $r$.

\subsection*{The mass average method} \label{sec:MASSAV}

An equivalent way of getting O($a$) improved lattice data from standard
Wilson simulations consists in comparing the two sets of formulae that
are obtained by only changing sign to $m_q$ in eqs.~(\ref{PPEXPMS})
and~(\ref{APEXPMS}). Taking into account the $\gamma_5$-chirality properties
of the continuum correlators listed in eqs.~(\ref{PP}) to~(\ref{PPPSI})
and~(\ref{AP}) to~(\ref{APPSI}), it is immediate to verify that lattice
correlators computed with fixed value of $r$ and opposite excess
quark mass, $m_q=M_0-M_{\rm{cr}}$, if linearly combined with a relative
sign equal to the ${\cal{R}}_5$-parity of the multi-local lattice operator
under consideration, yield quantities that are free of O($a$) discretization
effects. In the case at hand one gets
\beqn  &&\frac{1}{2}\Big{[}\langle
P^b(x)P^b(0)\rangle\Big{|}_{(r,m_q)} +
\langle P^b(x)P^b(0)\rangle\Big{|}_{(r,-m_q)}\Big{]}
=\nonumber\\ &&= \; \zeta^{PP}_{PP}(r)\langle P^b(x)
P^b(0)\rangle\Big{|}^{\rm{cont}}_{(m_q)} +{\rm{O}}(a^2)\, , \label{MAVPP}
\\\nonumber\\
&&\frac{1}{2}\Big{[}\langle\tilde{\partial}_\mu A^b_\mu(x)P^b(0)
\rangle\Big{|}_{(r,m_q)}-
\langle\tilde{\partial}_\mu A^b_\mu(x)P^b(0)\rangle\Big{|}_{(r,-m_q)}\Big{]}=
\nonumber\\&&= \; \zeta^{AP}_{AP}(r)\langle\partial_\mu
A^b_\mu(x)P^b(0)\rangle\Big{|}^{\rm{cont}}_{(m_q)}+{\rm{O}}(a^2)\, .
\label{MAVAP}\eeqn
Eqs.~(\ref{MAVPP}) and~(\ref{MAVAP}) are perfectly equivalent to
eqs.~(\ref{WAVPP}) and~(\ref{WAVAP}), respectively, as it is immediately
seen by using the lattice identities
\beqn &&\langle P^b(x) P^b(0)\rangle\Big{|}_{(r,-m_q)}=
\langle P^b(x) P^b(0)\rangle\Big{|}_{(-r,m_q)} \label{IDENTPPR}\\
&&-\langle\tilde{\partial}_\mu A^b_\mu(x) P^b(0)\rangle\Big{|}_{(r,-m_q)}=
\langle \tilde{\partial}_\mu A^b_\mu(x) P^b(0)\rangle\Big{|}_{(-r,m_q)}\, ,
\label{IDENTAPR} \eeqn
which are direct consequences of~(\ref{IDENTPPM}) and~(\ref{IDENTAPM}).

\appendix
\section*{Appendix B - Reflection positivity and transfer matrix}
\renewcommand{\thesection}{B}
\label{APPB}

In the continuum Euclidean theory the Wightman reconstruction
theorem is based on the axiom of reflection
positivity~\cite{OSCH}, i.e.\ on the existence of an anti-linear
operator, $\Theta$, which transforms an arbitrary functional of
the fields, $F_{+}$, with support in a region, $R$, of the
half-space $t>0$ into a functional, $\Theta F_{+}$, with support
in the reflected region, $\theta R$ (with the definition $\theta
x= \theta({\bf{x}},t)=({\bf{x}},-t)$), such that \beq \langle
F_{+}\, \Theta F_{+}\rangle\geq 0\, .\label{POS}\eeq This
condition implies the existence of a positive definite,
self-adjoint Hamiltonian. For a gauge theory the theorem is proved
by choosing the gauge $A_0=0$ (temporal gauge).

A similar theorem holds on the lattice~\cite{OS,IY,MON}. However,
one has to distinguish between link and site reflection
positivity. The associated reflection operations act differently
on lattice points: the former, $\theta_{\ell}$, refers to
reflection across a hyperplane which cuts time links in half
(e.g.\ the links $({\bf x},0)-({\bf x},a)$), while the latter,
$\theta_{s}$, to reflection across a time hyperplane which passes
through sites (e.g.\ the sites $({\bf x},0)$). Explicitly in the
above examples one has \beq
\theta_{\ell}({\bf{x}},t)=({\bf{x}},-t+a)\qquad
\theta_{s}({\bf{x}},t)=({\bf{x}},-t)\, .\label{TETA} \eeq The
validity of both site reflection positivity (eq.~(\ref{POS}) with
$\Theta=\Theta_{s}$) and link reflection positivity
(eq.~(\ref{POS}) with $\Theta=\Theta_{\ell}$) is a sufficient
condition for the existence of an Hermitean positive lattice
(single-link) transfer matrix. Viceversa, if such a transfer
matrix exists, one can prove that both types of reflection
positivity hold. If only one kind of reflection positivity is
valid (e.g.\ link reflection positivity, which typically holds
under weaker conditions on the action parameters than site
reflection positivity) one can still prove the existence of a
positive transfer matrix acting over a time translation of two
lattice sites (to be interpreted as the square of the single-link
transfer matrix). Naturally in the limit $a\rightarrow 0$ the
condition of link (or site) reflection positivity alone is
perfectly sufficient to extract continuum physics from lattice
data.

As in the continuum theory all these statements are proved in the temporal
gauge, which in the lattice corresponds to set all the link variables
associated to temporal links equal to the identity element of the gauge group
(except at one set of links, which can be conveniently chosen to be
$({\bf x},0)-({\bf x},a)$).

For completeness we report the explicit action of the anti-linear reflection
operation on the fundamental fields of the theory. $\Theta_{s}$ and
$\Theta_{\ell}$ act on $\bar\psi(x)$, $\psi(x)$ and $U_\mu(x)$ in the same
way, the only difference is how the arguments they depend on get reflected.
With the subscript $_{s}$ or $_{\ell}$ understood one has~\footnote{In the two
argument notation $U_\mu(x)\equiv U(x,x+a\hat\mu)$ the action of $\Theta$
reads $\Theta [U(x,y)] =U^*(\theta x,\theta y)$, $y=x+a\hat\mu$.}
\beqn
&\Theta[U_0(x)]=U_0^{T}(\theta x - a\hat{0})\, , &\quad
\Theta[U_k(x)]=U_k^{*}(\theta x)\, ,\, k=1,2,3\label{TUUD}\\
&\hspace{-1.cm}
\Theta[\psi(x)]=\bar\psi(\theta x)\gamma_0\, , &\quad
\Theta[\bar\psi(x)]=\gamma_0\psi(\theta x)\label{TPPB}\eeqn
and on monomials of fermionic fields
\beq
\Theta[f(U)\psi(x_1)\ldots\bar\psi(x_n)]=
f^\star(\Theta [U])\Theta[\bar\psi(x_n)]\ldots\Theta[\psi(x_1)]\, ,
\label{TMON}\eeq
$f(U)$ being a functional of link variables only. As usual the definition
\beq
\Phi^\dagger(\theta x)\equiv\Theta[\Phi(x)]
\label{DEFDAG}\eeq
is employed in this work.

\subsection*{Site reflection positivity for Wilson fermions}
\label{sec:RPWF}

In ref.~\cite{MP} it was proved that site reflection positivity holds for
Wilson fermions, provided
\beq
|8r+2aM_0|>6 \, , \quad r=1 \, .\label{RM}
\eeq
The explicit construction of the transfer matrix of the lattice theory was
carried out in ref.~\cite{LUS3}.

The proof of site reflection positivity given in~\cite{MP} for $r=1$
can be extended to $r=-1$ by noting that $\Theta_s$ commutes with
${\cal{R}}_5$. In fact, let $F_+(\bar\psi,\psi,U)$ be a functional of
the field variables defined at positive Euclidean times. Consider the
expectation value $\langle F_+\,\Theta_s F_+\rangle$. The change of
integration variables induced by the transformation~(\ref{PSIBAR})
leads to the relation
\beqn
&&\langle F_+(\bar\psi,\psi,U)\, \Theta_s F_+(\bar\psi,\psi,U)
\rangle\Big{|}_{(-1,M_0)}=\nonumber\\
&&=\langle F_+(-\bar\psi\gamma_5,\gamma_5\psi,U) \,
\Theta_s F_+(-\bar\psi\gamma_5,\gamma_5\psi,U)\rangle\Big{|}_{(1,-M_0)}\, .
\label{POSRONE}\eeqn
Thanks to reflection positivity which holds for $r=1$, one concludes that
the r.h.s.\ of eq.~(\ref{POSRONE}) is non-negative, provided (eq.~(\ref{RM})
with $M_0\rightarrow -M_0$)
\beq
|8-2aM_0|>6\, .\label{RM1}
\eeq
The inequality~(\ref{RM1}) is precisely the condition~(\ref{RM}), only
written for $r=-1$. This shows that Wilson fermions obey site reflection
positivity for both $r=1$ and $r=-1$ under the condition
$|8r+2aM_0|>6\, ,r=\pm 1$. Link reflection positivity was discussed
in sect.~\ref{sec:CRP}.

\subsection*{Site reflection positivity for tm-LQCD}
\label{sec:RPTM}

The situation for tm-LQCD is similar to that of the standard Wilson theory
summarized above. The validity of site reflection positivity was indirectly
proved in~\cite{FSW}, where along the lines of the work of
ref.~\cite{LUS3} the transfer matrix for the action~(\ref{TMQCDCHI})
was constructed under the condition ($\omega=\omega_m+\omega_r$)
\beq
|8r+2aM_{\rm{cr}}(r)+2am_q\cos\omega|>6\, , \quad r=1\, .\label{TMR}
\eeq
{}From this result the validity of site reflection positivity for the
form~(\ref{PHYSCHI}) of the tm-LQCD action can be proved by noting that
$\Theta_{s}$ commutes with the chiral fermion field
transformation~(\ref{PSICHI}). One finds, in fact, under the
condition~(\ref{TMR})
\beqn
&&\langle F_+(\bar\psi_{\rm{ph}},\psi_{\rm{ph}},U)\,
\Theta_s F_+(\bar\psi_{\rm{ph}},\psi_{\rm{ph}},U)
\rangle\Big{|}_{(r,m_q)}^{(\omega_r=\omega,\,\omega_m=0)}=\nonumber\\
&&=\langle F_+(\bar\psi,\psi,U) \,
\Theta_s F_+(\bar\psi,\psi,U)\rangle
\Big{|}_{(r,m_q)}^{(\omega_r=0,\,\omega_m=\omega)}\geq 0
\label{POSTM}\eeqn
with
\beq
\psi=e^{-i\omega\gamma_5\tau_3/2}\psi_{\rm{ph}}\quad
\bar\psi=\bar\psi_{\rm{ph}}e^{-i\omega\gamma_5\tau_3/2}\, ,
\label{CHIRPSI}\eeq
where the last inequality follows from the validity of site reflection
positivity for the action~(\ref{TMQCDCHI}).

\section*{Appendix C - Proof of eq.~(\ref{PPAR})}
\renewcommand{\thesection}{C}
\label{sec:APPC}

In this Appendix we want to show that the $\gamma_5$-chiralities of the
operators appearing in the Symanzik expansion~(\ref{SCHSYM}), or in
the analogous expansion~(\ref{SCHSYMT}) if tm-LQCD is employed,
satisfy the equation
\beq
P_{{\cal{R}}_5}[O]+P_{{\cal{R}}_5}[O_{\ell}]+n_{\ell}
=1\,\,{\mbox{mod}}\,(2)\, .\label{PARAP}\eeq
Eq.~(\ref{PARAP}) is an immediate consequence of the dimensional
relation~(\ref{DIMO}) and the formula
\beq
P_{{\cal{R}}_5}[O]+d[O]=P_{{\cal{R}}_5}[O_\ell]+d[O_\ell]
\,\,{\mbox{mod}\,(2)}\, ,\label{PDPDL}\eeq
which we prove below. {}From these equations one gets, in fact, successively
\beqn
&&P_{{\cal{R}}_5}[O]+P_{{\cal{R}}_5}[O_{\ell}]+n_{\ell} =
P_{{\cal{R}}_5}[O]+d[O]+P_{{\cal{R}}_5}[O_{\ell}]-d[O_{\ell}]+1=\nonumber\\
&&=P_{{\cal{R}}_5}[O]+d[O]-P_{{\cal{R}}_5}[O_{\ell}]-d[O_{\ell}]+1
\,\,{\mbox{mod}\,(2)}=1 \,\,{\mbox{mod}\,(2)}\, .\label{PROVED}\eeqn
In order to prove eq.~(\ref{PDPDL}) we note that with the definition
\beq
\Delta_5[O]=P_{{\cal{R}}_5}[O]+d[O]
\label{DELTO}\eeq
eq.~(\ref{PDPDL}) becomes
\beq
\Delta_5[O]-\Delta_5[O_\ell]=0
\,\, {\mbox{mod}\,(2)}\, .\label{DD}\eeq
Eq.~(\ref{DD}) is a direct consequence of the following symmetry
considerations.

\vskip .2cm
\noindent 1) - The Wilson and tm-LQCD actions (eqs.~(\ref{FERACT})
and~(\ref{PHYSCHI}), respectively), and {\it a fortiori}
the formal continuum action, are invariant under the combined transformation
${\cal{R}}_5\times {\cal{D}}_d$, where~\footnote{In this Appendix,
for notational uniformity, we will always denote the quark fields by
$\psi$ and $\bar \psi$ (as in eq.~(\ref{FERACT})), even when we refer
to the tm-LQCD action~(\ref{PHYSCHI}).}
\begin{equation}
{\cal{D}}_d : \left \{\begin{array}{lll}
U_\mu(x)&\rightarrow U_\mu^\dagger(-x-a\hat\mu) \\\\
\psi(x)&\rightarrow e^{3i\pi/2} \psi(-x)  \\\\
\bar{\psi}(x)&\rightarrow e^{3i\pi/2} \bar{\psi}(-x) \;\, .
\end{array}\right . \label{FIELDT} \end{equation}
The invariance of the pure gauge part of the action is obvious. For the
fermionic part invariance follows by first noting that under ${\cal{D}}_d$
(recall the definitions~(\ref{LATCOVDER}))
\begin{eqnarray}
&& \bar\psi(x) \sum_\mu \gamma_\mu (\nabla_\mu + \nabla_\mu^*) \psi(x)  \,
\rightarrow \,
\bar\psi(-x) \sum_\mu \gamma_\mu (\nabla_\mu + \nabla_\mu^*) \psi(-x) \;\, ,
 \nonumber \\
&& \bar\psi(x) \sum_\mu \nabla_\mu^* \nabla_\mu \psi(x) \,
\rightarrow \,
 - \, \bar\psi(-x) \sum_\mu \nabla_\mu^* \nabla_\mu \psi(-x) \;\, ,
 \nonumber \\
&& \bar\psi(x) \psi(x) \, \rightarrow \, - \, \bar\psi(-x) \psi(-x)
 \;\, .
\label{D_FERM_ACT}
\end{eqnarray}
The minus sign in front of the chirally non-invariant terms is compensated
by the transformation ${\cal{R}}_5$. Finally the space-time arguments are
brought back to their initial values by the trivial change of space-time
integration variables $-x \rightarrow x$ in the action. In the formal
continuum theory an analogous argument goes through. In fact, the continuum
version of the transformation ${\cal{D}}_d$, besides changing the sign of all
the space-time coordinates, has simply the effect of multiplying each local
term ${\cal{L}}_i$ in the action density by the factor
$(-1)^{d[{\cal{L}}_i]}$, where $d[{\cal{L}}_i]$ is the naive dimension
of ${\cal{L}}_i$.

\vskip .2cm
\noindent 2) - Performing the change of functional integration variables
induced by the transformation ${\cal{R}}_5\times {\cal{D}}_d$, one can
prove that the lattice correlators of the type we are interested in obey
the formula
\beq
\langle O(x_1,x_2,\ldots ,x_n)\rangle
\Big{|}_{(r,m_q)}^{(\omega)}=(-1)^{\Delta_5[O]}
\langle O(-x_{1},-x_{2},\ldots ,-x_{n})\rangle\Big{|}_{(r,m_q)}^{(\omega)}
\label{OPTRA}
\eeq
with $\omega=0$ in the case of the standard Wilson action~(\ref{FERACT}).
The relation~(\ref{OPTRA}) directly follows from our definition of $\Delta_5$
(eq.~(\ref{DELTO})) and the way ${\cal{D}}_d$ (eq.~(\ref{FIELDT})) acts on
the fundamental fields in $O$.

Similarly, all continuum operators contributing to the Symanzik
expansions~(\ref{SCHSYM}) or~(\ref{SCHSYMT}) of the expectation value of
$O(x_1,x_2,\ldots ,x_n)$ must obey eq.~(\ref{OPTRA}) with a value of
$\Delta_5$ equal (mod\,(2)) to that of $O$. In other words all terms in the
Symanzik expansion of $O$ must have the same $(-1)^{\Delta_5}$-parity as $O$.
This is just our thesis, eq.~(\ref{DD}).

\appendix
\section*{Appendix D - Discretization effects in the determination of
$M_{\rm{cr}}$}
\renewcommand{\thesection}{D}
\label{sec:APPD}

In this Appendix we want to show that O($a$) discretization effects
in the determination of $M_{\rm{cr}}(r)$ do not jeopardize the proof
of O($a$) improvement of averaged correlators given in the text.
We give the proof in the case of standard Wilson fermions. The generalization
to tm-LQCD is straightforward.

\subsection*{The problem}
\label{sec:OADE}

In PT the analytic dependence of lattice correlators on $a$ is
completely and explicitly computable. Thus an unambiguous
expression of $M_{\rm{cr}}(r)$ can be given to any order in
$g^2_0$. The situation is different non-perturbatively. An
estimate of $M_{\rm{cr}}(r)$ can be obtained by using, for
instance, PCAC or the observation that (in the presence of
S$\chi$SB) the pion mass is expected to vanishes at the chiral
point (modulo cutoff effects, see sect.~\ref{sec:CRMASSB}).
If the
the pion mass is employed to determine $M_{\rm{cr}}(r)$,
one has to proceed by extracting from the large $t$ behaviour of
the two-point correlator~\footnote{In this appendix we label
lattice expectation values by using, besides $r$, the bare quark
mass. As in the rest of the paper, the ubiquitous $g_0^2$
dependence is understood.} \beq G_{\pi}(t;r,M_0)=a^3\sum_{\rm{\bf
x}} \langle P^b({\bf x},t) P^b(0)\rangle\Big{|}_{(r,M_0)}  \, ,
\label{P5P5A} \eeq a lattice measurement of the pion mass,
$m_\pi(r,M_0)$. A sensible estimate of the critical mass, which
will be referred to as $M^{e_1}_{\rm{cr}}(r)$ and obeys \beq
M^{e_1}_{\rm{cr}}(r)=-M^{e_1}_{\rm{cr}}(-r)\, , \label{MCRHAT}
\eeq 
can thus be obtained as discussed in sect.~\ref{sec:CRMASSB}.

If PCAC (see e.g.\ eq.~(\ref{WTI})) is employed to measure
$M_{\rm{cr}}(r)$, one would get another, equally good,
non-perturbative estimate of it, say $M^{e_2}_{\rm{cr}}$, given by
the solution of the equation \beq \label{critmassPCAC}
\bar{M}(r,M^{e_2}_{\rm{cr}}(r)) - M^{e_2}_{\rm{cr}}(r) =0 \, .
\eeq The new estimate is again odd in $r$ (see eq.~(\ref{MBAR})).

For the sake of this argument we may summarize the situation by
formally writing for any such estimate \beq \label{DELDE}
M^{e}_{\rm{cr}}(r) = M_{\rm{cr}}(r) +\delta^{e}(r)\, , \eeq with
\beq \delta^{e}(r)=-\delta^{e}(-r)={\rm{O}}(a)\, . \label{DELHAT}
\eeq The difference $M_{\rm{cr}}(r)-M^{e}_{\rm{cr}}(r)$ is
obviously odd under $r\rightarrow -r$, because $M_{\rm{cr}}(r)$ as
well as anyone of its estimates is odd in $r$ (see
sect.~\ref{sec:SF}).

The problem we want to address is whether O($a$) ambiguities in the
knowledge of the critical mass can invalidate the argument for O($a$)
improvement developed in sect.~\ref{sec:IQCD} leading to eqs.~(\ref{WILAV})
and~(\ref{MASSAV}) (or to the similar ones valid in the case of tm-LQCD).
We show below that eqs.~(\ref{DELDE}) and~(\ref{DELHAT}) are indeed all is
needed for eqs.~(\ref{WILAV}) and~(\ref{MASSAV}) to be valid.

This result should not come as a surprise, since effectively we are here
merely claiming that all the O($a$) cutoff effects, which generically are
present before averaging, get canceled in $WA$'s of correlators and derived
quantities. Among these cutoff effects there are, indeed, also those arising
from O($a$) errors on $M_{\rm{cr}}(r)$. They are not in any sense special and
actually stem from the presence of O($a$) cutoff effects
in the correlators employed to estimate the value of the critical mass.

\subsection*{The proof}
\label{sec:ARG}

Once the values of $r$ ($>0$), $g_0^2$ and the bare lattice mass
$M_0=M_0^{+}$ are chosen, the Symanzik expansion of the
expectation value of the operator $O$ takes the form (see
eq.~(\ref{SCHSYM}) in sect.~\ref{sec:IQCD}) \beqn &&\langle O
\rangle\Big{|}_{(r,M_0^+)} =
[\zeta^{O}_{O}(r)+am_q^+\xi^{O}_{O}(r)]\langle O \rangle
\Big{|}^{\rm{cont}}_{(m_q^+)}+
\nonumber\\
&&+ a\,\sum_{\ell}(m_q^+)^{n_{\ell}}\eta^{O}_{O_{\ell}}(r)
\langle O_{\ell}\rangle\Big{|}^{\rm{cont}}_{(m_q^+)}+{\rm{O}}(a^2)\, ,
\label{SCHSYMP}\eeqn
where
\beq
m_q^+\equiv M^+_0-M_{\rm{cr}}(r)
\label{MQP}\eeq
is the continuum mass at the subtraction point $a^{-1}$. $m_q^+$ is defined
in terms of the (unknown) value of the critical mass, $M_{\rm{cr}}(r)$.

We want to average this expectation value with the similar expression
obtained by changing sign to $r$ and adjusting $M_0$ to a suitable value,
$M_0\rightarrow M_0^{-}$, such that all O($a$) terms cancel in the sum.
We have shown in sect.~\ref{sec:IQCD} that this is indeed possible with
the choice
\beq
M_0^- -M_{\rm{cr}}(-r)=M_0^+ -M_{\rm{cr}}(r)=m_q^+\, .
\label{MPMM}\eeq
The ``practical'' problem with this prescription is that, as we just
recalled, one can only get an ``estimate'', $M^{e}_{\rm{cr}}(r)$, of
the critical mass with the properties summarized in eqs.~(\ref{DELDE})
and~(\ref{DELHAT}).

Now the question is: given $M_0^+$, which value of the bare lattice mass
do we have to take in practice, when we go over to $-r$, in order to get
rid of O($a$) terms?

Clearly the only possible answer (and, as we shall see, the right one)
is to consider the expectation value of $O$ computed with $-r$ and a
bare mass, ${M'}^-_0$, given by
\beq
{M'}^-_0\equiv M_0^+ -M^{e}_{\rm{cr}}(r)+M^{e}_{\rm{cr}}(-r)
=M_0^+ -2 M^{e}_{\rm{cr}}(r)\, ,
\label{HATMIN}\eeq
where the last equality follows from the $r$-parity properties of
$M^{e}_{\rm{cr}}(r)$. The Symanzik expansion of $\langle O
\rangle |_{(-r,{M'}_0^-)}$ takes the form
\beqn
&&\langle O \rangle\Big{|}_{(-r,{M'}_0^-)} =
[\zeta^{O}_{O}(-r)+am_q^-\xi^{O}_{O}(-r)]\langle O \rangle
\Big{|}^{\rm{cont}}_{(m_q^-)}+
\nonumber\\
&&+ a\,\sum_{\ell}(m_q^-)^{n_{\ell}} \eta^{O}_{O_{\ell}}(-r)
\langle
O_{\ell}\rangle\Big{|}^{\rm{cont}}_{(m_q^-)}+{\rm{O}}(a^2)\, ,
\label{SCHSYMM}\eeqn with the definition \beqn
m_q^-={M'}_0^--M_{\rm{cr}}(-r)\, . \label{MQMDEF}\eeqn For future
use we note the chain of equalities \beqn
&&m_q^-={M'}_0^--M_{\rm{cr}}(-r)=M_0^+ -2
M^{e}_{\rm{cr}}(r)-M_{\rm{cr}}(-r)= \nonumber\\&&=M_0^+  -2
M^{e}_{\rm{cr}}(r)+M_{\rm{cr}}(r) =M_0^+
-M_{\rm{cr}}(r)-2\delta^{e}(r)=\nonumber\\&&=m_q^+-2\delta^{e}(r)\,
, \label{MQMIN}\eeqn where eqs.~(\ref{HATMIN}), (\ref{DELDE})
and~(\ref{DELHAT}) have been employed. 

Clearly the O($a$) mismatch between $M_0^-$ (eq.~(\ref{MPMM})) and
${M'}_0^-$ (eq.~(\ref{HATMIN})), as well as that between $m_q^+$ and $m_q^-$
matters only for the first term of the expansions~(\ref{SCHSYMP})
and~(\ref{SCHSYMM}), because all the other terms have an explicit factor of
$a$ in front of them. Thus summing~(\ref{SCHSYMP}) and~(\ref{SCHSYMM}) all
O($a$) contributions that are multiplied by the ($r$ odd) Symanzik
coefficients $\xi$ and $\eta$ cancel and we get (see eq.~(\ref{ZETA}))
\beq
\langle O \rangle\Big{|}_{(r,M_0^+)}+
\langle O \rangle\Big{|}_{(-r,{M'}_0^-)}=
\zeta^{O}_{O}(r)
\Big{[}\langle O \rangle\Big{|}^{\rm{cont}}_{(m_q^+)}
+\langle O \rangle\Big{|}^{\rm{cont}}_{(m_q^-)}\Big{]}
+{\rm{O}}(a^2)\, .
\label{WILAPP}
\eeq
At this point, recalling the expressions~(\ref{MQP}) of $m_q^+$
and~(\ref{MQMIN}) of $m_q^-$, we can put the r.h.s.\ of the previous
equation in the form
\beqn
\hspace{-1.cm}&&\zeta^{O}_{O}(r)
\Big{[}\langle O \rangle\Big{|}^{\rm{cont}}_{(m_q^+)}+\langle O
\rangle\Big{|}^{\rm{cont}}_{(m_q^-)}\Big{]}+{\rm{O}}(a^2)=\nonumber\\
\hspace{-1.cm}&&=\zeta^{O}_{O}(r)
\Big{[}\langle O \rangle\Big{|}^{\rm{cont}}_{(M_0^+-M_{\rm{cr}}(r))}
+\langle O \rangle\Big{|}
^{\rm{cont}}_{(M_0^+-M_{\rm{cr}}(r)-2\delta^{e}(r))}\Big{]}+{\rm{O}}(a^2)\, .
\label{WILNEW}
\eeqn
In this expression the two continuum correlators are evaluated at
``slightly'' different values of the excess quark mass, namely
$M_0^+-M_{\rm{cr}}(r)$ and $M_0^+-M_{\rm{cr}}(r)-2\delta^{e}(r)$. Neglecting
O($a^2$) terms, they can be immediately brought to the common value of the
mass parameter
\beq
m_q=M_0^+-M_{\rm{cr}}(r)-\delta^{e}(r)=M_0^+-M^{e}_{\rm{cr}}(r)
= {M'}_0^- - M_{\rm cr}^e(-r)
\label{PHYSMA}\eeq
by a simple Taylor expansion of the continuum correlators in powers of
$\delta^{e}(r)$. The terms with the mass derivative of continuum correlators
cancel out and one gets
\beq
\frac{1}{2}\Big{[}\langle O \rangle\Big{|}_{(r,m_q+M^{e}_{\rm{cr}}(r))}+
\langle O \rangle\Big{|}_{(-r,m_q-M^{e}_{\rm{cr}}(r))}\Big{]}=
\zeta^{O}_{O}(r)\langle O \rangle\Big{|}^{\rm{cont}}_{(m_q)}
+{\rm{O}}(a^2)\, ,
\label{FINFOR}
\eeq
where only well defined, numerically accessible inputs, like
$M^{e}_{\rm{cr}}(r)$ and the excess quark mass, $m_q$, appear.

We end by noticing that, had we used the excess quark mass to label lattice
correlators (as we did in the text after sect.~\ref{sec:IQCD}), the
formula~(\ref{FINFOR}) would have actually read
\beq
\frac{1}{2}\Big{[}\langle O \rangle\Big{|}_{(r,m_q)}+
\langle O \rangle\Big{|}_{(-r,m_q)}\Big{]}=
\zeta^{O}_{O}(r)\langle O \rangle\Big{|}^{\rm{cont}}_{(m_q)}
+{\rm{O}}(a^2)\, .
\label{FINFOR1}
\eeq
The result of this elaborated discussion is that O($a$) improved
correlators are obtained using the formula~(\ref{FINFOR1}) - which is
identical to the $WA$ prescription~(\ref{WILAV}) - with $m_q$ given by
eq.~(\ref{PHYSMA}).

\appendix
\section*{Appendix E - Improvement of the baryonic spectrum}
\renewcommand{\thesection}{E}
\label{APPE}

In this Appendix we want to give a quick argument to show that the
procedure for O($a$) improvement proposed in this paper
(eq.~(\ref{MASSIMP}) and its tm-LQCD counterpart) works also for the baryonic
sector of the hadronic spectrum. For concreteness we will concentrate on the
spin 1/2 and 3/2 states that can be formed with an $SU(2)_{\rm{f}}$ flavour
doublet of quarks.

First of all we note that all baryonic energies can be extracted from spin
averaged two-point correlators of the type~(\ref{HHFT}) and~(\ref{HHFTK}),
for the untwisted and twisted Wilson formulations, respectively. The (bare)
interpolating operators for spin 1/2 and 3/2 baryonic states are of the form
\beq
[\Phi_{h,\alpha}^{f}](x) \; = \; [\epsilon_{ijl}
\psi_{\alpha,i}^f {\cal M}^{jl}_{h}](x) \, ,
\label{BIO}
\eeq
where $f$ and $\alpha$ are free flavour and Dirac indices,
$i,j,l$ represent (contracted) color indices and ${\cal M}^{jl}_{h}$
is a bosonic operator with definite transformation properties
under ${\cal R}_5$. For instance, for proton ($f=1$, $\psi^1 \equiv u$)
or neutron ($f=2$, $\psi^2 \equiv d$) states we can take
\beq \label{BNUCL}
{\cal M}^{jl}_{p,n}\; =\,i\bar \psi^{\rm c}_j\,\gamma_5\tau_2\,\psi_l
\eeq
with $\bar\psi^{\rm c}$ denoting the charge conjugated of $\bar\psi$.
Similarly, the (bare) interpolating operator for the spin 3/2 baryon
$\Delta^{++}$ is of the form~(\ref{BIO}) with $f=1$ and
\beq \label{BDELT}
{\cal M}^{jl}_{\Delta, k}\,=\, \bar\psi^{\rm c}_j\,{1\over 2}
(1+\tau_3) \gamma_k \, \psi_l\, , \quad\quad\quad k=1,2,3 \, .
\eeq
In all cases the bi-local operator $\Phi_h(x)\Phi_h^\dagger(0)$ (with all
Dirac indices contracted) has definite (positive) ${\cal R}_5$-parity.
Since this property is preserved also in presence of mixing, we can apply
the considerations of sect.~\ref{sec:HSME} (or the analogous ones valid
for tm-LQCD) and conclude that also baryonic masses are improved by the
usual $WA$ procedure.

Moreover, in the tm-LQCD formulation with action~(\ref{PHYSCHI}) the
masses of the baryonic states~(\ref{BIO}) are even functions of the
twisting angle $\omega$, hence automatically free of O($a$) corrections
if $\omega = \pm \pi/2$. To prove this result it is sufficient to check
that the argument presented in sect.~\ref{sec:OPTR} for the $\omega$-parity
of masses applies to the states under consideration here. This
is actually so, because the interpolating operators~(\ref{BIO}) are such
that the spin averaged product $\Phi_h(x)\Phi_h^\dagger(0)$ has obviously a
definite positive formal parity (eq.~(\ref{OPARGEN})) and this property is
preserved under renormalization. Incidentally, $\Phi_h$ itself
could be chosen to have definite formal parity, if one decides to
multiply the operator in the r.h.s.\ of~(\ref{BIO}) by the projectors
$(1\pm\gamma_0)/2$.

\appendix
\section*{Appendix F - Proof of eq.~(\ref{PARSTATT})}
\renewcommand{\thesection}{F}
\label{APPF}

A detailed proof of eq.~(\ref{PARSTATT}) requires repeating in the case of
tm-LQCD the line of arguments that in a parity invariant theory leads from
the way ${\cal{P}}$ acts on the elementary fields entering the Lagrangian
density (see eq.~(\ref{PAROP})) to the notion of parity for the eigenstates
of the Hamiltonian. Observing that in the case at hand
(action~(\ref{PHYSCHI})) the spurionic symmetry
${\cal{P}}^{\rm{sp}}\equiv{\cal{P}}\times (\omega\to -\omega)$
($({\cal{P}}^{\rm{sp}})^2=1\!\!1$) replaces ${\cal{P}}$, one can proceed
as follows.

1) To composite bare operators formal parity labels are attributed according
to the naive prescription induced by the field transformation~(\ref{PAROP}).

2) A basis of m.r.\ operators can always be constructed, consisting of
operators endowed with a definite parity label. In fact, to
comply with the symmetry ${\cal{P}}^{\rm{sp}}$ the mixing pattern
will have to be such that (bare) operators with opposite formal parity
get mixed with coefficients whose relative $\omega$-parity is negative.

This procedure is equivalent to assign a parity label via eq.~(\ref{OPARGEN}).
If one wishes to employ this method, one should start by assigning parity to
local operators (e.g.\ quark bilinears) or simple products of local operators
with non-zero (for suitable choices of the space-time arguments) vacuum
expectation value and then move on to progressively more complicated cases.

The situation here is exactly analogous to what happens with the
spurionic symmetry ${\cal{R}}_5^{\rm{sp}}$, where eq.~(\ref{KEYREL})
is exploited to attribute a definite $\gamma_5$-chirality,
$P_{{\cal{R}}_5}[O]$, actually the naive one, to any (gauge invariant)
m.r.\ composite operator, $O$.

3) For the vacuum state we have in general
\beq
\widehat{\cal{P}} |\Omega\>\Big{|}_{(r,m_q)}^{(\omega)} =
\eta_{\Omega}(\omega) |\Omega\>\Big{|}_{(r,m_q)}^{(-\omega)} \, ,
\label{PARVAC}
\eeq
with $\eta_{\Omega}(\omega)\eta_{\Omega}(-\omega)=1$ (from
$\widehat{\cal{P}}^2=1\!\!1$~\cite{WEIN}) and $|\eta_{\Omega}(\omega)|^2=1$
(from the vacuum normalization condition). In view of these results
one can always decide to simply take the phase of the vacuum state such
that $\eta_{\Omega}(\omega)=1$ for any value of $\omega$.
If one does not want to do so, one will have to keep track of the vacuum
phase in all the forthcoming formulae.

4) A basis for the states of the Fock space is constructed by applying
creation operators to the vacuum (cyclic vector). The elements of the basis
are highly non-trivial and schematically can be written in the form
\beq
|h,{\bf k};\alpha\>_{(r,m_q)}^{(\omega)}=
a_{h\alpha}^\dagger({\bf k}) |\Omega\>\Big{|}_{(r,m_q)}^{(\omega)}\, ,
\label{ONEPAR}\eeq
where, according to 1) and 2), the creation operators
$a_{h\alpha}^\dagger({\bf k})$ ($\alpha$ is an index which labels the
states with the same unbroken quantum numbers $h$ and three-momenta
${\bf k}$) will satisfy the relation
\beq
\widehat{\cal{P}}a_{h\alpha}^\dagger({\bf k})\widehat{\cal{P}} =
\eta_{h\alpha}a_{h\alpha}^\dagger(-{\bf k}) \, .
\label{OPERPAR}\eeq
where $\eta_{h\alpha}$ ($ = \pm 1$) is by construction an
$\omega$- and ${\bf{k}}$-independent label.

5) The eigenstates of the transfer matrix $\widehat{T}(\omega,r,m_q)$
(see eq.~(\ref{EVTM})) will be expressed as linear combinations of the
states~(\ref{ONEPAR}) of the kind
\beqn
&&|h,n,{\bf k}\>\Big{|}^{(\omega)}_{(r,m_q)} = \sum_\alpha
c_{n\alpha}(\omega,r,m_q)|h,{\bf k};\alpha\>_{(r,m_q)}^{(\omega)}=\nonumber\\
&&=\sum_\alpha c_{n\alpha}(\omega,r,m_q)a_{h\alpha}^\dagger({\bf k})
|\Omega\>\Big{|}_{(r,m_q)}^{(\omega)}\, ,
\label{ONEPAREIG}
\eeqn
with appropriate coefficients, $c_{n\alpha}(\omega, r, m_q)$, to ensure
normalizability of the state and consistency with the spurionic invariance
${\cal{P}}^{\rm{sp}}$.

In particular, consistency with the spurionic symmetry ${\cal{P}}^{\rm{sp}}$
implies the operator relation
\beq
\widehat{\cal{P}}\,\Big{[}|h,n,{\bf k}\>
\<h,n,{\bf k}|\Big{]}_{(r,m_q)}^{(\omega)}\,\widehat{\cal{P}}=
\Big{[}|h,n,-{\bf k}\>\<h,n,-{\bf k}|\Big{]}_{(r,m_q)}^{(-\omega)}\, ,
\label{SATT}\eeq
which is proved below in subsect.~\ref{sec:PROO}. This relation can only
hold if the coefficients $c_{n\alpha}(\omega,r,m_q)$ have (up to an
irrelevant $\alpha$-independent phase, which we set equal to
unity~\footnote{This phase can always be readsorbed in a physically
unconsequential redefinition of the overall phase of the state
$|h,n,{\bf k}\>|^{(\omega)}_{(r,m_q)}$. See also the comment at the end
of this section.}) a definite parity, $s_{n\alpha}=\pm 1$,
under $\omega \to -\omega$, i.e.\ only if they satisfy the relation
\beq
c_{n\alpha}(\omega,r,m_q)=s_{n\alpha}c_{n\alpha}(-\omega,r,m_q)
\label{COEFFO}\eeq
with
\beq
\eta_{h,n} \equiv  \eta_{h\alpha} s_{n\alpha}
\label{ETAHN}\eeq
an $\alpha$-independent label.

6) For the states~(\ref{ONEPAREIG}) the relation
\beq
\widehat{\cal{P}}|h,n,{\bf k}\>_{(r,m_q)}^{(\omega)}=
\eta_{h,n}|h,n,-{\bf k}\>_{(r,m_q)}^{(-\omega)} \, ,
\label{PARSTATDEF}
\eeq
is immediately seen to be a consequence of eqs.~(\ref{OPERPAR}),
(\ref{COEFFO}) and~(\ref{ETAHN}). By construction (see eq.~(\ref{ETAHN}))
$\eta_{h,n}$ can only take the values $\pm 1$. The
relation~(\ref{PARSTATDEF}) is nothing but eq.~(\ref{PARSTATT})
of sect.~\ref{sec:PTME}.

Whether the parity label $\eta_{h,n}$ is $1$ or $-1$ can be decided
numerically by the method discussed in sect.~\ref{sec:DWPC}. The resulting
value of $\eta_{h,n}$ - at least for values of the bare parameters for which
quantities of O($a^2$) and O($a^0$) can be unambiguously distinguished -
is to be identified with the parity quantum number of the state
corresponding to $|h,n,{\bf k}\>_{(r,m_q)}$ in the continuum limit.
By construction $\eta_{h,n}$ is independent of the three-momentum
of the state, but does depend on the excitation level $n$ because ${\cal P}$
is not an exact symmetry of tm-LQCD. There may exist, in fact, states with
opposite parities but the same unbroken tm-LQCD quantum numbers.
For instance in the vacuum sector one encounters the vacuum itself
(with $\eta_\Omega=1$), the neutral pion (with  $\eta_{\pi^0}=-1$), etc.

We observe in conclusion that one could decide to separately alter
the phases of the states $|h,n,{\bf k}\>_{(r,m_q)}^{(\omega)}$ and
$|h,n,-{\bf k}\>_{(r,m_q)}^{(-\omega)}$ with respect to what one gets
from the above construction. If one does so, then track of
this modification should be kept in all the relevant equations, in
particular in eq.~(\ref{PARSTATDEF}). The only consequence of this
choice is that all lattice formulae relating quantities with opposite
values of $\omega$, or with values of $\omega$ differing by $\pi$,
become unnaturally and unnecessarily cumbersome.

\subsection{Proving eq.~(\ref{SATT})}
\label{sec:PROO}

Let us finally come to the proof of eq.~(\ref{SATT}). For this purpose
it is convenient to consider the general two-point correlator
\beq
G_{\Phi_h\Xi_h}({\bf k},t;\omega,r,m_q)=
a^3\sum_{\rm{\bf x}}e^{i{\bf{k\cdot x}}}
\langle \Phi_h({\bf x},t)\,
\Xi_h^{\dagger}({\bf 0},0)\rangle\Big{|}_{(r,m_q)}^{(\omega)}
\, , \label{PXFTK}
\eeq
where $\Phi_h$ and $\Xi_h$ are two arbitrary local operators with
formal parities $\eta_{\Phi_h}$ and $\eta_{\Xi_h}$ and equal
unnbroken quantum numbers $h$. Performing the change of variables induced
by~(\ref{PAROP}), one obtains
\beq
G_{\Phi_h\Xi_h}({\bf k},t;\omega,r,m_q)=\eta_{\Phi_h}\eta_{\Xi_h}
G_{\Phi_h\Xi_h}(-{\bf k},t;-\omega,r,m_q)\, .
\label{PHXI}\eeq
By comparing the spectral representations of the two sides of eq.~(\ref{PHXI})
and using eq.~(\ref{EPOMKAP}), we can get the set of relations
\beqn
&\hspace{-0.8cm} \<\Omega|\Phi_h({\bf 0},0)|h,n,{\bf k}\>
\<h,n,{\bf k}|\Xi_h^{\dagger}({\bf 0},0)|\Omega\>
\Big{|}_{(r,m_q)}^{(\omega)}=\nonumber \\
&\hspace{-0.8cm} =\eta_{\Phi_h}\eta_{\Xi_h}
\<\Omega|\Phi_h({\bf 0},0)|h,n,-{\bf k}\>
\<h,n,-{\bf k}|\Xi_h^{\dagger}({\bf 0},0)|\Omega\>
\Big{|}_{(r,m_q)}^{(-\omega)}\, ,\label{RNPMKM}\eeqn
thanks to the leverage we have from the dependence of $G_{\Phi_h\Xi_h}$
on $t$. Inserting factors $\widehat{\cal{P}}^2=1\!\!1$ on the l.h.s.\
of eq.~(\ref{RNPMKM}) yields
\beqn
&\hspace{-0.8cm} \<\Omega| \widehat{\cal{P}} \, \Phi_h({\bf 0},0)
\widehat{\cal{P}} |h,n,{\bf k}\>
\<h,n,{\bf k}| \widehat{\cal{P}} \, \Xi_h^{\dagger}({\bf 0},0)
\widehat{\cal{P}} |\Omega\>\Big{|}_{(r,m_q)}^{(\omega)} =
\nonumber \\
&\hspace{-0.8cm} = \<\Omega|\Phi_h({\bf 0},0)|h,n,-{\bf k}\>
\<h,n,-{\bf k}|\Xi_h^{\dagger}({\bf 0},0)|\Omega\>
\Big{|}_{(r,m_q)}^{(-\omega)}\, .\label{RNPMKM_2}\eeqn
If we now combine eq.~(\ref{PARVAC}) with eq.~(\ref{RNPMKM_2})
and exploit the arbitrariness of the operators $\Phi_h$ and $\Xi_h$,
we arrive at the relation~(\ref{SATT}).

\end{document}